\documentclass{aa}  
\usepackage[breaklinks=true,colorlinks,citecolor=blue]{hyperref}
\usepackage{graphicx}
\usepackage{enumerate}
\usepackage{epsfig}
\usepackage{color}
\usepackage{txfonts}
\usepackage{natbib}
\usepackage{multirow}
\usepackage{ulem} 
\usepackage{media9}

\usepackage{hyperref}
\definecolor{ao}{rgb}{0.0, 0.5, 0.0}

\newcommand{\bea}	{\begin{array}}
\newcommand{\eea}	{\end{array}}
\newcommand{\beq}	{\begin{equation}}
\newcommand{\eeq}	{\end{equation}}
\newcommand{\ben}	{\begin{eqnarray}}
\newcommand{\een}	{\end{eqnarray}}
\newcommand{\bsq}	{\begin{mathletters}}
\newcommand{\esq}	{\end{mathletters}}

\begin{document}

%\title{In-situ mergers of thick disc Milky Way globular clusters: I. Mergers, tidal interactions, mass exchange}
\title{Mergers, tidal interactions, and mass exchange in a population of disc globular clusters: II. Long-term evolution}

\titlerunning{Globular clusters mergers}

\author{Alessandra Mastrobuono-Battisti$^{1}$, Sergey Khoperskov$^{2}$, Paola Di Matteo$^{2}$, Misha Haywood$^{2}$}
\authorrunning{A. Mastrobuono-Battisti et al.}

\institute{Max-Planck-Institut f\"ur Astronomie, K\"onigsthul 17, D-69117, Heidelberg, Germany \and GEPI, Observatoire de Paris, PSL Research University, CNRS, Place Jules Janssen, F-92195 Meudon Cedex, France}

\date{Received ; accepted }
 
\abstract{
Globular clusters (GCs), the oldest stellar systems observed in the Milky Way, 
have for long been considered single stellar populations.
As such, they provided an ideal laboratory to understand stellar dynamics and primordial star formation processes. 
However, during the last two decades, observations unveiled their real, complex nature. 
Beside their pristine stars, GCs host one or more helium enriched and possibly younger stellar populations whose
formation mechanism is still unknown.
Even more puzzling is the existence of GCs showing star by star iron spreads. 
%This property can be explained if the stars formed from differently metal enriched gas.
Using detailed $N$-body simulations we explore the hypothesis that these anomalies in metallicity could be the result of
mutual stripping and mergers between a primordial population of disc GCs. 
In the first paper of this series we proved, both with analytical arguments and short-term $N$-body simulations, that disc GCs have larger fly-by and close encounter rates with respect to halo clusters. 
These interactions lead to mass exchange and even mergers that form new GCs, possibly showing metallicity spreads. 
Here, by means of long-term direct $N$-body simulations, we provide predictions on the dynamical properties of GCs that underwent these processes. 
The comparison of our predictions with available and future observational data could provide insights on the origin of GCs and on the Milky Way build-up history as a whole. 
%
%In the first paper of this series we proved, both with analytical arguments and short-term $N$-body simulations, that GCs confined in this region of the Galaxy have larger fly-by and close encounter rates with respect to halo clusters. 
%These interactions lead to mass exchange and even mergers that form new GCs, possibly showing metallicity spreads. 
%Here, by means of long-term simulations, we provide predictions on the dynamical properties of massive GCs that underwent mergers or strippings. 
%The comparison of these predictions with available and future observations (e.g. Gaia, MOONS) will finally shed light on the origin of 
%GCs and on the Milky Way build-up history as a whole.  
}


\maketitle

%%%%%%%%%%%%%%%%% BODY OF PAPER %%%%%%%%%%%%%%%%%%

\section{Introduction}
Globular clusters (GCs) have been for long described as simple, monolithic objects,
composed by stars born at the same time with the same chemical composition.
This picture changed when high resolution spectroscopic and photometric data unveiled the presence of multiple stellar populations in almost every observed GC
\citep[see e.g.][]{Gra04, Car07, Kay08, Car09a, Car10, Pan10, Mil10, Mil12, Mil13, Gra12, Car15}.
The presence of multiple populations
with star by star abundance variations in light elements appears to be ubiquitous in Galactic and extragalactic GCs \citep{Gra12}.
However, a small fraction of Galactic, massive GCs are even more peculiar, showing a significant spread in iron content \citep[see Table 10 in ][for a partial summary]{Mar15}.
The first GC where this spread has been observed is $\omega$Cen \citep{Nor95}, a cluster strongly suspected to be
the remnant nucleus of a disrupted dwarf galaxy \citep{Fr93, Di99, Hu00,BF03, Bo08}. 
{ While the iron spread found in some clusters \citep[M22, ][]{HHM77, Ma09, Ma11, Ma12, Ma13, Lee15} is still controversial \citep{Muc15, Lee16},  the latest works confirm previous finding and the number of clusters where this peculiarity is detected is increasing with time \citep{Ma18}}. 
Recent analysis added M22 \citep{HHM77, Ma09, Ma11, Ma12, Ma13, Lee15},  M2 \citep{Pi12, La13, Mil15},
M54 \citep{Sa95, Be08, Ca10}, NGC 1851 \citep{Mi09, CGL10, Yong08, CGL11}, NGC 5286 \citep{Na13, Mar15}, NGC 5824 \citep{Sa12, DC14}, Terzan 5 \citep{Fe09, MMF14}, M19 \citep{Jo15} and NGC 6934 \citep{Ma18}  to the list of clusters with iron spreads. 
Such clusters have the typical multiple sequences in their HR diagrams and
anomalies in light element abundances, as well as star by star spreads in iron content larger than $0.1$ dex.
The origin of multiple populations is still unknown, and could be the result of a secondary star formation event from gas lost by first population fast rotating massive stars  \citep{De07}, AGB stars \citep{Dan04,Ve01} or massive binaries \citep{DM09}. 
This process is however unable to produce iron spreads, adding another complication to the GC formation scenario.
\cite{Bek16} and \cite{Gav16} proposed that clusters born at slightly different times ($\sim300$ Myr) could merge forming GCs with the observed metallicity spreads.
 %%%%%%%%%%%FIGURE%%%%%%%%%%%
\begin{figure*}
\centering
\includegraphics[width=0.45\textwidth]{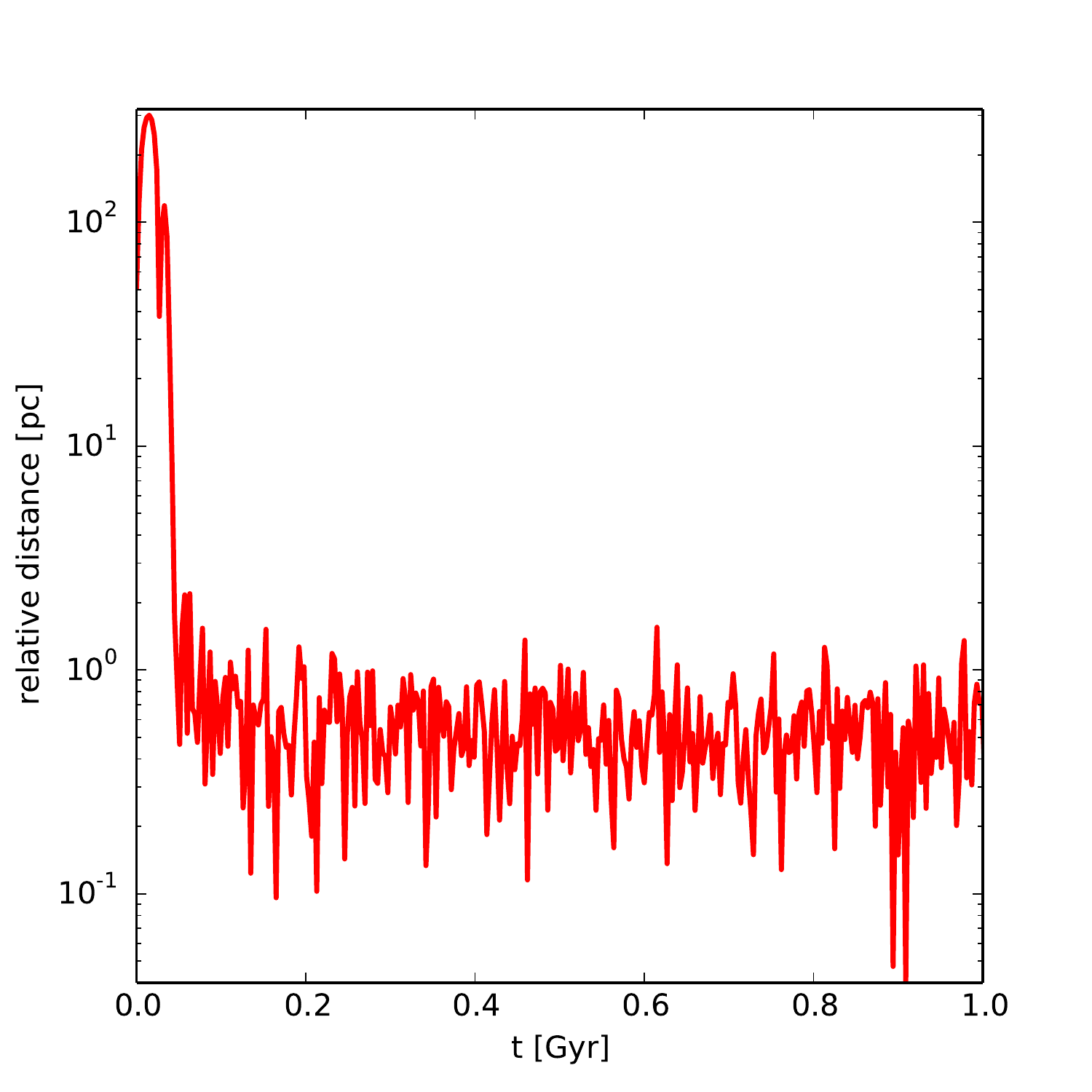}~\includegraphics[width=0.45\textwidth]{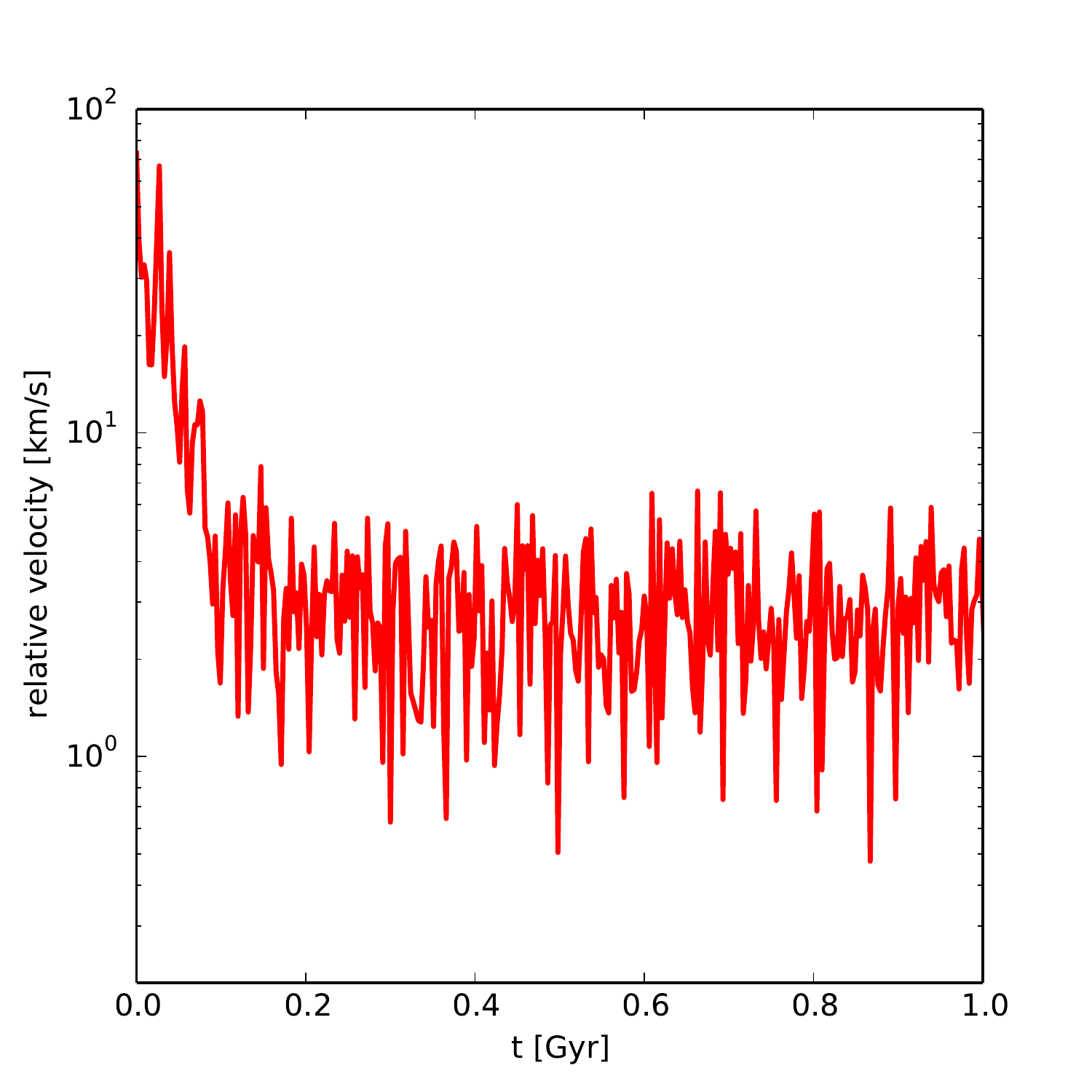}
\caption{The first Gyr of evolution of the relative distance (left panel) and velocity (right panel) between the two C2 clusters that merge and form  CM2. After two bounces, the clusters merge very quickly and become a single GC in less than 0.5Gyr.}\label{fig:reldv}
\end{figure*}
 %%%%%%%%%%%%%%%%%%%%%%%%%%%
 %%%%%%%%%%%FIGURE%%%%%%%%%%%
\begin{figure*}
\centering
\includegraphics[width=0.2535\textwidth, trim=0.3cm 1.38cm 5cm 1cm, clip=true]{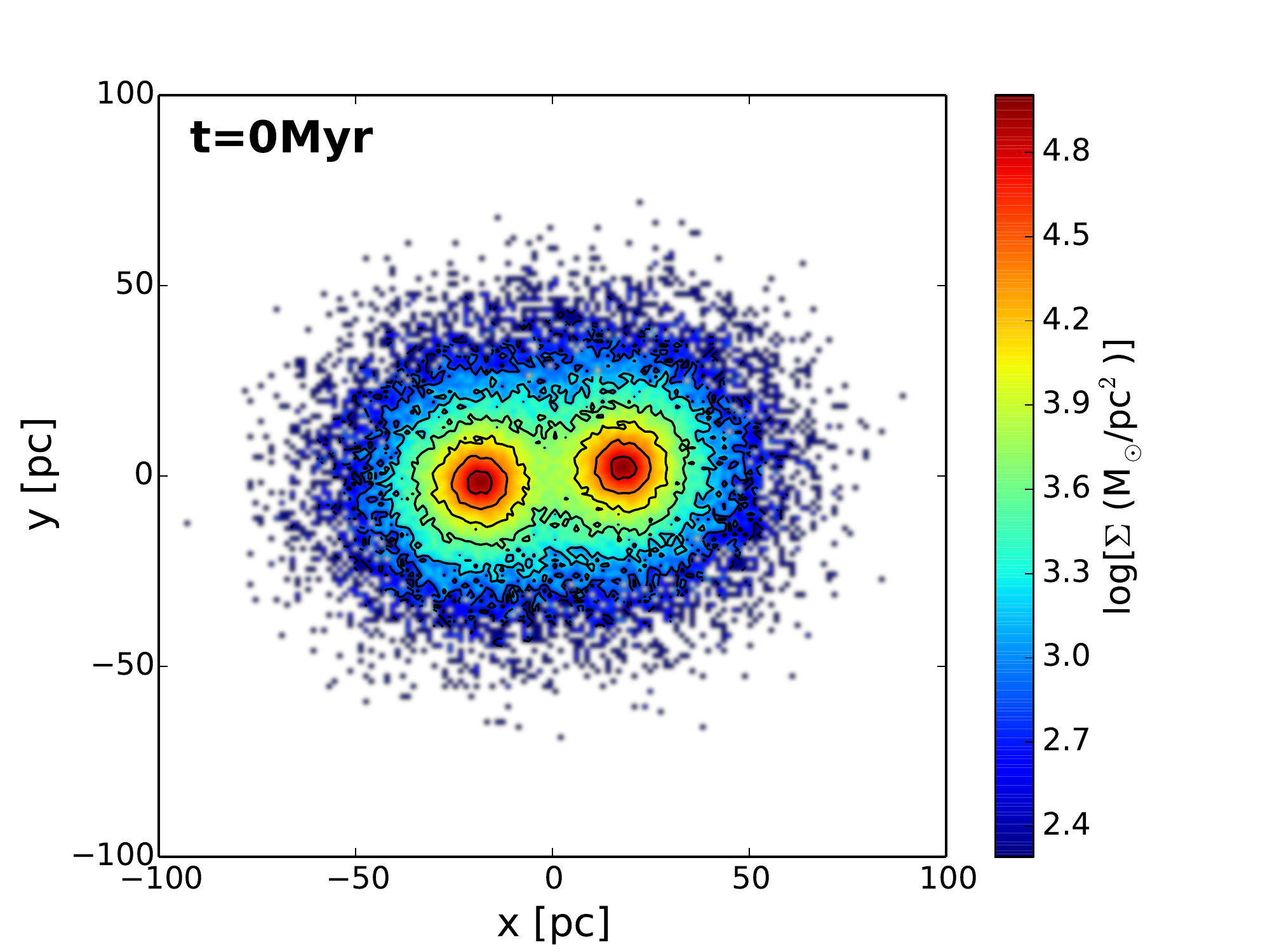}~\includegraphics[width=0.2535\textwidth, trim=0.3cm 1.38cm 5cm 1cm, clip=true]{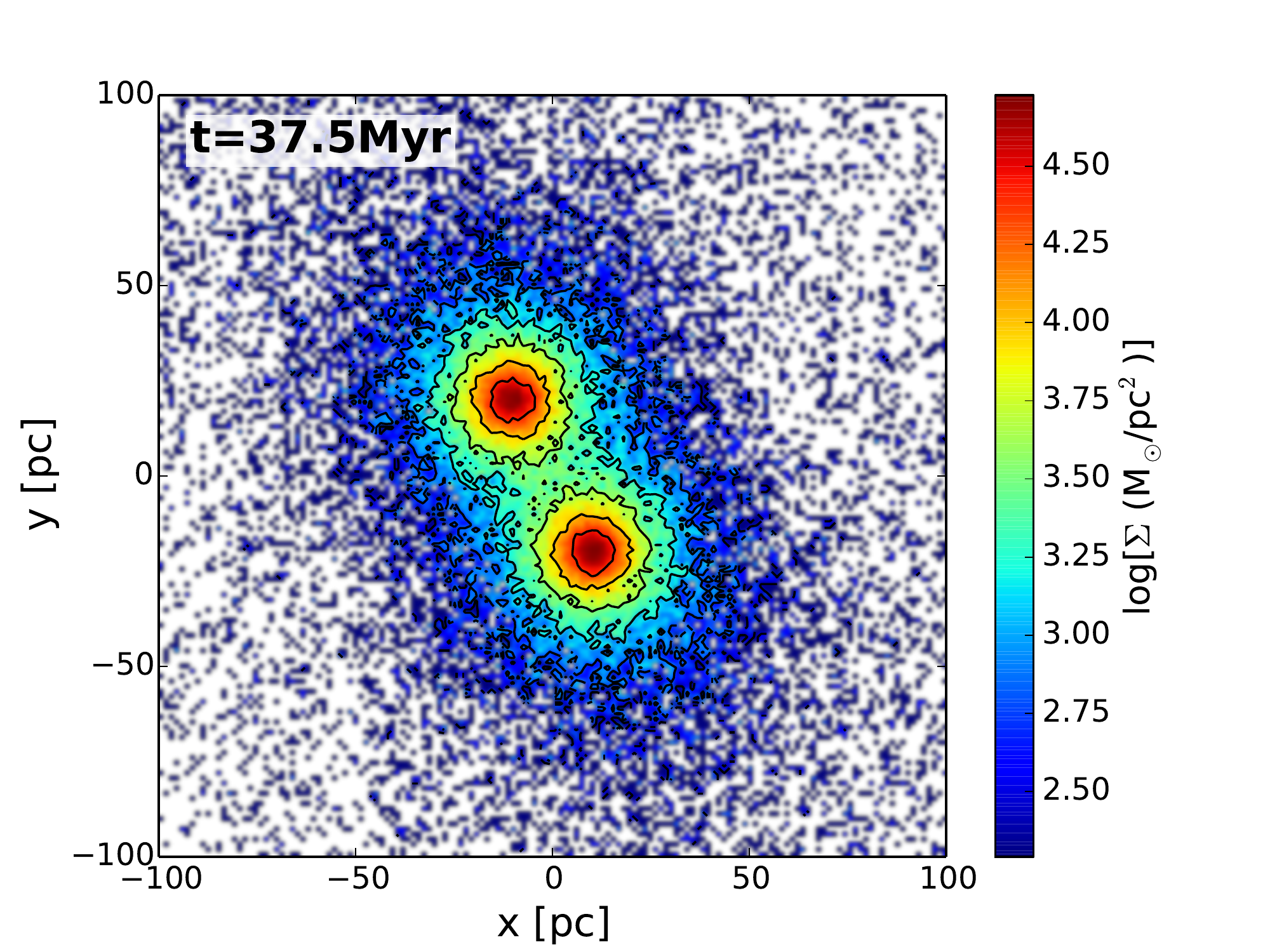}~\includegraphics[width=0.3155\textwidth, trim=0.35cm 1.38cm 1cm 1cm, clip=true]{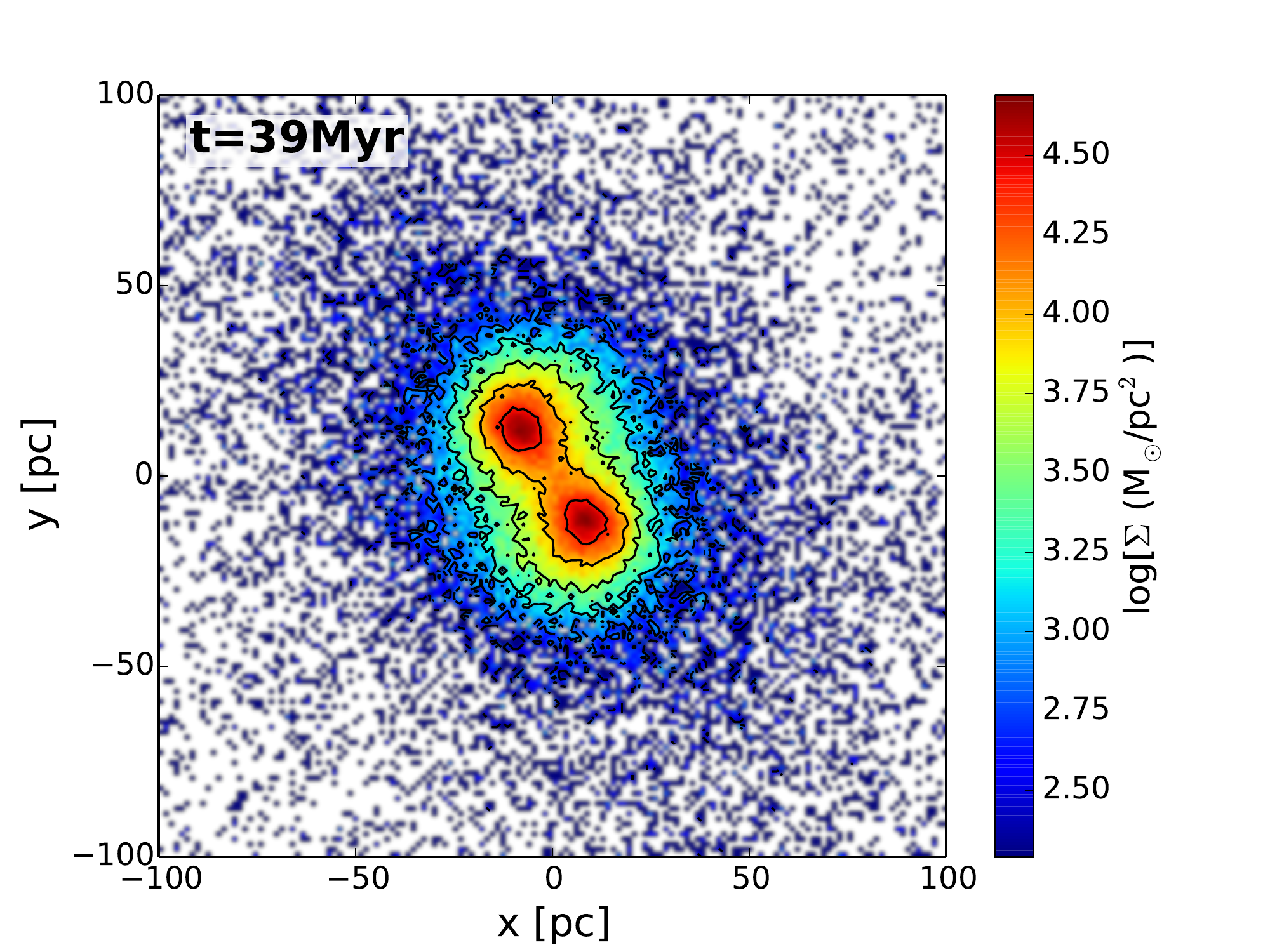}\\
\includegraphics[width=0.253\textwidth, trim=0.3cm 1.4cm 5cm 1cm, clip=true]{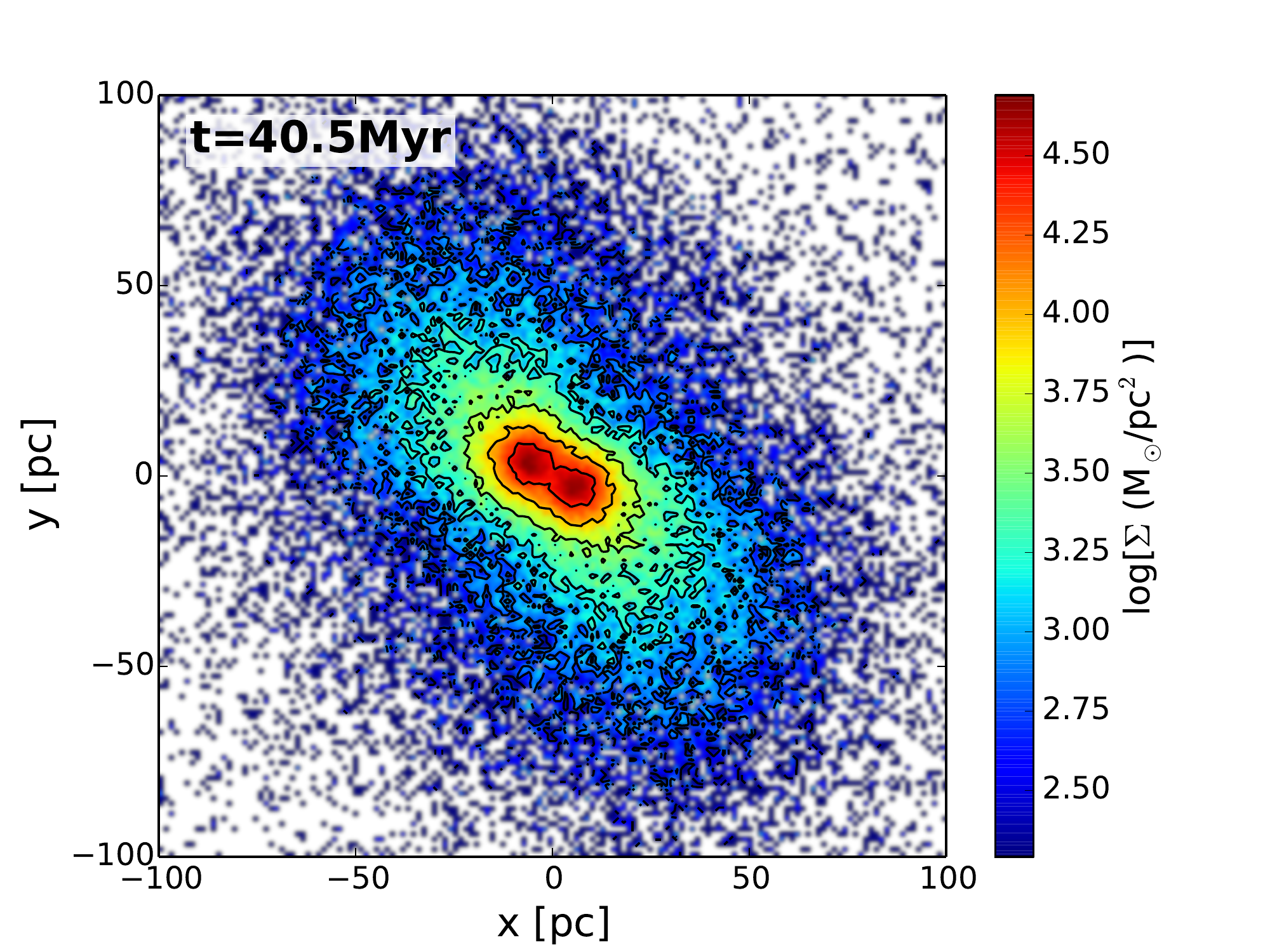}~\includegraphics[width=0.253\textwidth, trim=0.3cm 1.4cm 5cm 1cm, clip=true]{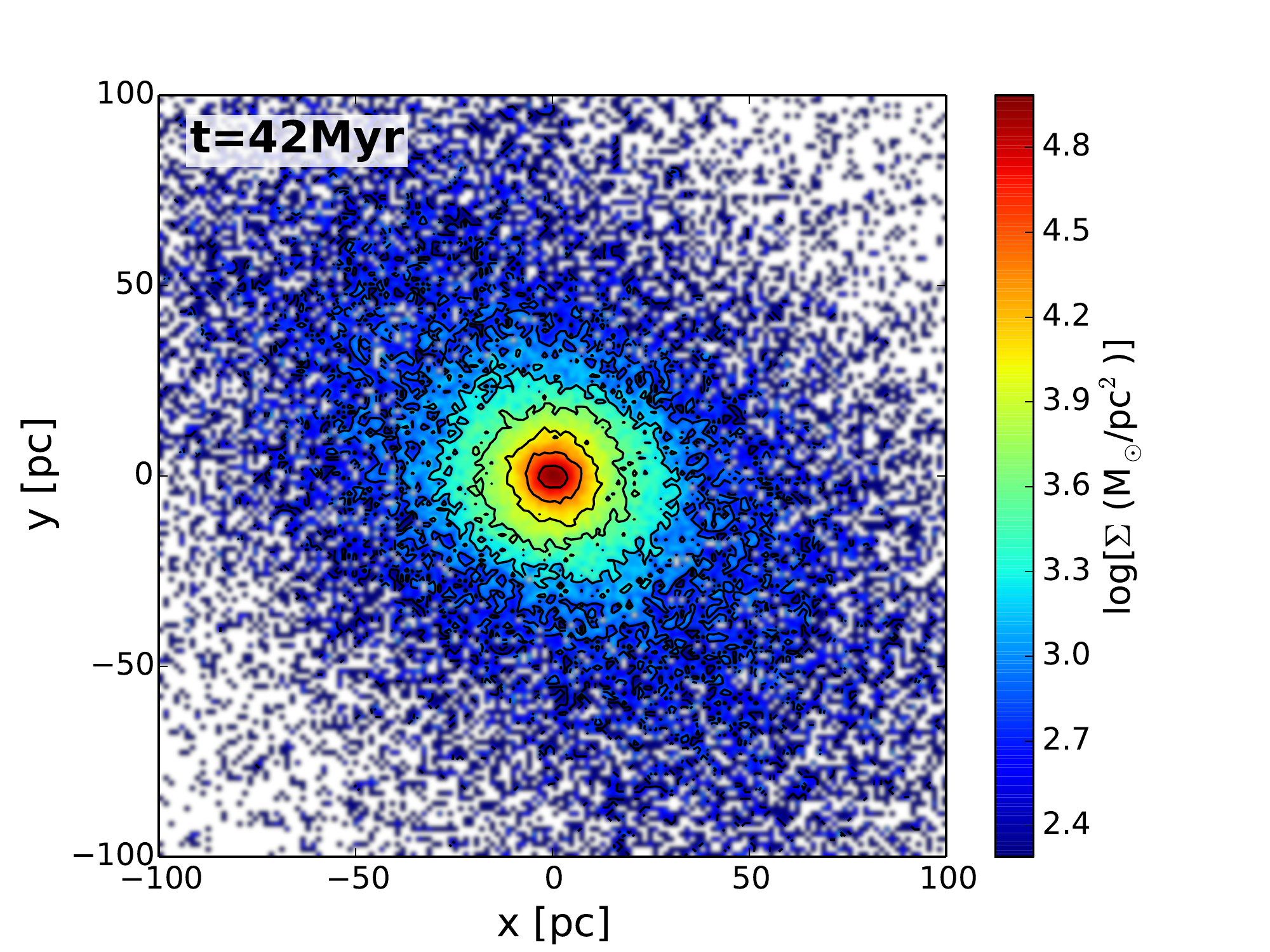}~\includegraphics[width=0.315\textwidth, trim=0.3cm 1.4cm 1cm 1cm, clip=true]{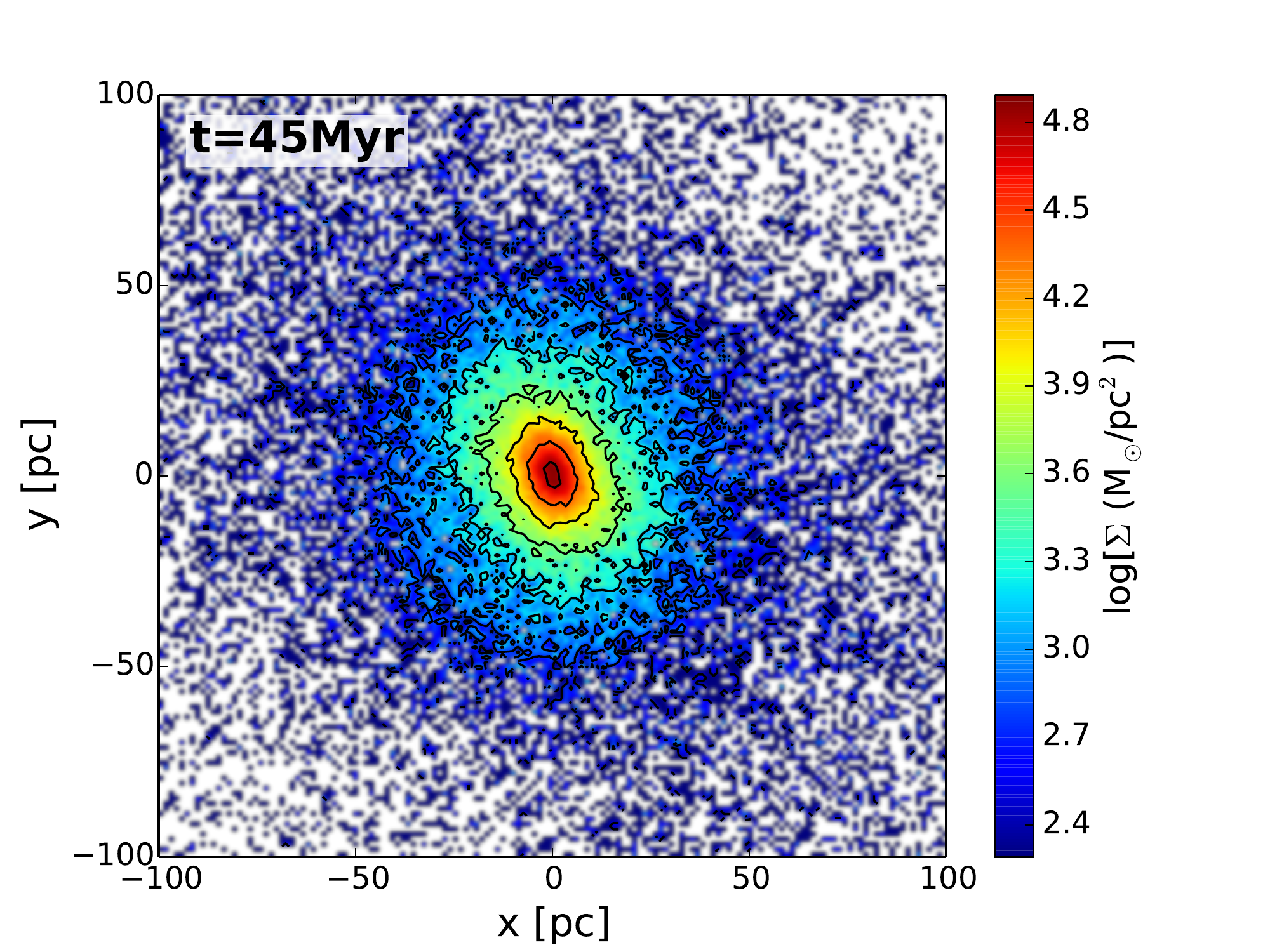}\\
\includegraphics[width=0.253\textwidth, trim=0.3cm 1.4cm 5cm 1cm, clip=true]{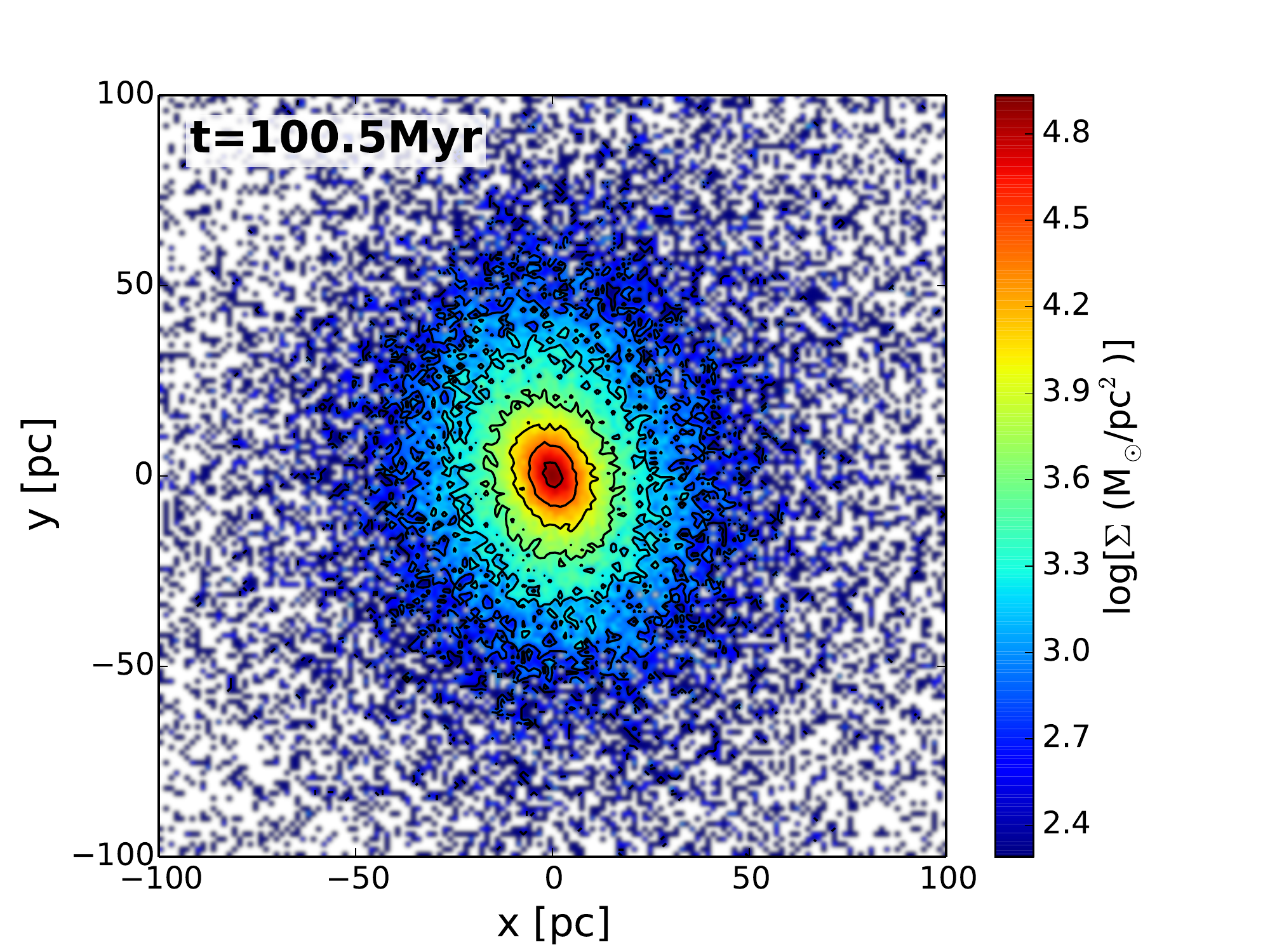}~\includegraphics[width=0.253\textwidth, trim=0.3cm 1.4cm 5cm 1cm, clip=true]{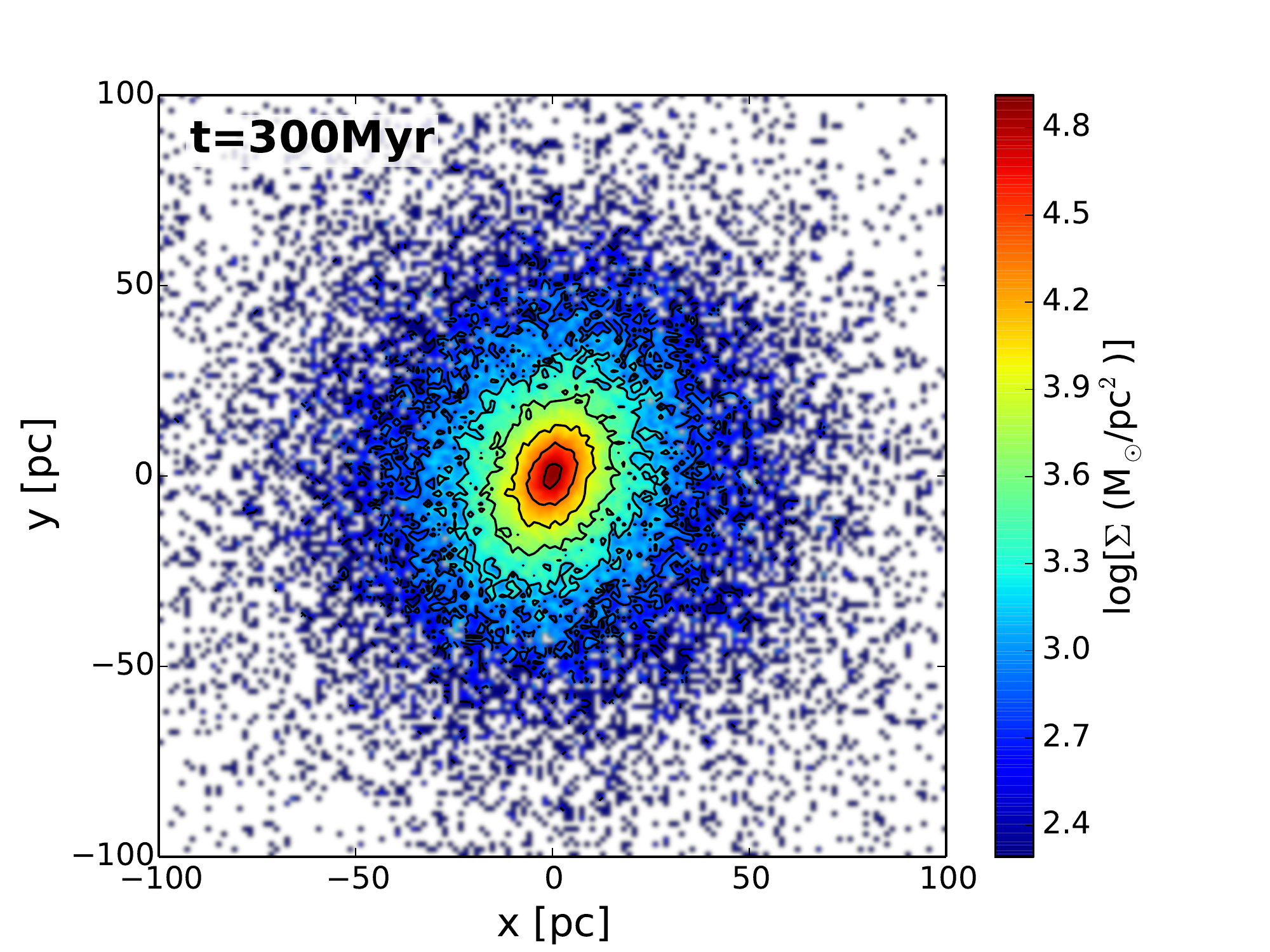}~\includegraphics[width=0.315\textwidth, trim=0.3cm 1.4cm 1cm 1cm, clip=true]{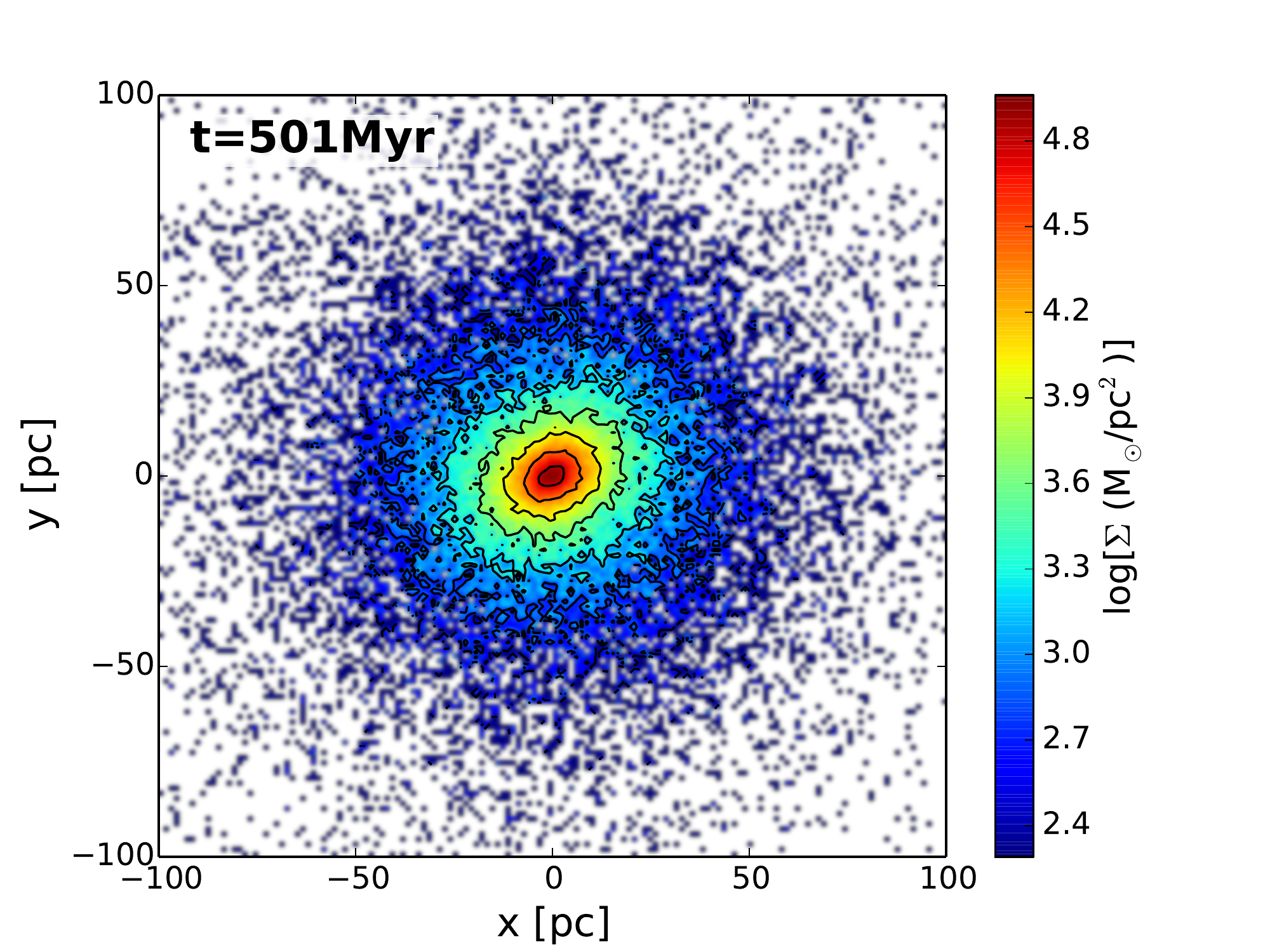}\\
\hspace{0.1cm}\includegraphics[width=0.253\textwidth, trim=0.3cm 0cm 5cm 1cm, clip=true]{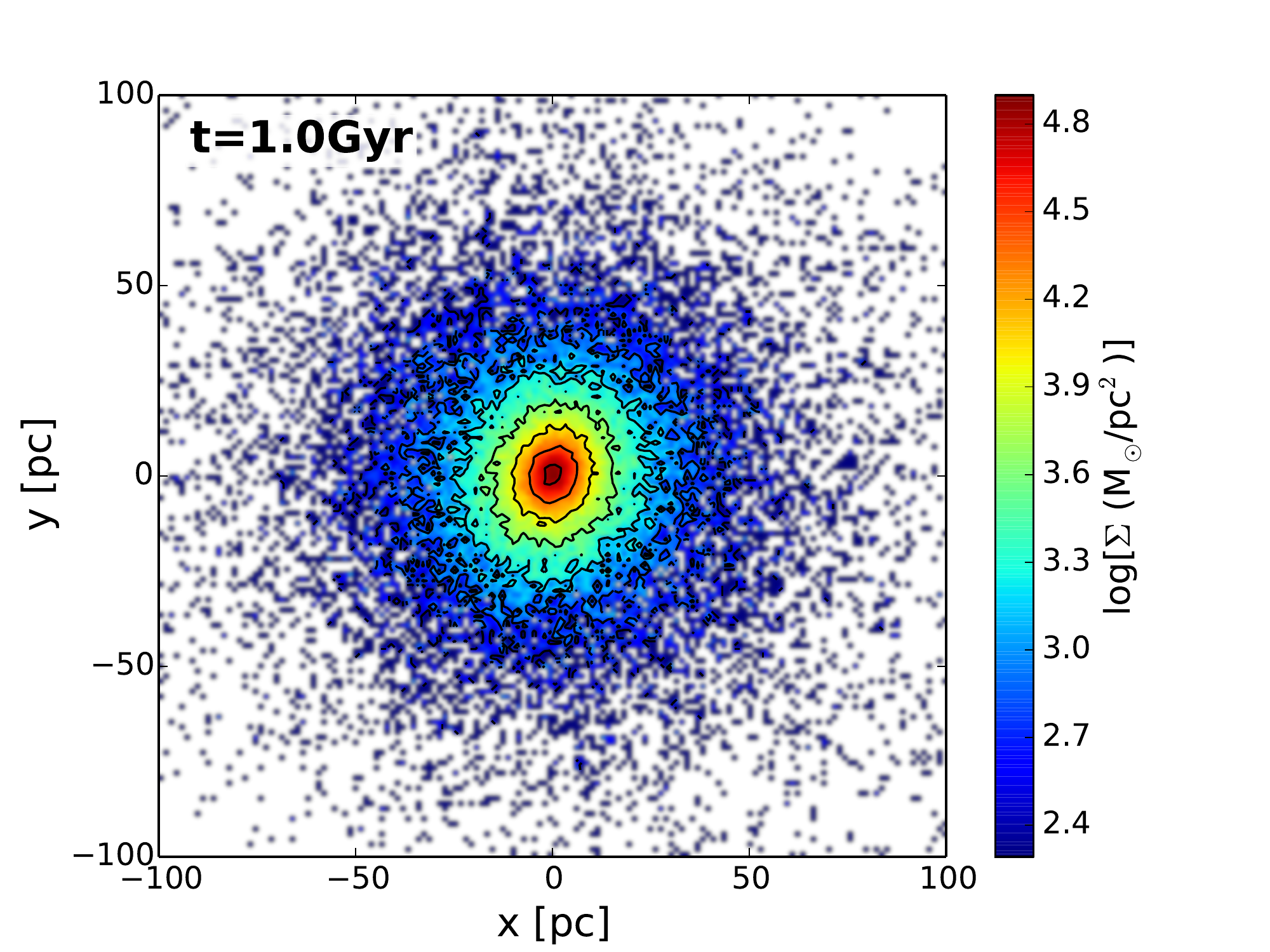}~\includegraphics[width=0.253\textwidth, trim=0.3cm 0cm 5cm 1cm, clip=true]{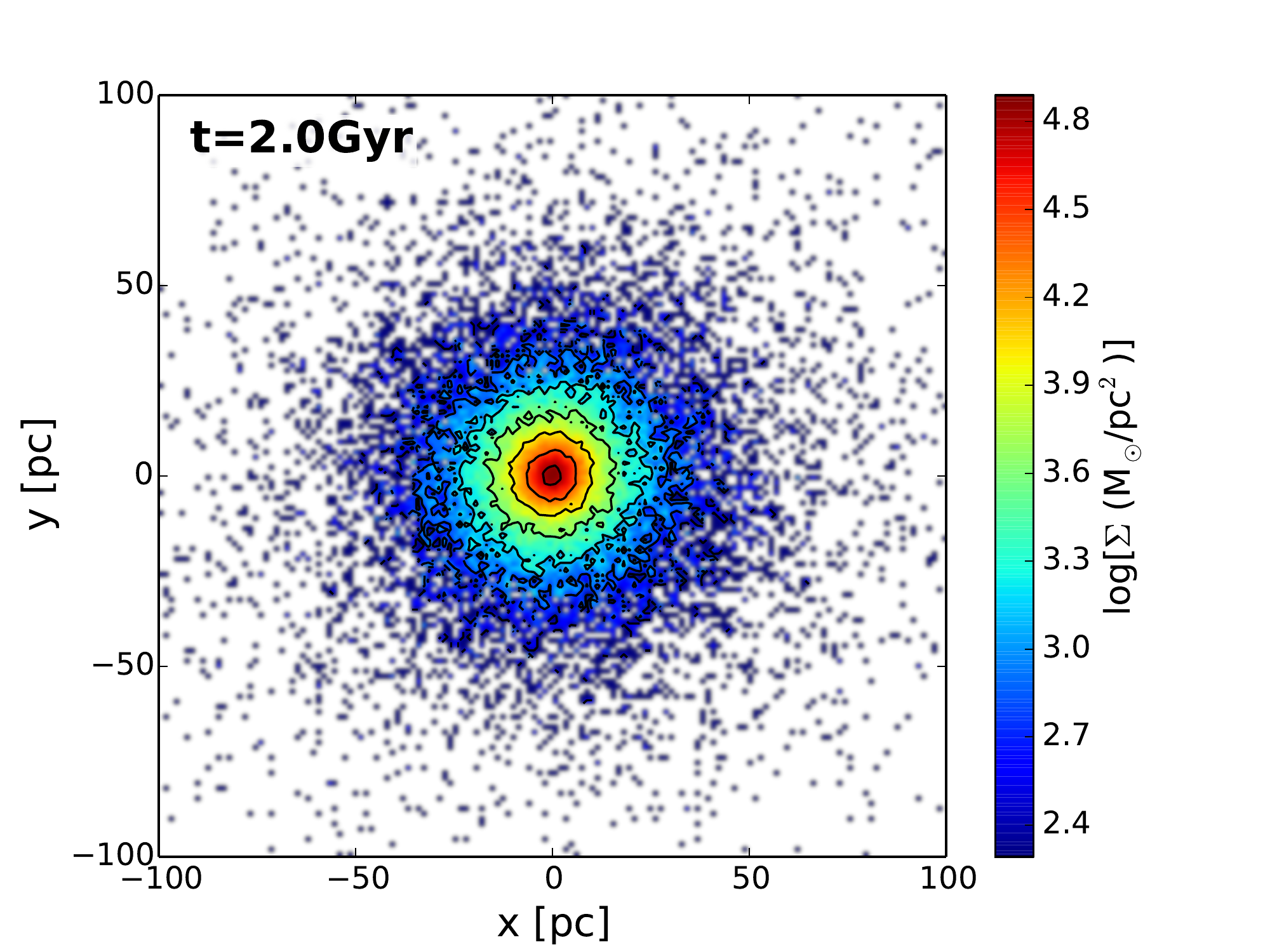}~\includegraphics[width=0.315\textwidth, trim=0.3cm 0cm 1cm 1cm, clip=true]{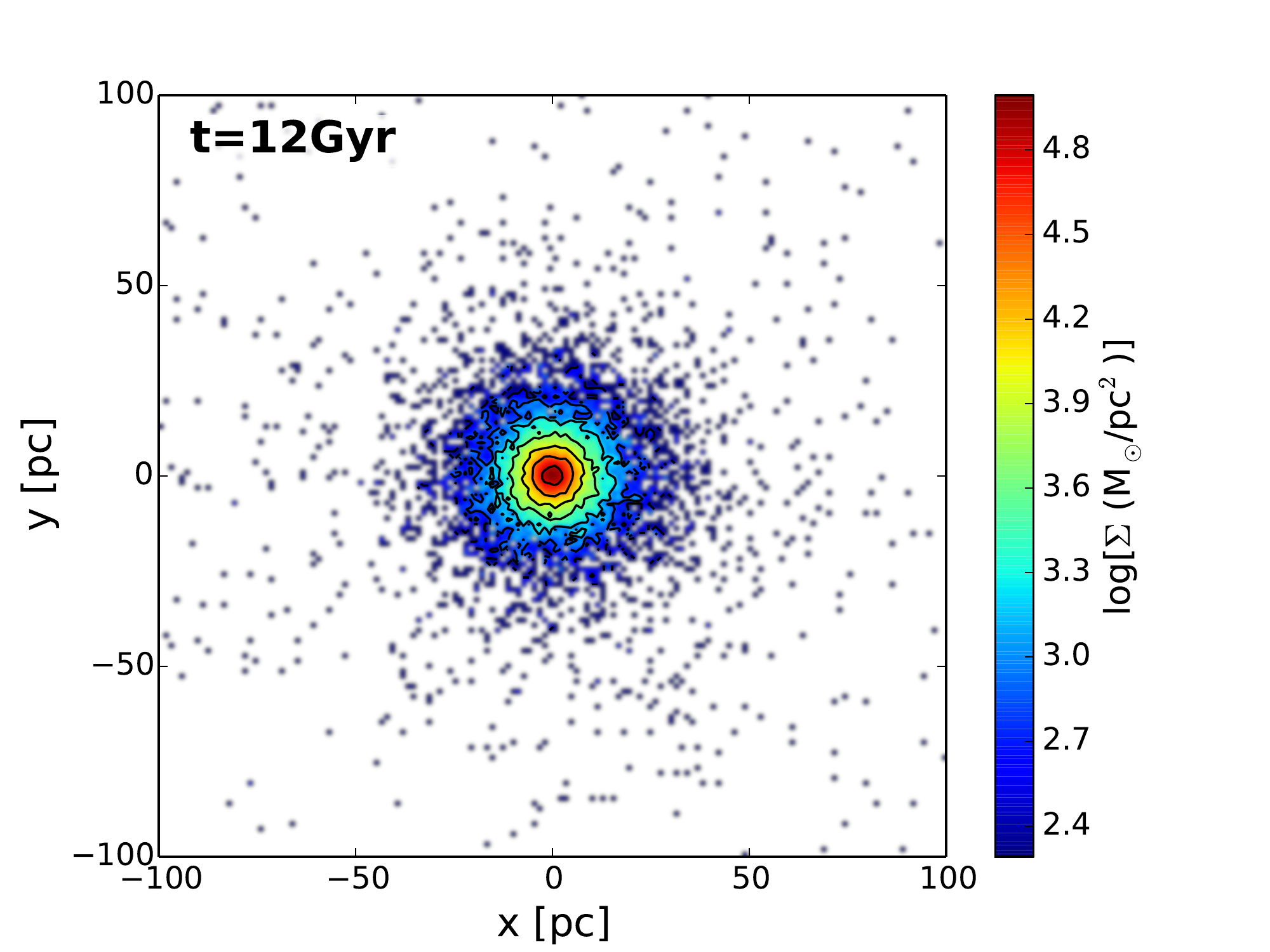}
\caption{Isodensity contour maps of two C2 GCs that merge and form the CM2 cluster. The projection of the system on the $xy$ plane is shown at different times. The merger is complete after less than 1 Gyr. The composite CM2 cluster becomes less massive and more compact with time. The system is plotted with respect to its center of density.}\label{fig:merger}
\end{figure*}
%%%%%%%%%%%FIGURE%%%%%%%%%%%
%%%%%%%%%%%%%%%%%%%FIGURE%%%%%%%%%%%%%%%
\begin{figure*}
\centering
\centering
\includegraphics[width=0.3\textwidth, trim=0.3cm 0.cm 4.7cm 1cm, clip=true]{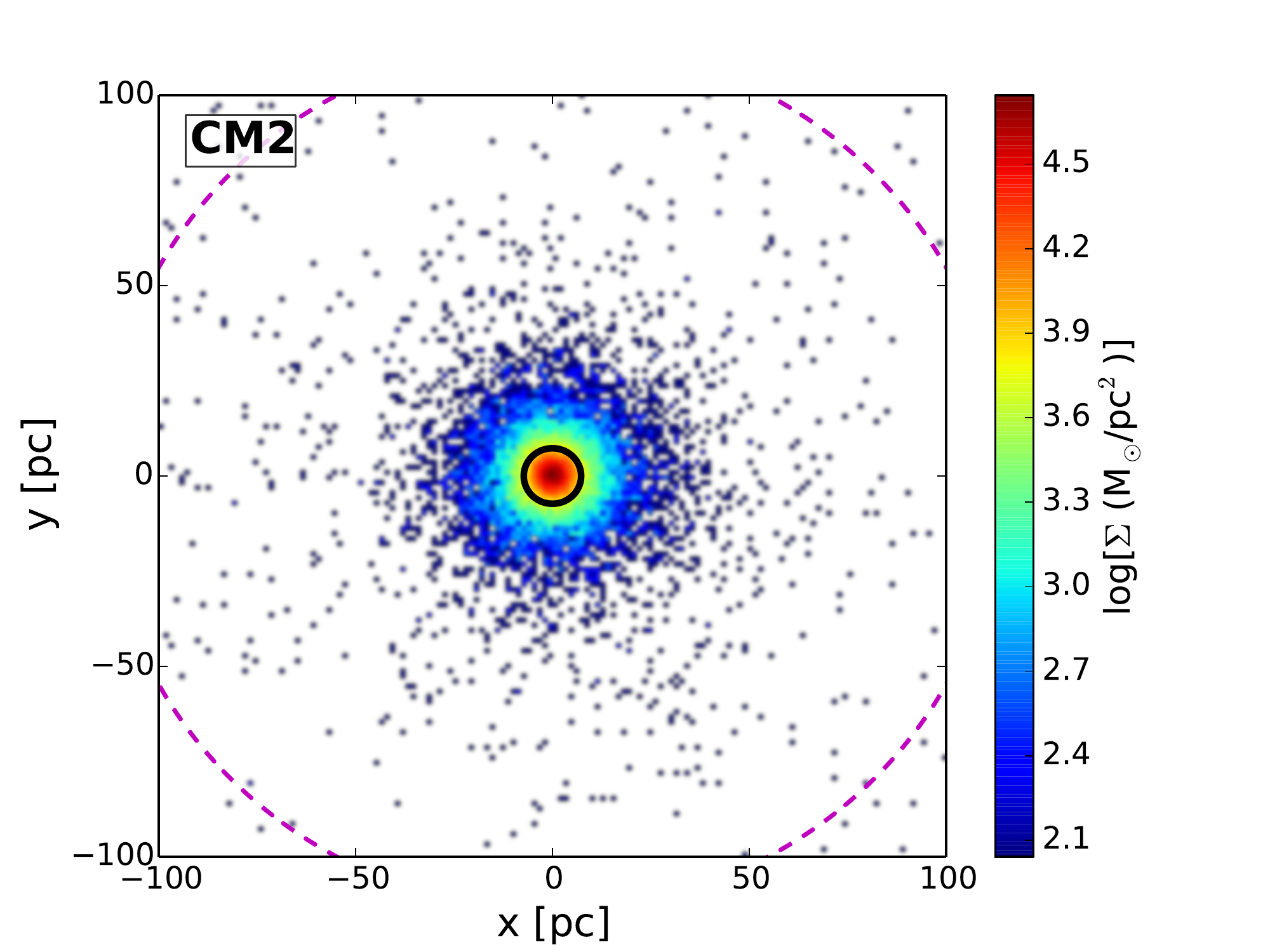}~\includegraphics[width=0.3\textwidth, trim=0.3cm 0cm 4.7cm 1cm, clip=true]{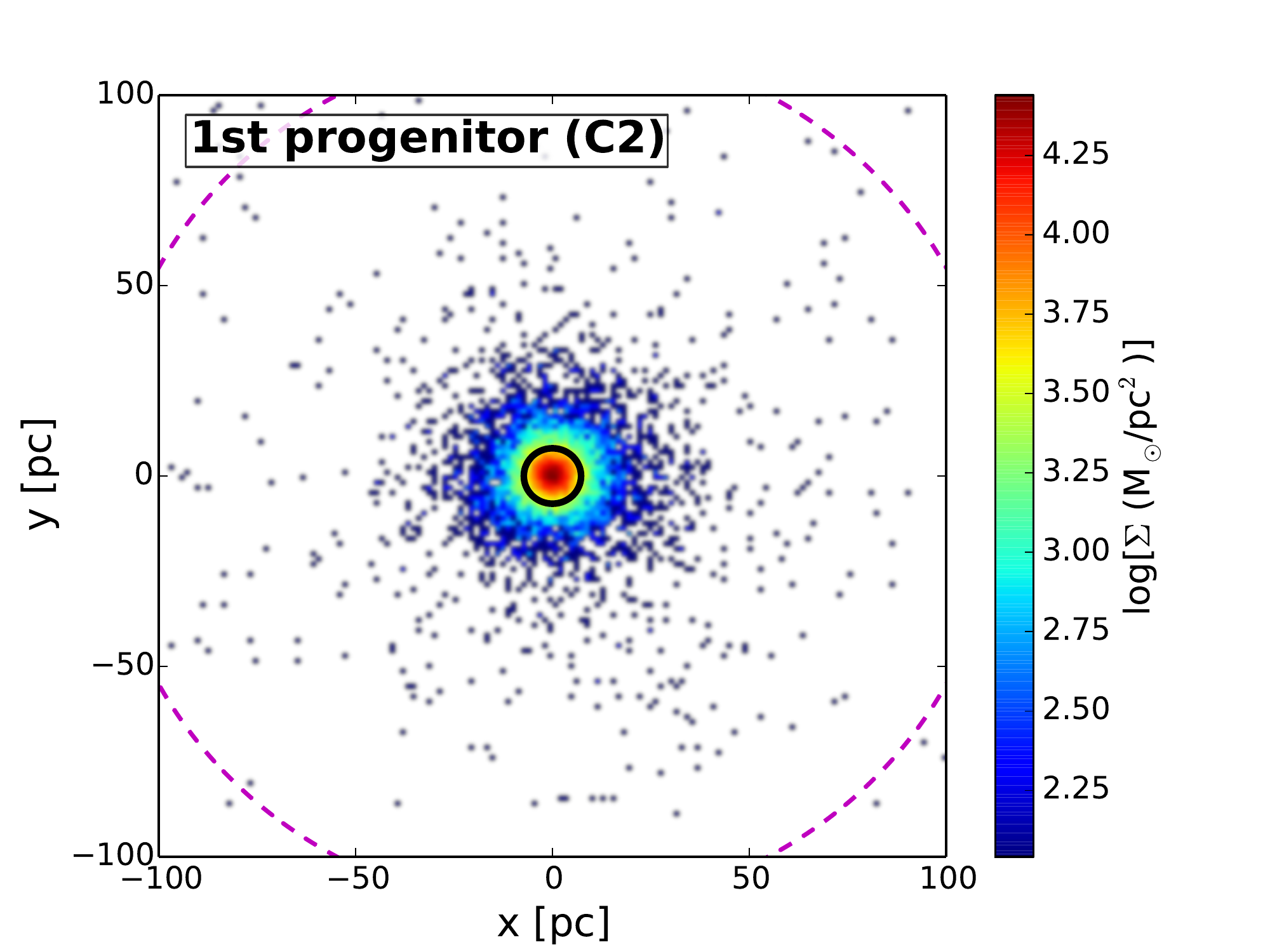}~\includegraphics[width=0.37\textwidth, trim=0.3cm 0cm 1cm 1cm, clip=true]{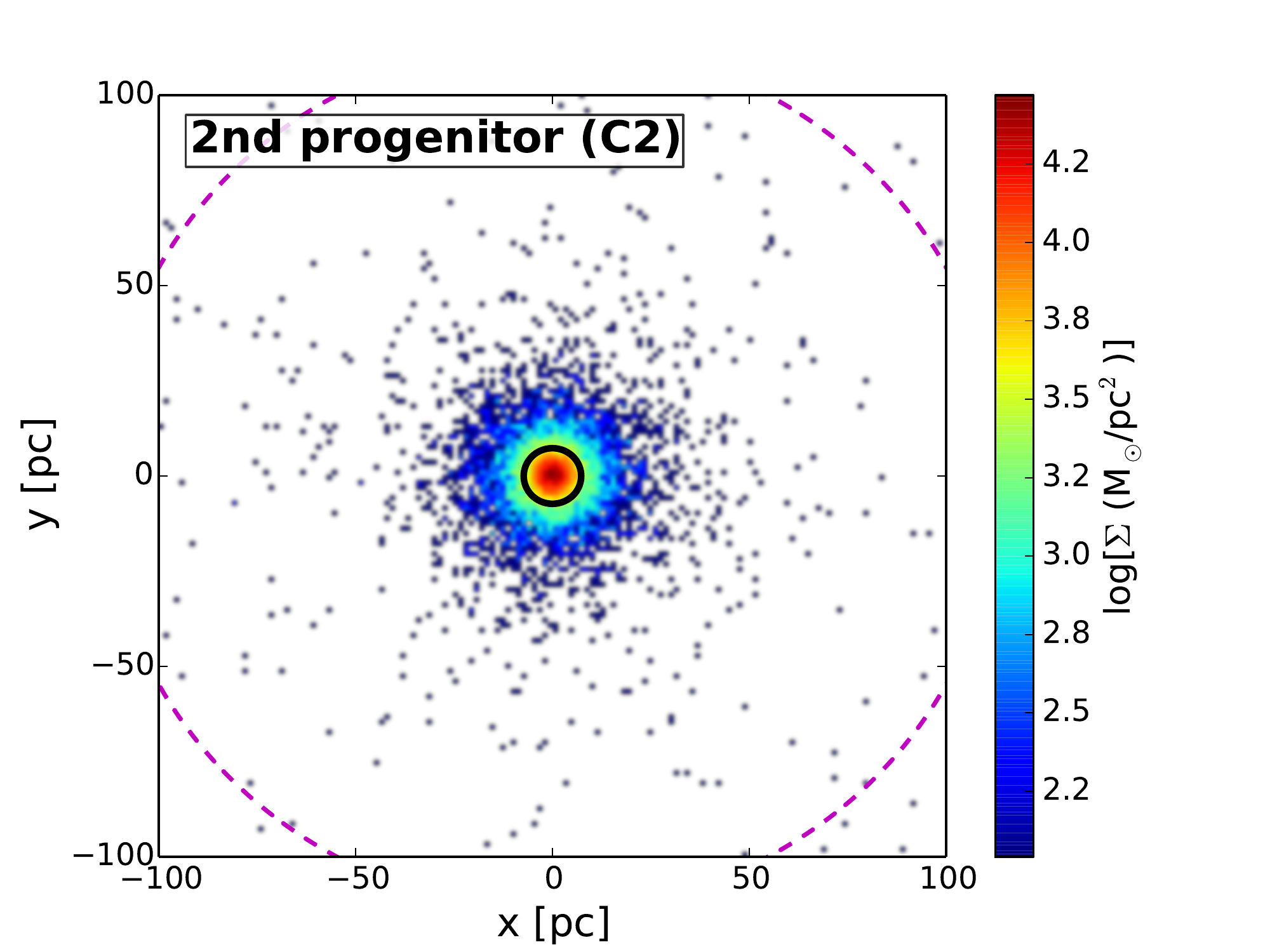}
\includegraphics[width=0.3\textwidth, trim=0.3cm 0.cm 4.7cm 1cm, clip=true]{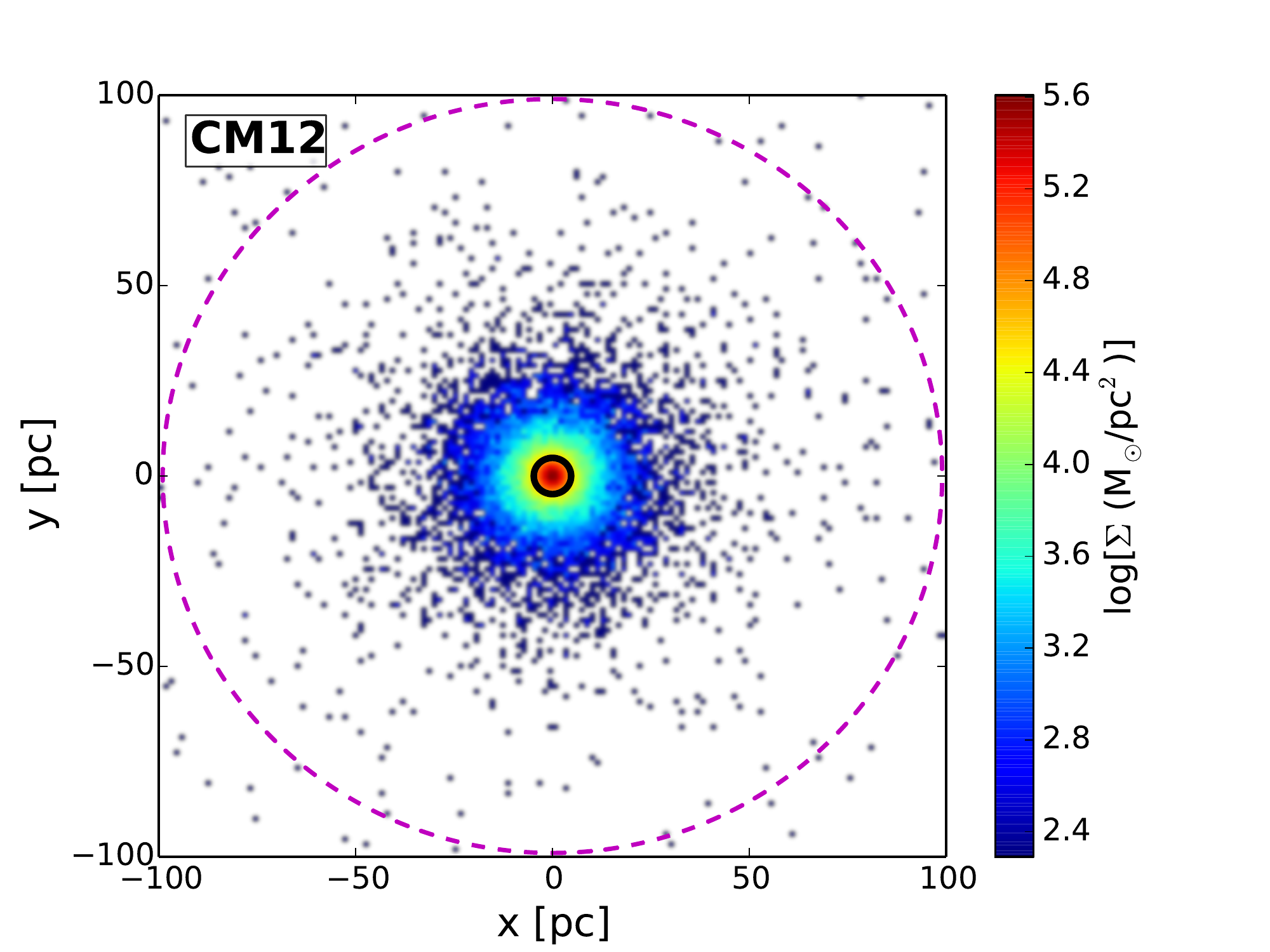}~\includegraphics[width=0.3\textwidth, trim=0.3cm 0cm 4.7cm 1cm, clip=true]{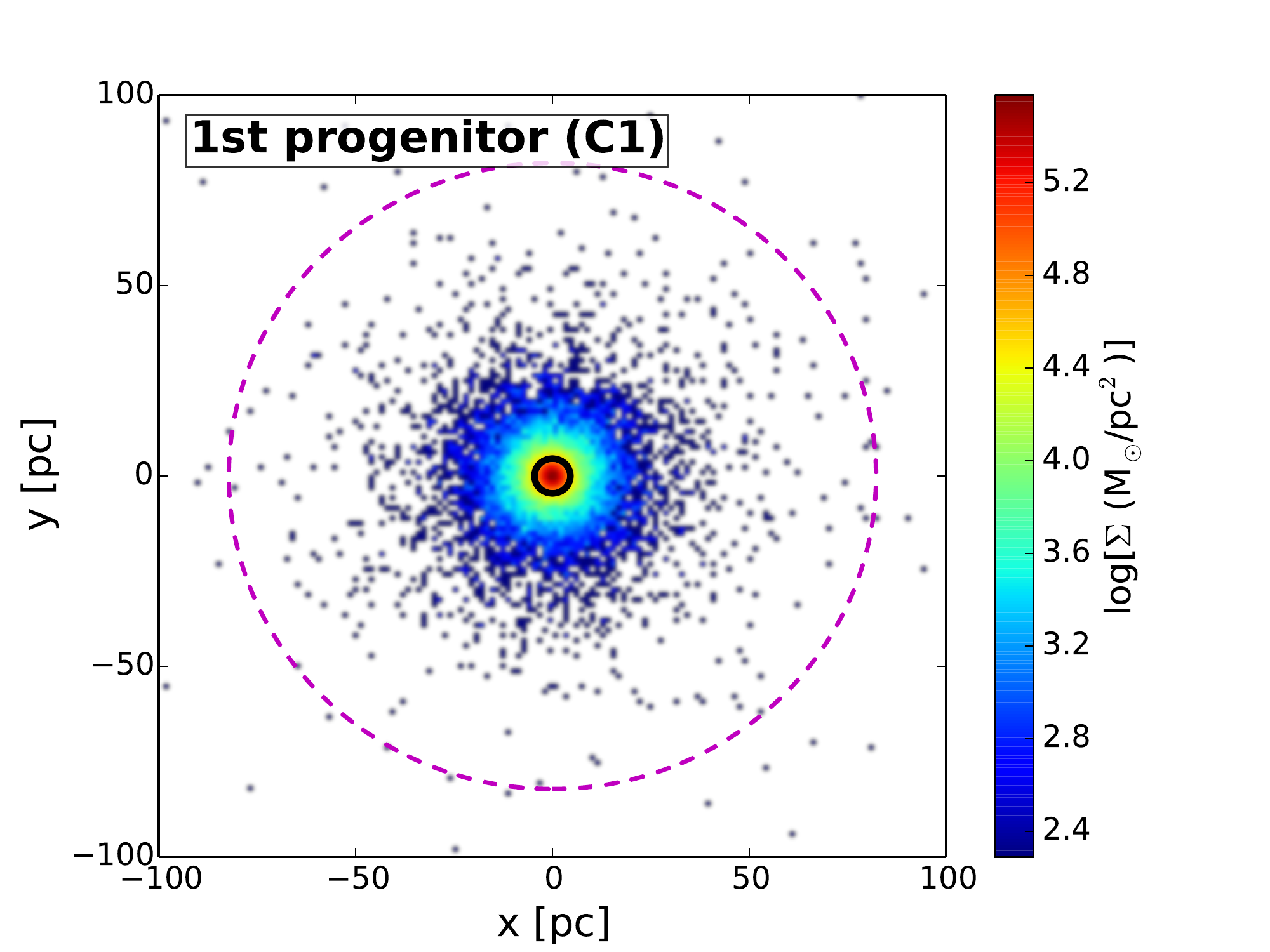}~\includegraphics[width=0.37\textwidth, trim=0.3cm 0cm 1cm 1cm, clip=true]{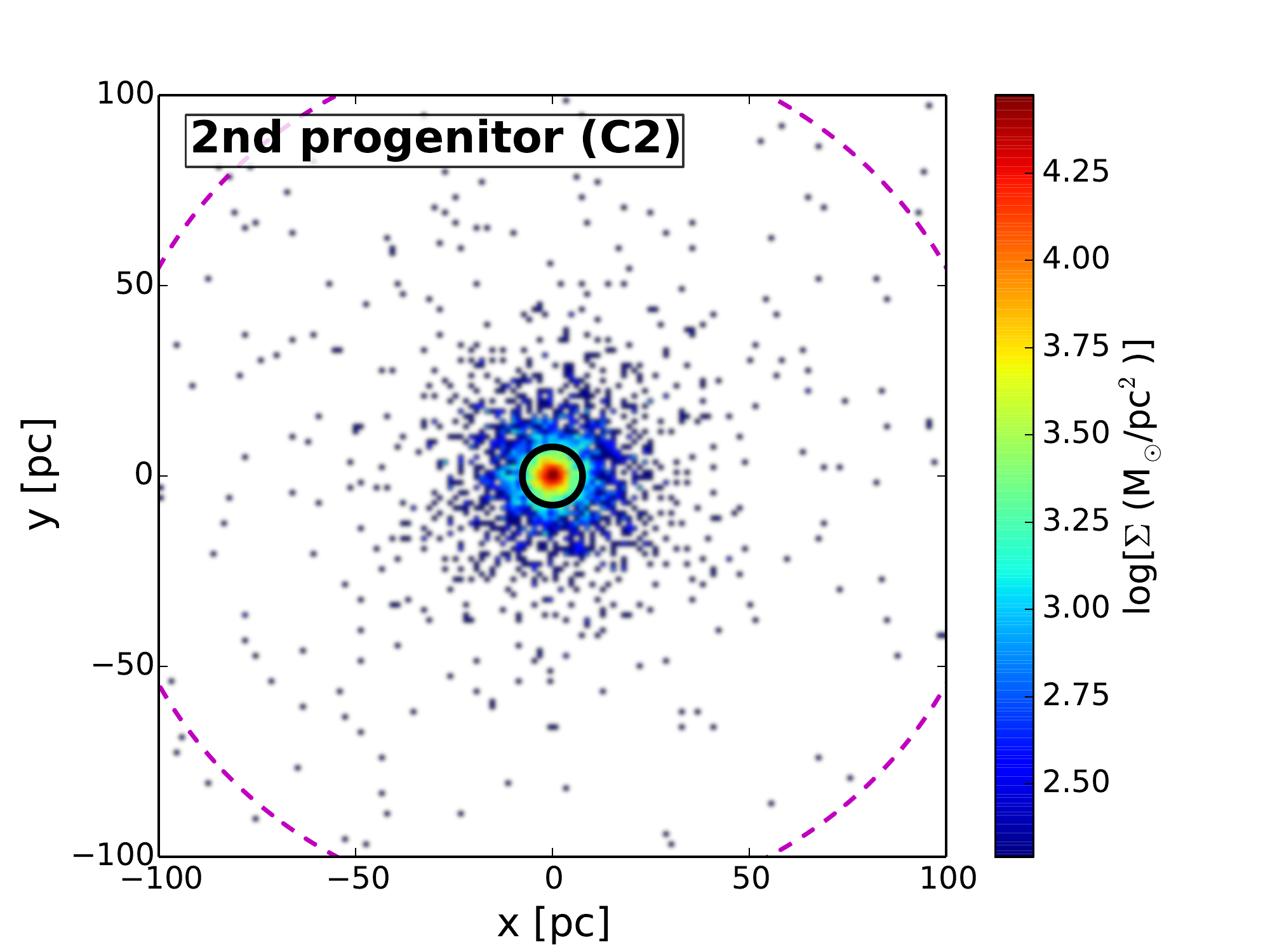}
\caption{The final density maps for CM2 and CM12 { and for the stellar component that comes from each progenitor (C2+C2 or C1+C2) are shown in the upper and lower panels respectively. The panel on the left column shows the entire cluster, the central panel is for the stars coming from the first progenitor and the right panel for those coming from the second progenitor} that merge to form the new cluster. The black solid line and the purple dashed lines represent the half mass and tidal radius of each system respectively.
While for CM2 the two sub-clusters have similar densities and spatial distributions, for CM12 the C1 component  is less dense and extended than the one inherited from C2.}\label{fig:gcdiscCM12}
\end{figure*}
%%%%%%%%%%%%%%%%%%%%%%%%%%%%%%%%%%%%%%
According to the authors, these mergers can only take place in the halo or nucleus of dwarf galaxies, because of the shallower gravitational potential that characterises these environments.
The dwarf galaxies, along with their pristine or merged GCs, are later accreted by the Milky Way.  
{ A similar process has been suggested by \cite{VdB96} to explain Galactic GCs with composite color-magnitude diagrams, and explored in detail with $N$-body simulations by \cite{AS13} to understand the multimetallic clusters in the Antennae galaxy.  More recently,  \cite{HDG17} numerically investigated the possible merger origin of light-elements anomalies in  intermediate-age clusters in the Magellanic Clouds. Mergers between  young clusters, born in a bound binary configuration, have been studied in several $N$-body works with particular focus on the kinematics the cluster resulting from the process \citep[see e.g.][]{Mak91,DO98,Po07, Pr16,Ar17}.}
%However, contrarily to the expectations of this scenario,  a large fraction of the GCs showing an iron spread (...) are on disc orbits or very close to the Galactic center rather than in the halo, where accreted clusters are expected to be found.
The age difference found between the metal poor and metal rich populations in Terzan 5 and its chemical similarity with the Galactic bulge populations \citep[12 Gyr and 4.5 Gyr][]{Fer16}, however, suggest  an in-situ formation for this cluster.
In \citep[K18 hereafter]{K18}), we analytically estimated that the rate of physical encounters between thick disc GCs in the Galaxy is up to 1.8 Gyr$^{-1}$. This rate is mostly determined by the number density and  relative velocities of the clusters. 

The encounters might result in full mergers or partial contaminations between two or more clusters. The short term evolution of a full initial population of disc GCs,  
followed by means of $N$-body simulations run with a Tree-code \citep{Kho14}, shows how, while many clusters are destroyed in the field, loosing their stars in the disc, part of them survive and can interact with each other. 
%When using a population of clusters of $10^6M_\odot$ we did not observe any merger, and we had very limited mutual contaminations. 
%However, when increasing the mass of the clusters to
%$10^7M_\odot$ we found two major mergers and several stripping events that produced clusters with up to 50\% contamination 
%from a second progenitor.
In particular, when simulating a population of 128 disc GCs each of $10^7M_\odot$, we observed two major mergers and several stripping events that produced clusters with up to 50\% contamination from a second progenitor.
After $1.5$ Gyr of evolution, the polluted and merged clusters are fully mixed and their populations have spatial densities and total masses comparable to those of the current massive GCs in the Milky Way (see K18). 
For the first time{, in K18, }we deem the hypothesis that the observed iron spreads could be the result of mergers and mass exchange happening in the primordial Galactic disc.
Here we follow up these results, studying the long-term evolution of the interacting clusters using direct $N$-body simulations.
While in K18 we explored only the first $1.5$ Gyr of evolution of a full population of resolved clusters, here we follow the long-term evolution of pairs of GCs modelled as self-consistent $N$-body systems, initialising the simulation at the moment of the closest encounter and aiming at studying in detail the final ($\approx 12$ Gyr) dynamical and structural properties of the result of the merger (or mass exchange).
The paper is organised as follows.
In Sect. \ref{sec:MIC} we describe the models, the initial conditions and set-up of our simulations, whose results are illustrated in Sect. \ref{sec:res}
and discussed in Sect. \ref{sec:disc}. In Sect. \ref{sec:concl} we draw our conclusions.

\section{Models and initial conditions}\label{sec:MIC}

\subsection{Galactic potential and globular cluster models}\label{sec:models}
Our Galactic model is the same as used in K18 and it consists of a dark matter halo, as well as of a thin and of a thick disc. The functional forms of these components are taken from \cite{AS91}
while the relative parameters are given by \cite[][(Model II)]{PDMH17}. This model reproduces a number of observables, including
the stellar density at the solar vicinity, thin and thick disc scale lengths and heights, rotation curve, and the absolute value of the perpendicular force
as a function of distance to the Galactic centre.
We consider the bulge as part of the Galactic disc  \citep[see][and references therein]{Dim16} and, therefore, we do not include any classical spheroid in our model.
The mass of the  modelled halo is $2.07\times10^{11}M_\odot$ and its scale length is $14$ kpc. The thick disc has a mass 
of  $3.91\times10^{10}M_\odot$ and radial and vertical scale lengths of $2$ kpc and $800$ pc,  while the mass of the thin disc is $3.68\times10^{10}$ M$_\odot$
and its scale radius is $4.8$ kpc and the scale height is $250$ pc.

We ran our simulations using two different \cite{King66} models for our GCs: i) a ``standard'' GC whose parameters are typical of current GCs \citep{H96}, i.e. core radius $r_c=1$ pc and tidal radius is $r_t=35$ pc, corresponding to 
an adimensional parameter $W_0=7$ and ii) a more extended cluster with lower central density, $r_c=4$ pc, $r_t=80$ pc and $W_0=6$ { with parameters similar to the most massive Galactic GC, $\omega$ Cen \citep{Me87}}. The second model corresponds to the one used in K18. 

We considered initial, relatively small masses of  $10^6~M_\odot$ (we identify as CS1 the standard model and as CS2  the more extended cluster), $3\times10^6~M_\odot$ (CS3 model)  and $10^7~M_\odot$ (C1 is for the standard model and C2 is for the more extended cluster, see Table \ref{tab1} for details and IDs used in the text). 
We used single mass particles and adopted $N=25584$ for CS1 and CS2 and CS3 and $N=51151$ for C1 and C2 (we used $N=85279$ for C2 in the interaction with CS3, to have same particle mass in both clusters).
%We simulated a system of one hundred GCs following the assumption that the current population ($\sim 50$ disc GCs) is the remnant of a larger initial GC system (see K18 for more details on this assumption).  We distributed the clusters on disc orbits using the method described in \cite{Ro09}, as done also in K18.
%{ The number of particles has been chosen accordingly to computational limitations. In each simulation all particles of both clusters have the same mass.}
%%%%%%%%%%%%%%%%%%%%%%%%%%
\subsection{$N$-body simulations set-up} \label{sec:nbody}
%\item  Model C1 is more concentrated and less extended while model C2 is less dense and with a larger tidal radius. We adopted total masses of $10^6M_\odot$, $3\times10^6M_\odot$ or $10^7M_\odot$.  The remaining 98 clusters where kept in each simulation as point mass clusters. The clusters were simulated for 12Gyr.
%\item We followed the evolution of several couples and we found that, in the cases analysed, $10^6M_\odot$ clusters never merge. 
{ 
While K18 simulated the short-term ($1.5$\,Gyr) evolution of a full population of resolved disc GCs, here we focus on GCs pairs  that merge and follow the resulting composite system for 12Gyr. 
To find the potential mergers, we simulated a hundred disc clusters -- represented as point mass particles orbiting the Galactic disc --  and took record of the closest passages characterised by the lowest relative velocities\footnote{The { point-mass} simulations have been run by means of an NBSymple simplified, serial version.}.  
The observed disc GCs are currently $\sim 50$ disc GCs and, therefore, we assumed that those are the remnants of an initially larger population \citep[see][]{K15,Re17}. We distributed the clusters on disc orbits using the iterative method developed by \cite{Ro09} , which is able to construct equilibrium phase models of stellar systems in a fixed mass distribution.
We ran a simulation with 100 $10^6$ M$_\odot$ clusters, and another one using the same number of $10^7$ M$_\odot$ clusters.
{ In both cases the scale height and scale length of the disc GC systems we simulated are $1$ and $2$ kpc, respectively.}
By following the evolution of the positions and velocities of the point mass GCs with time, we selected pairs of clusters that encounter each other with mutual distances smaller than $50$ pc and relative velocities smaller than few hundreds km/s\footnote{Mergers require low relative velocity dispersions, however, we find that sufficiently massive clusters can mutually affects the respective orbits, such that the gravitational focusing 
becomes stronger than the effect of the Galactic potential.}. 
{ Among those close approaches we identified the ones with the smallest impact parameters and the lowest relative velocities, focussing on $10$ pairs of $10^6M_\odot$ clusters and $\sim10$ pairs of $10^7M_\odot$ clusters. 
The positions and velocities of the two interacting point mass clusters, taken $<1$ Myr before the closest encounter, have been adopted to set the initial conditions for the centre of mass of
different combinations of our $N$-body models (see Table \ref{tab1}).
In order to keep the orbits of the selected clusters unchanged, we retained the remaining 98 clusters in the original distribution, representing them as analytic \cite{P11} spheres with scale length equal to $20$ pc.  This choice is necessary to soften the encounters between massive particles.}

We ran our simulations by means of NBSymple \citep{CMBM11}, a direct and symplectic $N$-body code. NBSymple is an efficient hybrid code, that runs on GPU equipped machines. The several versions of the code -- single, double, emulated double precision and second, fourth or sixth order time integration symplectic methods -- have been used to explore different issues concerning the evolution of GCs in the Galactic potential \citep[see e.g.][]{Mas12, So12, Le14, Pe18}. In this work, for computational reasons, we used the emulated double precision to evaluate the acceleration on the GPUs and a second-order symplectic method (leapfrog) for the time integration.\\ }
We used a softening length $\varepsilon$ of $0.1$ pc to smoothen impacts between the stellar particles in the clusters.
The time step is chosen such that $\Delta t=\varepsilon^3/(GM)$, where $M$ is the total mass of a single of the clusters. 
Our choices, including the number of particles used to represent the clusters, are due to computational limitations as we aimed to $12$ Gyr long simulations.
The relative energy variation -- over $12$ Gyr -- is $\Delta E/E_0\leq 10^{-4}$, where $E_0$ is the initial energy of the system.
%%%%%%%%%%%TABLE%%%%%%%%%%%
\begin{table}
\centering
\begin{tabular}{lccccc}
\hline 
ID & C1 & C2 & CS1 & CS2 & CS3\tabularnewline
\hline 
$M_{0}(M_{\odot})$ & $1\times10^{7}$ & $1\times10^{7}$ & $1\times10^{6}$ & $1\times10^{6}$ & $3\times10^{6}$ \tabularnewline
$r_{c}$ (pc) & 1 & 4.4 & 1 & 4.4 & 1\tabularnewline
$r_{t}$ (pc) & 35 & 80 & 35 & 80 & 35\tabularnewline
$W_{0}$ & 7 & 6 & 7 & 6 & 7\tabularnewline
\hline 
\end{tabular}

\caption{Initial parameters and IDs of the adopted GC models.}\label{tab1}

\end{table}
%%%%%%%%%%%%%%%%%%%%%%%%%%
%%%%%%%%%%%TABLE%%%%%%%%%%%
\begin{table*}
\centering
\begin{tabular}{cccccccc}
\hline 
\multirow{2}{*}{Result} & \multirow{2}{*}{ID} & $x$  & $y$ & $z$ & $vx$  & $vy$  & $vz$\tabularnewline
 &  & kpc & kpc & kpc & km/s & km/s & km/s\tabularnewline
\hline 
\multirow{2}{*}{CM1} & C1 & $-1.50$ & $2.44$ & $0.105$ & $-125$ & $-187$ & $27.5$ \tabularnewline
 & C1 & $-1.47$ & $2.41$ & $0.124$ & $-152$ & $-138$ & $53.0$ \\[0.2cm]
\multirow{2}{*}{CM2} & C2 & $-1.33$ & $-0.226$ & $0.425$ & $-60.6$ & $-67.6$ & $28.3$  \tabularnewline
 & C2 & $-1.29$ & $-0.226$ & $0.461$ & -95.8 & $-101$ & $26.2$  \\[0.2cm]
\multirow{2}{*}{CM12} & C1 & $-1.33$ & $-0.226$ & $0.425$ & $-60.6$ & $-67.6$ & $28.3$  \tabularnewline
 & C2 & $-1.29$ & $-0.226$ & $0.461$ & -95.8 & $-101$ & $26.2$  \\[0.2cm]
Cont. C1 & C2 & $-0.321$ & $0.840$ & $0.149$ & $-9.51$ & $-8.99$ & $-309$ \tabularnewline
Cont. CS3 & CS3 & $0.159$ & $0.782$ & $0.505$ & $-17.1$ & $118.1$ & $6.56$ \tabularnewline
\hline 
\end{tabular}
\caption{Orbital initial conditions for the clusters involved in each simulated interaction.}\label{tab2}
\end{table*}
%%%%%%%%%%%%%%%%%%%%%%%%%%
%%%%%%%%%%%TABLE%%%%%%%%%%%
\begin{table*}
\centering
\begin{tabular}{lccccccccccc}
\hline 
ID & $M_f$ (M$_\odot$) & $M_{f,1}$ (M$_\odot$)  & $M_{f,2}$ (M$_\odot$) & $r_c$ (pc)& $r_h$ (pc)& $r_t$ (pc)& $b/a|_c$ & $c/a|_c$ & {$e$} & {$r_p$} (kpc) & {$r_a$} (kpc)\tabularnewline
\hline 
CM1   &  $9.4\times10^6$  & $4.7\times10^6$ & $4.7\times10^6$ & $1.0$ & $9.5$ & 82   & $0.85$ & $0.84$& {$0.7$} & {$0.68$} & {$6.9$}\tabularnewline
CM2   & $5.1\times10^6$  & $2.6\times10^6$  & $2.5\times10^6$ & $2.2$ & $7.3$ & 114 &$0.97$ & $0.84$ & {$0.9$} & {$0.12$} & {$1.84$}\tabularnewline
CM12 & $8.9\times10^6$  & $9.3\times10^5$  & $8.0\times10^6$ & $1.4$ & $4.7$ & 99   & $0.85$ & $0.84$ & {$0.6$} &{$0.17$} & {$2.2$}\tabularnewline
\hline 
\end{tabular}

\caption{{ Structural and orbital} parameters of the clusters resulting from the mergers at the end of the simulation (12Gyr). The final mass of the clusters ($M_f$) and the mass belonging to each of the progenitors ($M_{f,1}$ and $M_{f,2}$)
are listed together with the core, half-mass and tidal radii ($r_c$, $r_h$ and $r_t$). The axial ratios at the core radius ($b/a|_c$ and $c/a|_c$) and the eccentricity ($e$), perigalacticon ($r_p$) and apogalacticon ($r_a$) are also shown. }\label{tab3}

\end{table*}
%%%%%%%%%%%%%%%%%%%%%%%%%%

%%%%%%%%%%%FIGURE%%%%%%%%%%%
\begin{figure*}
\centering
\includegraphics[width=0.45\textwidth]{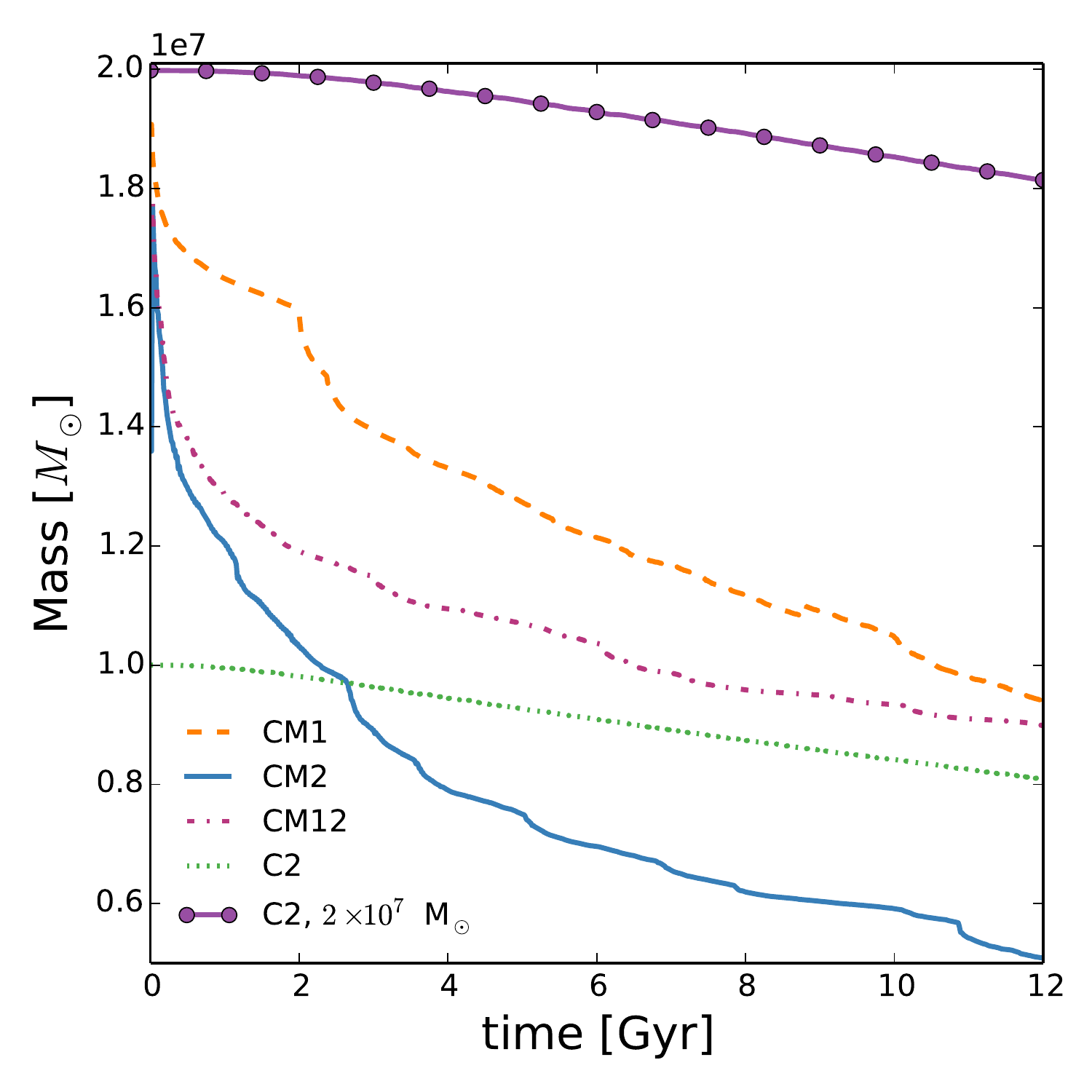}~\includegraphics[width=0.45\textwidth]{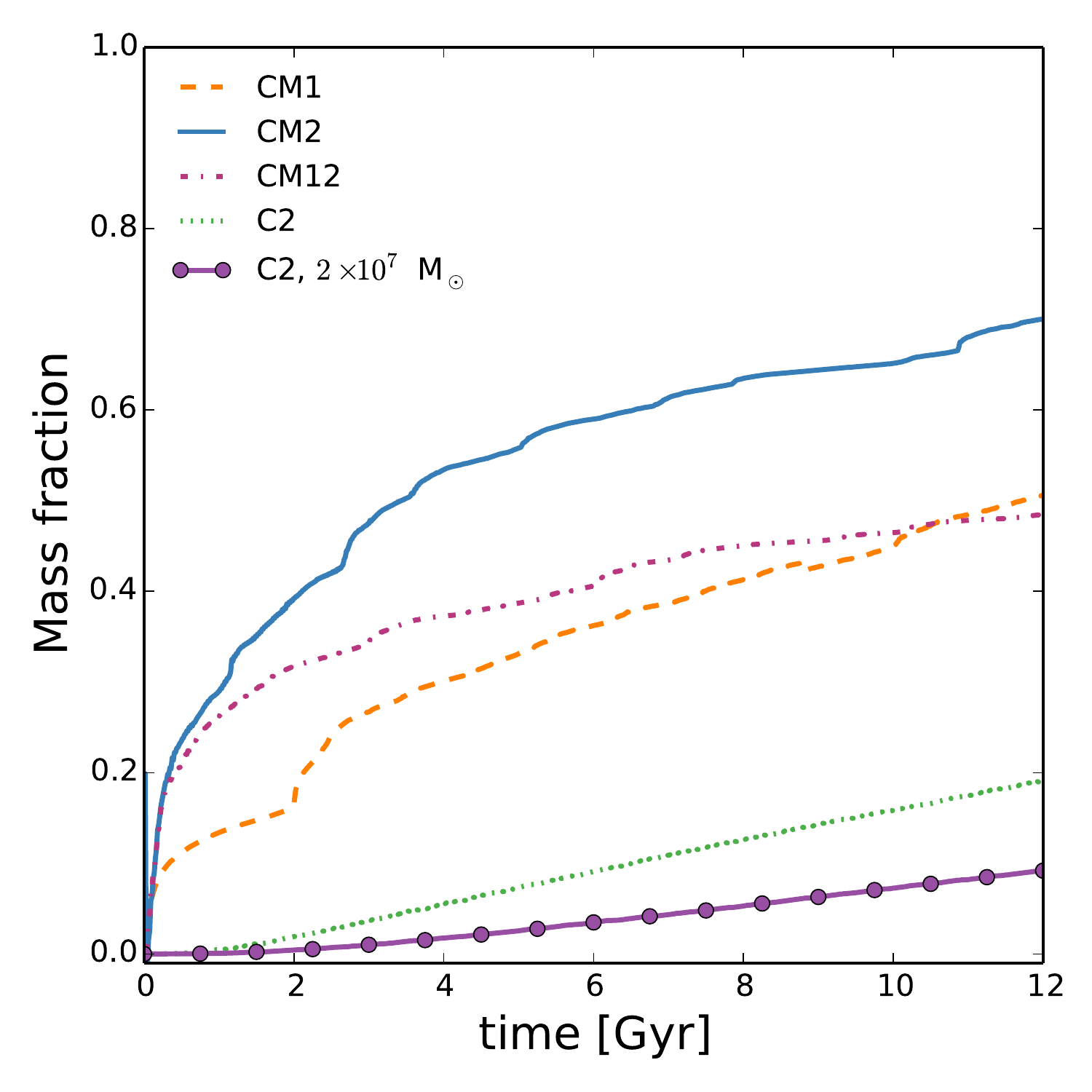}
\caption{Mass lost by the GCs during and after the merger. The orange dashed line is for CM1, the blue solid line is for CM2 and purple dot-dashed line is for CM12. The green dotted line { and the purple solid line with bullets are for the non-interacting C2-like clusters, respectively with $10^7$M$_\odot$ and $2\times10^7$M$_\odot$,} moving on the  orbit of CM2. The mass loss is large in the case of CM2, intermediate in the case of CM1 and CM12 and low in case of the isolated cluster.} \label{fig:massloss1}
\end{figure*}
%%%%%%%%%%%%%%%%%%%%%%%%%%
%%%%%%%%%%%%%%%%%%%FIGURE%%%%%%%%%%%%%%%
\begin{figure*}
\centering
\includegraphics[width=0.3\textwidth, trim=0.5cm 0.cm 4.7cm 1cm, clip=true]{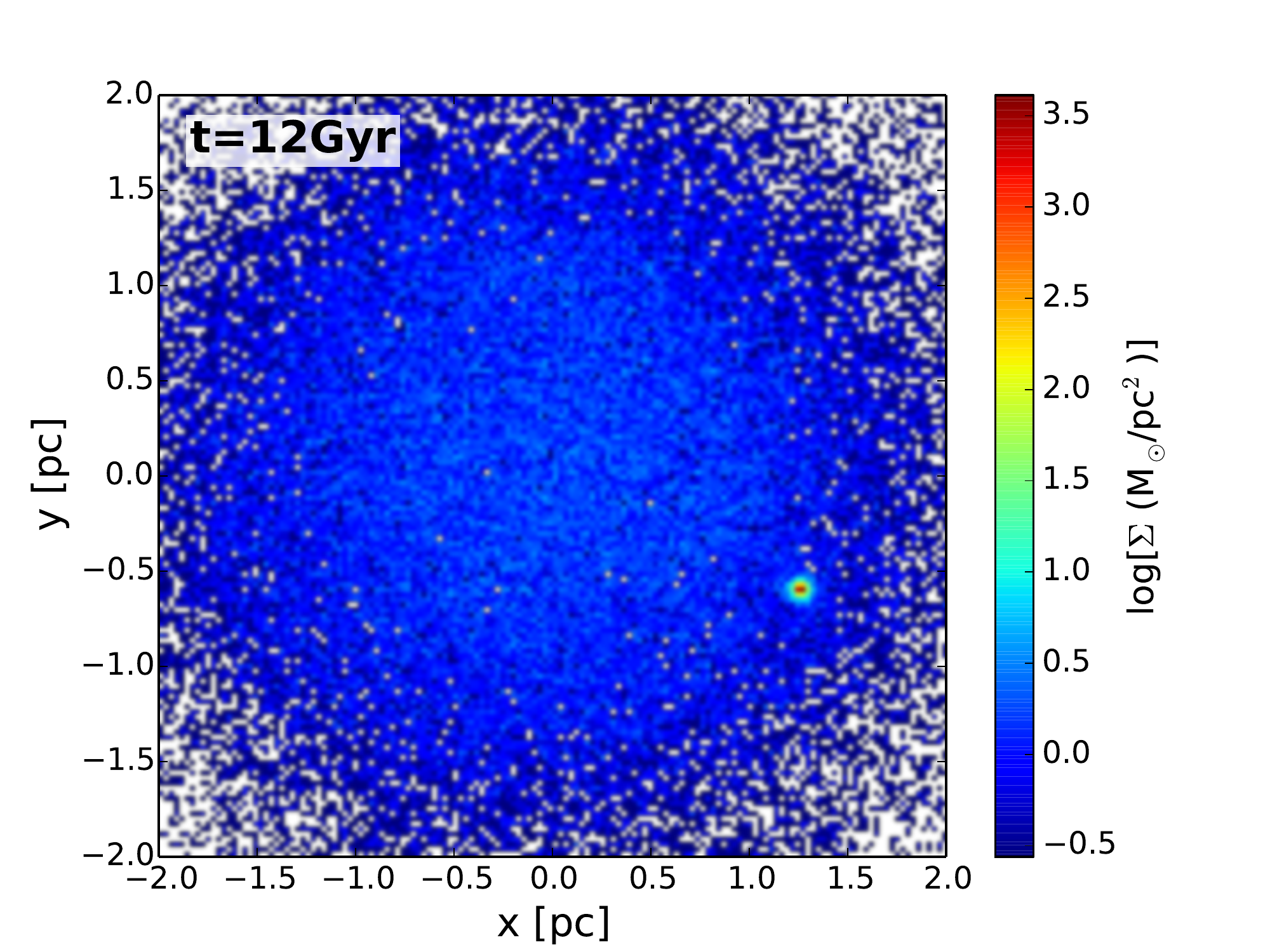}~\includegraphics[width=0.3\textwidth, trim=0.5cm 0cm 4.7cm 1cm, clip=true]{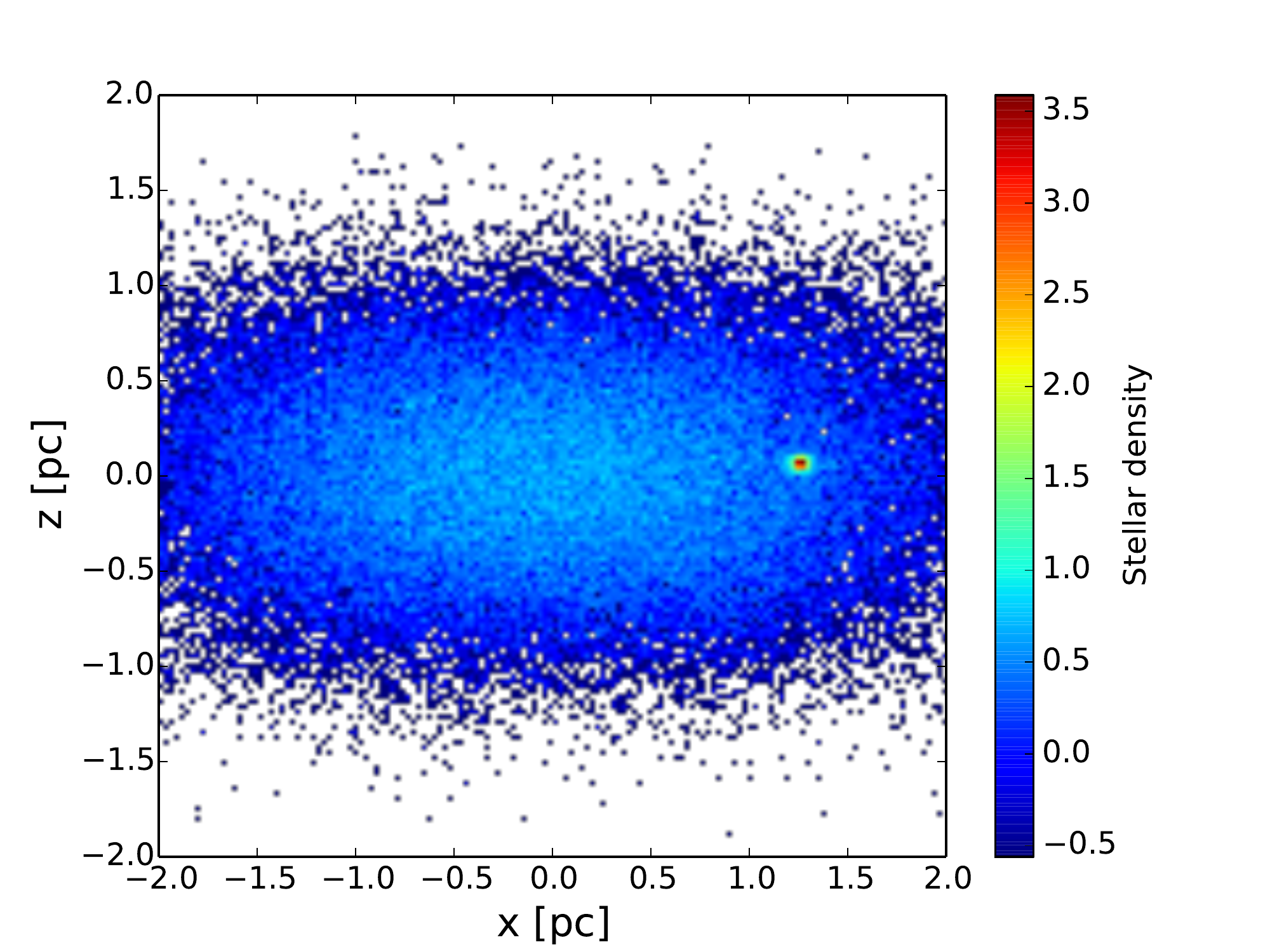}~\includegraphics[width=0.38\textwidth, trim=0.5cm 0cm 1cm 1cm, clip=true]{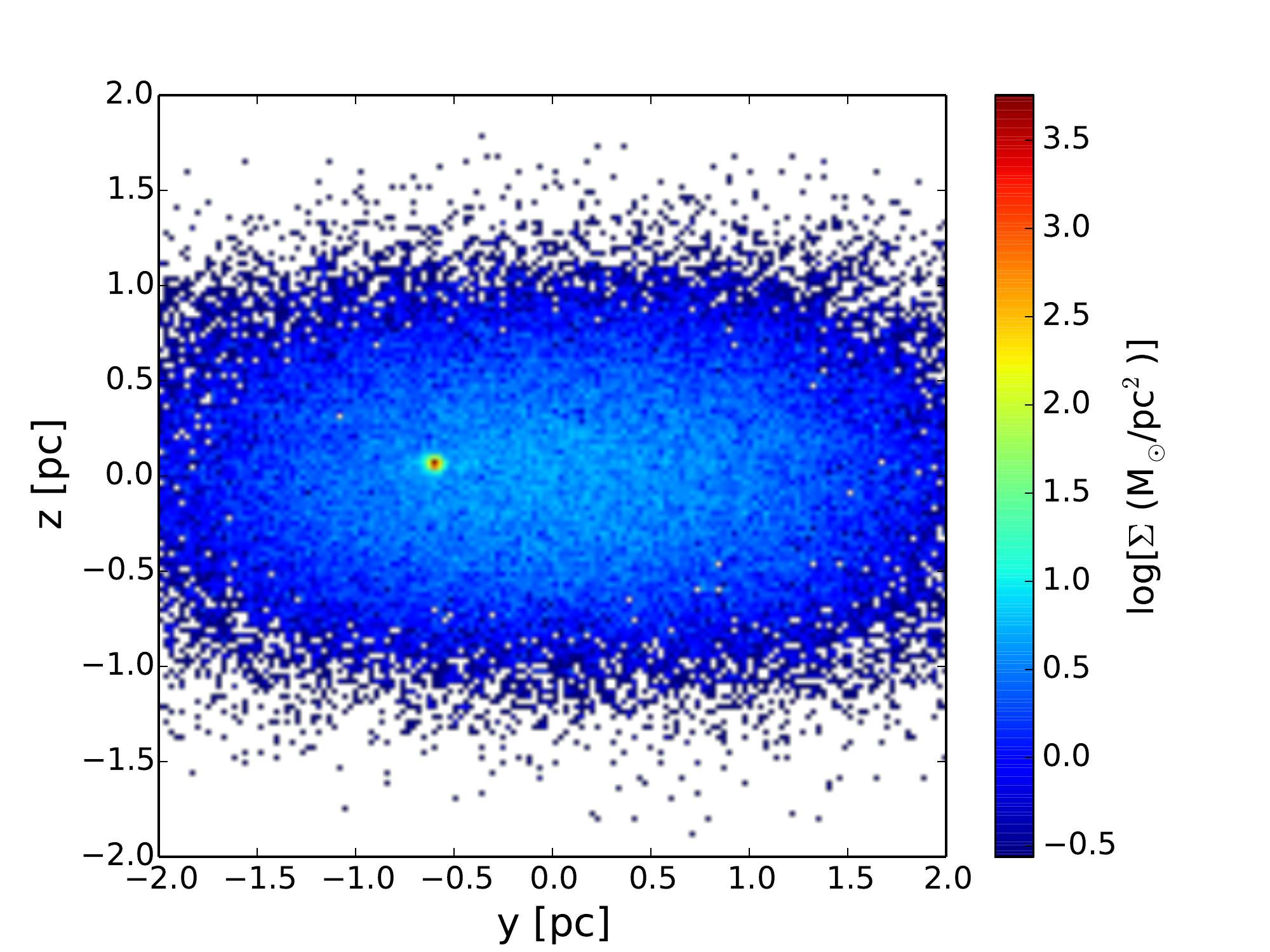}
\caption{Density maps of a  $2$kpc$\times2$kpc region after CM2 has evolved for 12Gyr. The origin is at the Galactic centre. Projections on the xy (left panel), xz (middle panel) and yz (right panel) planes are shown.
During and after the merger, the composite GC loses its stars along its orbit producing a low density background that mixes with and contributes to the thick disc population. The GC is the denser, approximately round overdensity visible in each of the panels. }\label{fig:gcdisc}
\end{figure*}
%%%%%%%%%%%%%%%%%%%%%%%%%%%%%%%%%%%%%%
\begin{figure}
\centering
\includegraphics[width=0.45\textwidth]{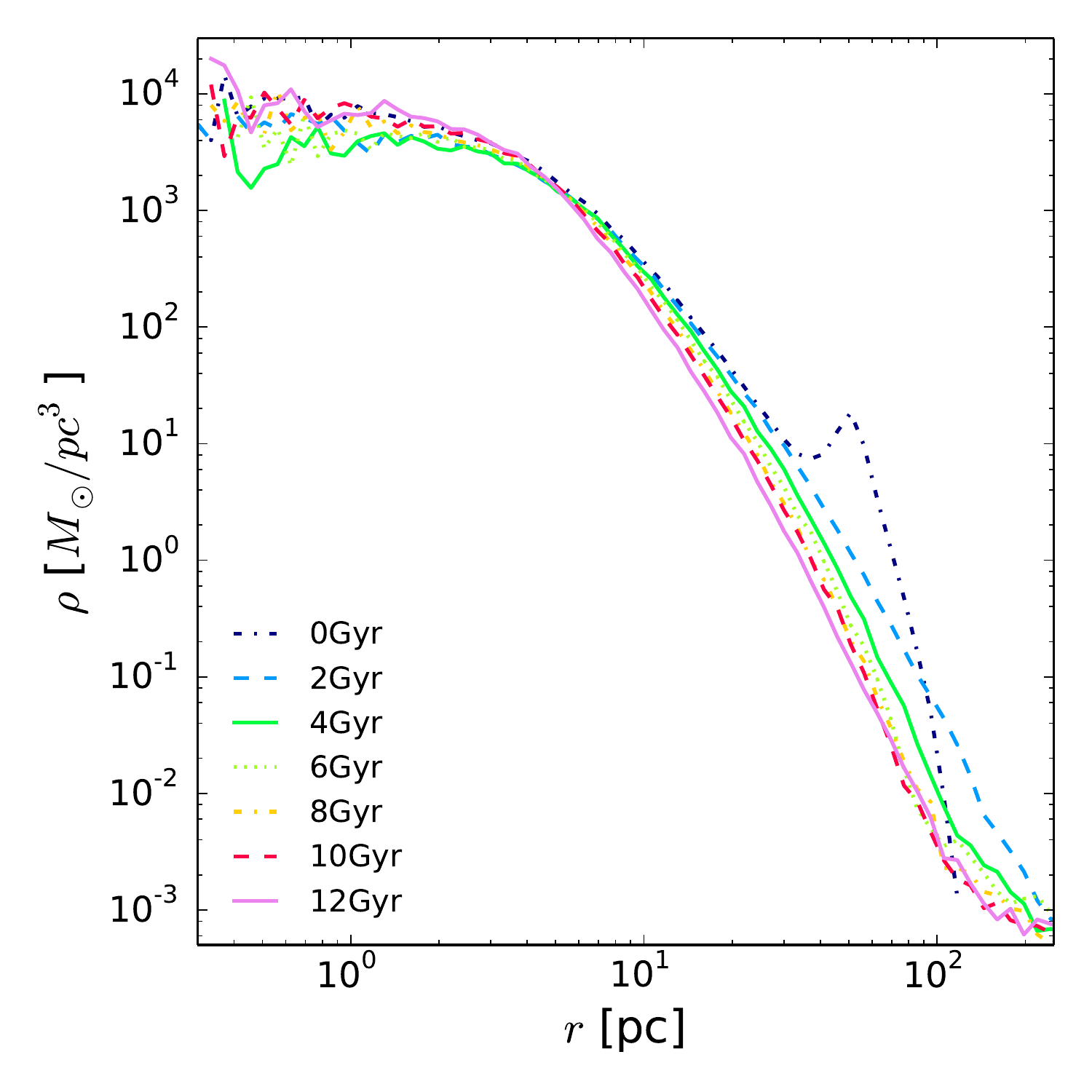}
\caption{Density evolution with time for CM2. This cluster results from the merger of two C2 clusters. The two populations quickly mix and show the same density immediately after the first phases of the merger. The density is shown at the beginning of the simulation and after 2, 4, 6, 8, 10 and 12 Gyr.}\label{fig:den_t}
\end{figure}
%%%%%%%%%%%FIGURE%%%%%%%%%%%
\begin{figure}
\centering
\includegraphics[width=0.45\textwidth]{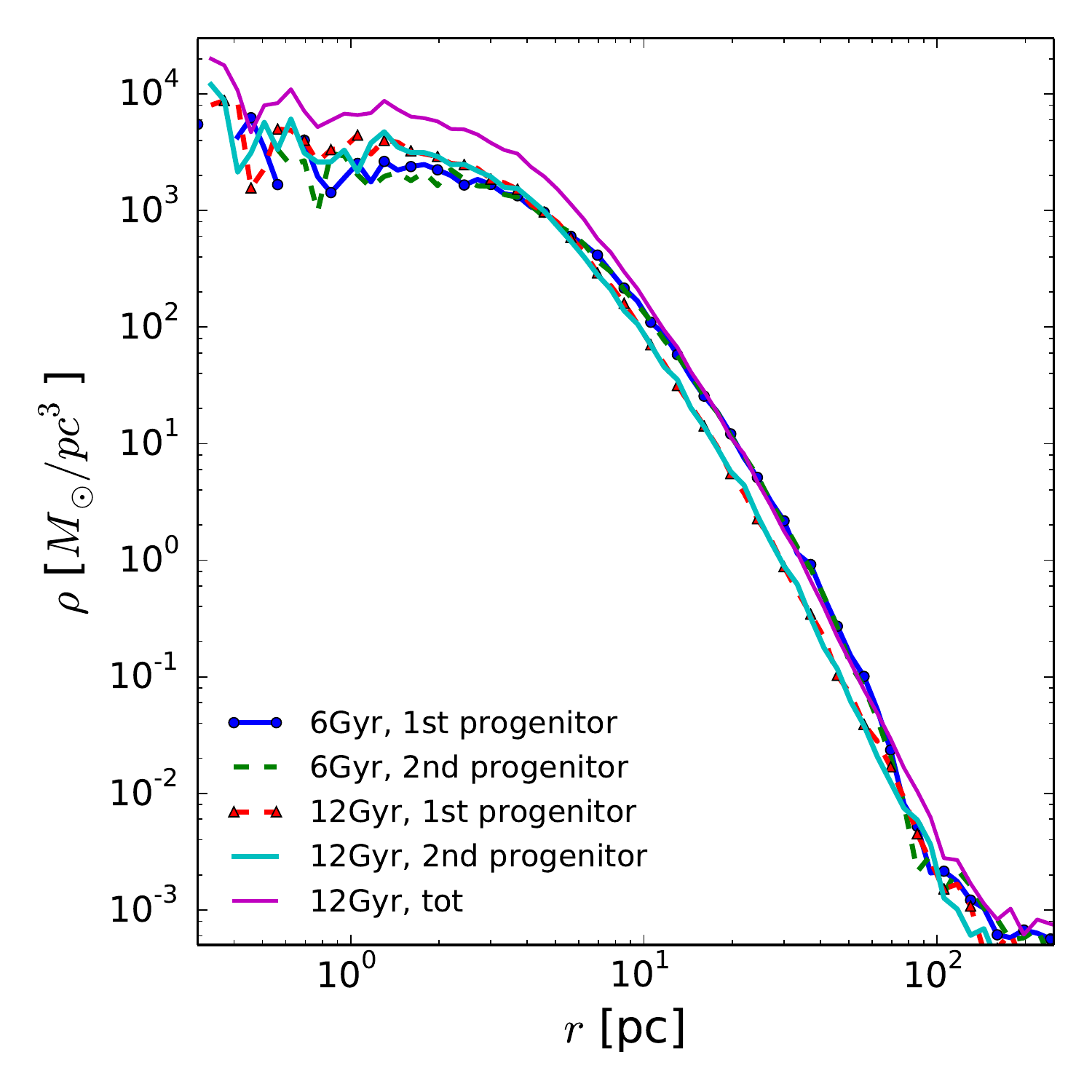}
\caption{Density of the two populations forming CM2 after 6 and 12 Gyr. The two populations are fully mixed and have the same density profile. The total density profile (solid purple line) is compatible with those of the massive Galactic GCs.} \label{fig:den_popsCM2}
\end{figure}
 %%%%%%%%%%%FIGURE%%%%%%%%%%%
\begin{figure}
\centering
\includegraphics[width=0.45\textwidth]{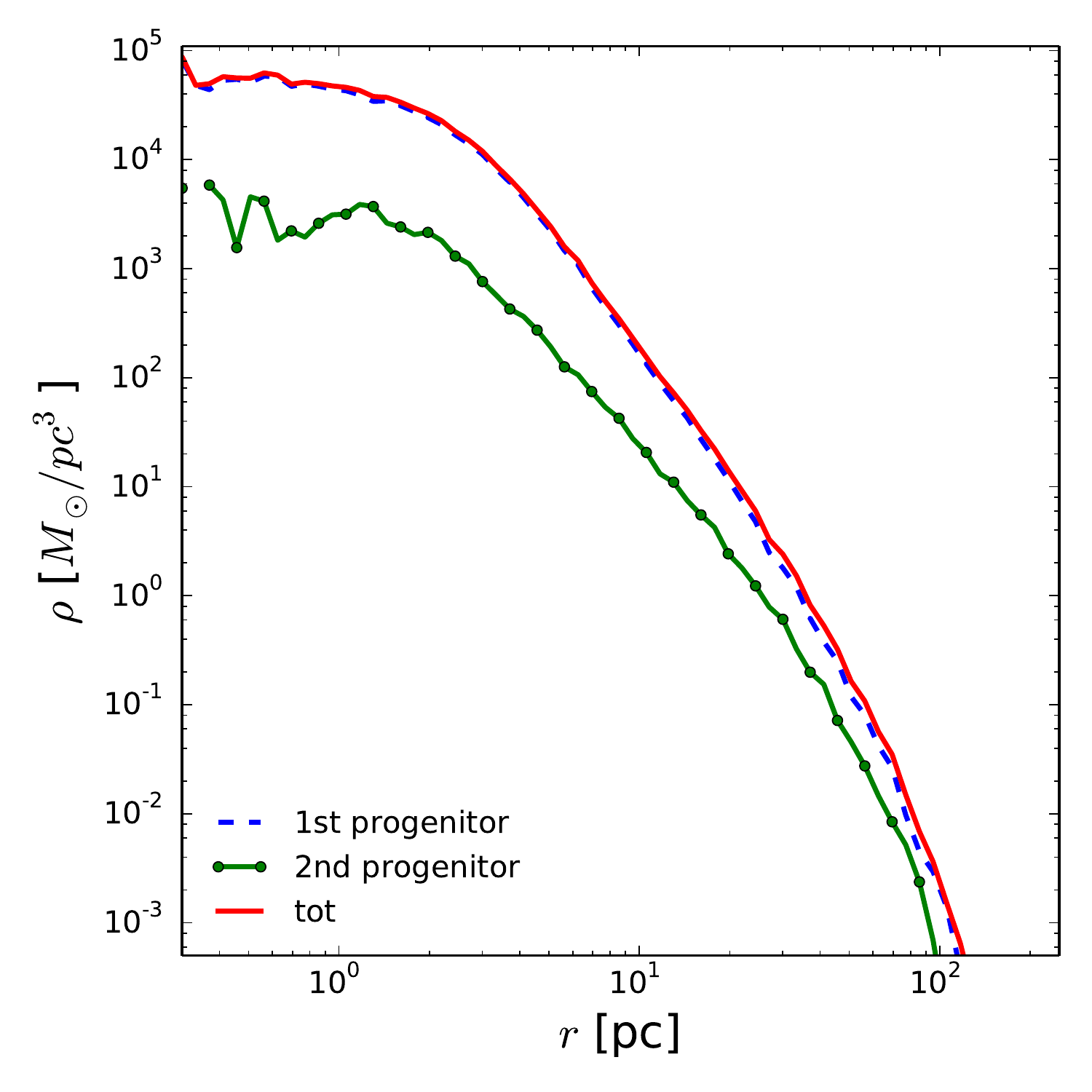}
\caption{Density of the two populations forming CM12 after 12 Gyr of evolution. The stars initially belonging to C1 (green solid line with dots) are less concentrated and extended than the stars inherited from C2 (dashed blue line). The final cluster is dense and massive (red solid line).}\label{fig:denCM12}
\end{figure}
%%%%%%%%%%%FIGURE%%%%%%%%%%%
\begin{figure*}
\centering
\includegraphics[width=0.45\textwidth]{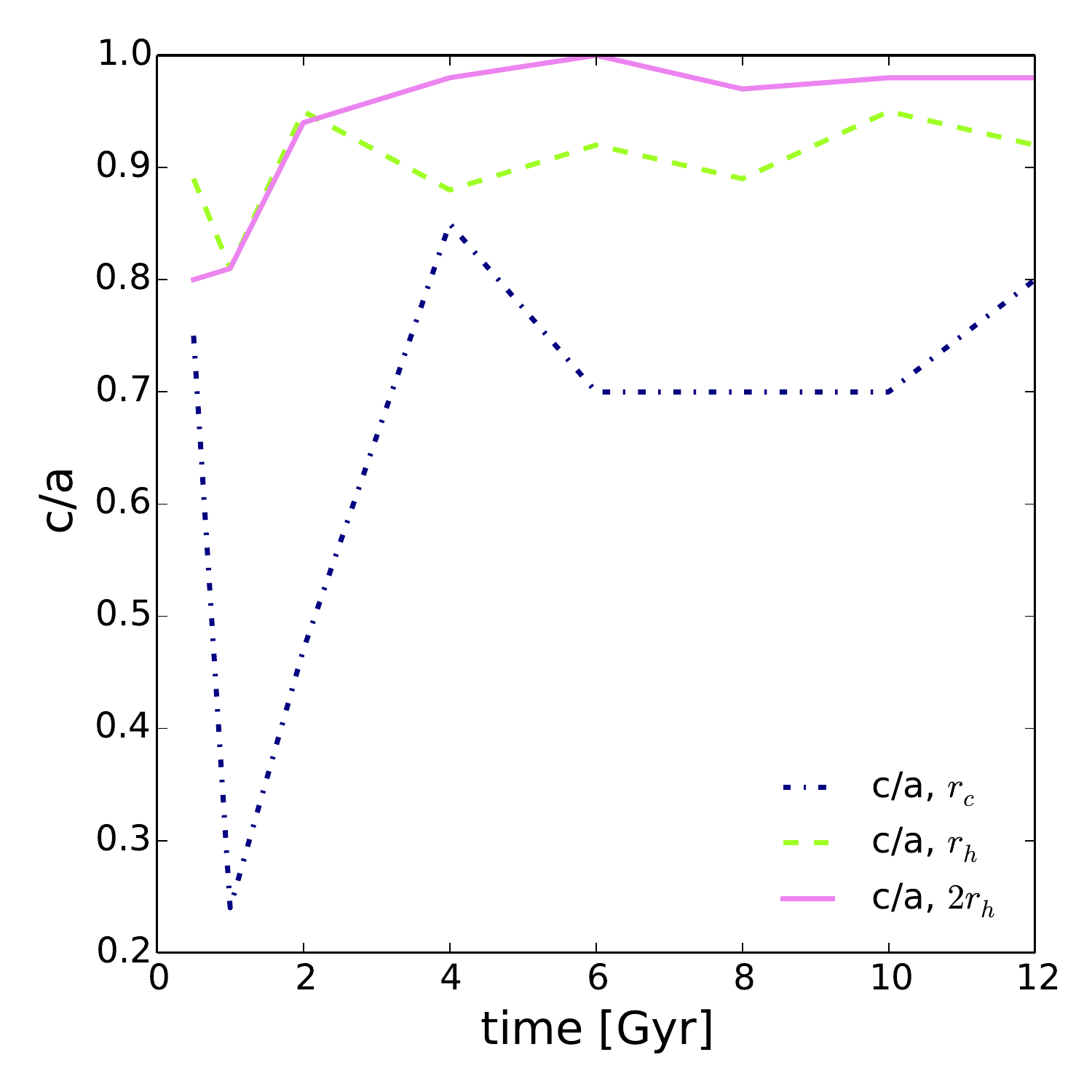}~\includegraphics[width=0.45\textwidth]{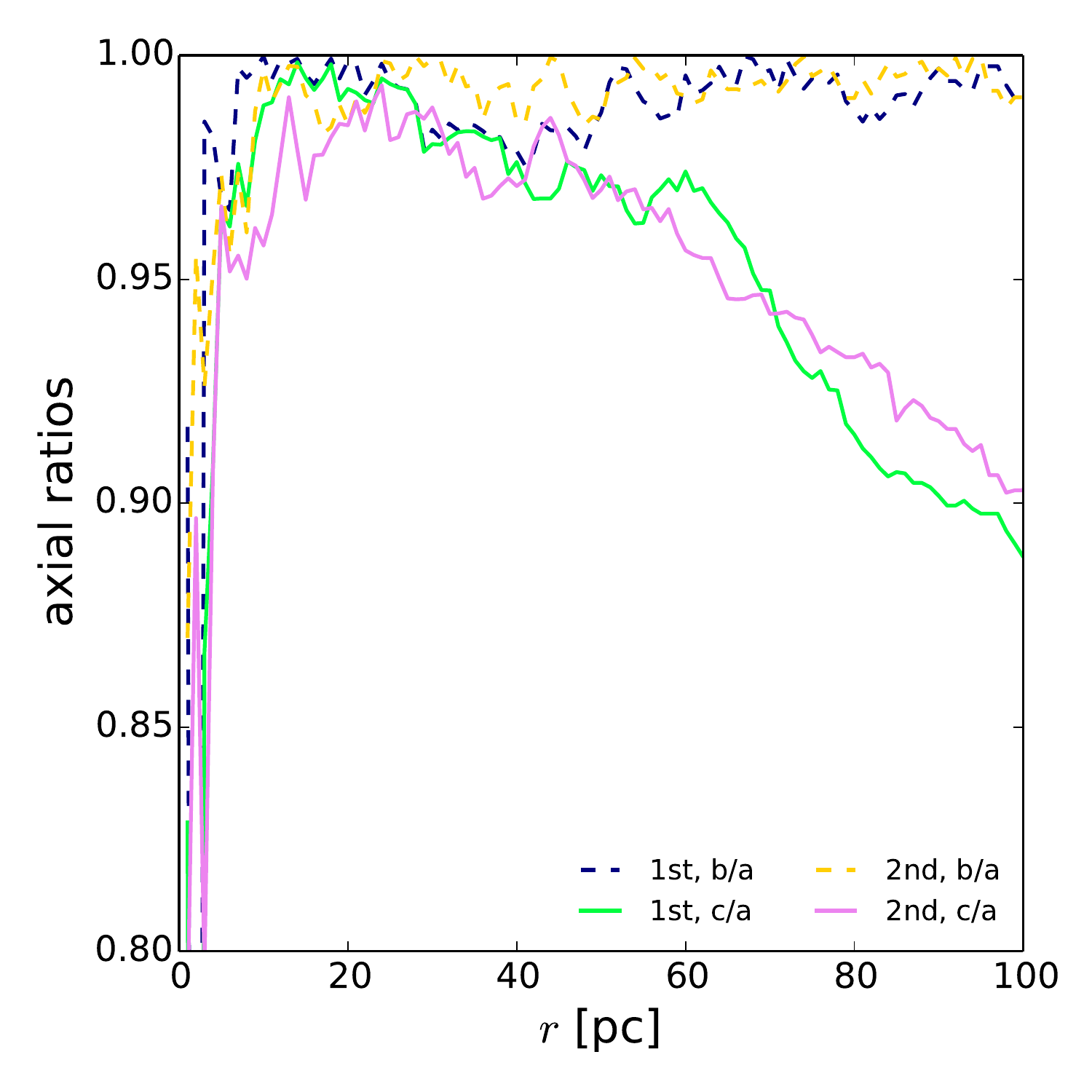}
\caption{Axial ratio $c/a$ evolution with time for CM2 at different radii (left panel). The axial ratios for the two { stellar components coming from the progenitors that form} CM2 (right panel).}\label{fig:ev_ax_ratios}
\end{figure*}
%%%%%%%%%%%FIGURE%%%%%%%%%%%
\begin{figure}
\centering
\includegraphics[width=0.45\textwidth]{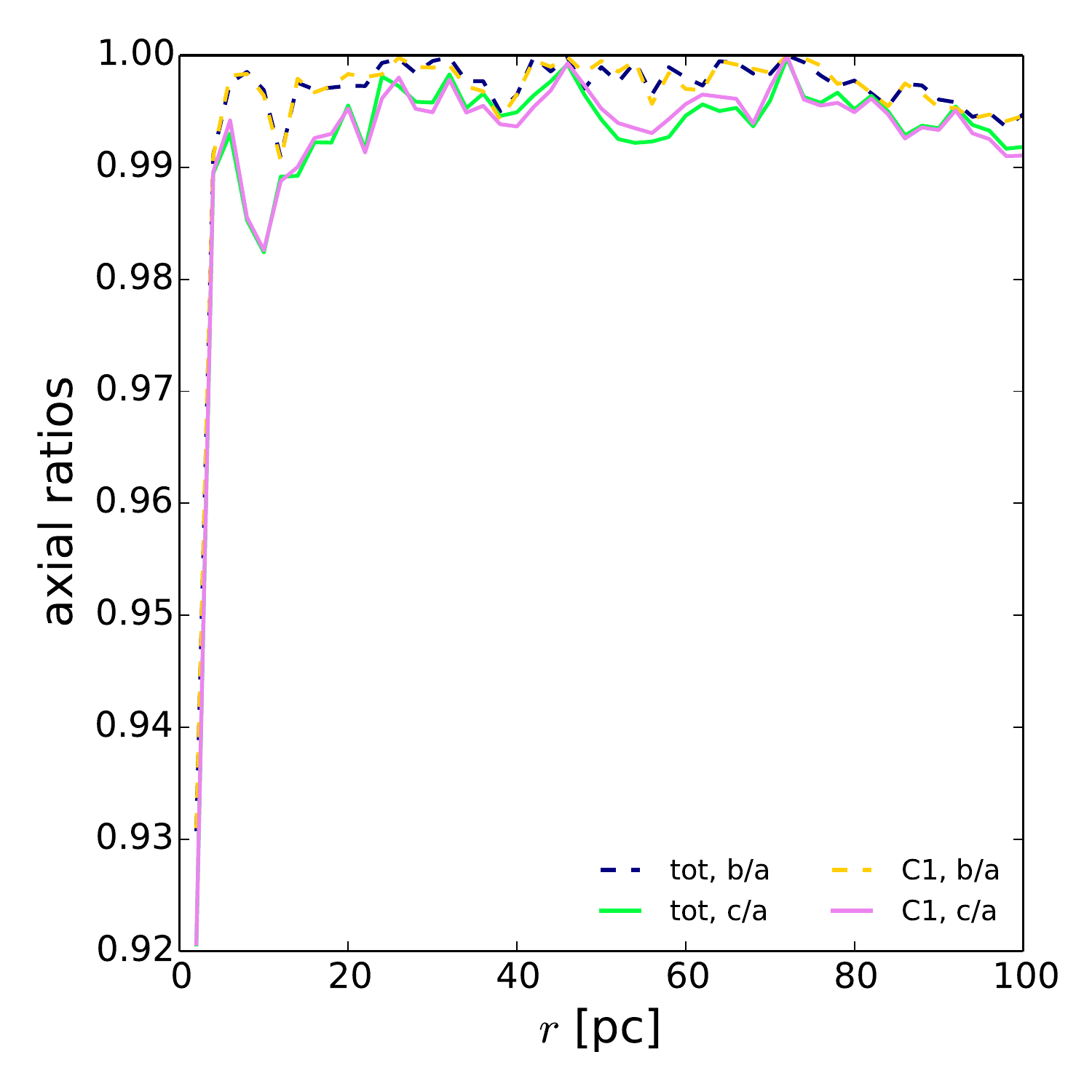}
\caption{The final axial ratios for the C1 including the particle stripped from CS3 (green solid and blue dashed lines) or excluding them (magenta solid and yellow dashed lines).}\label{fig:ax_ratios_stripping}
\end{figure}
 %%%%%%%%%%%%%%%%%%%%%%%%%%%

%%%%%%%%%%%FIGURE%%%%%%%%%%%
\begin{figure*}
\centering
\includegraphics[width=0.3\textwidth]{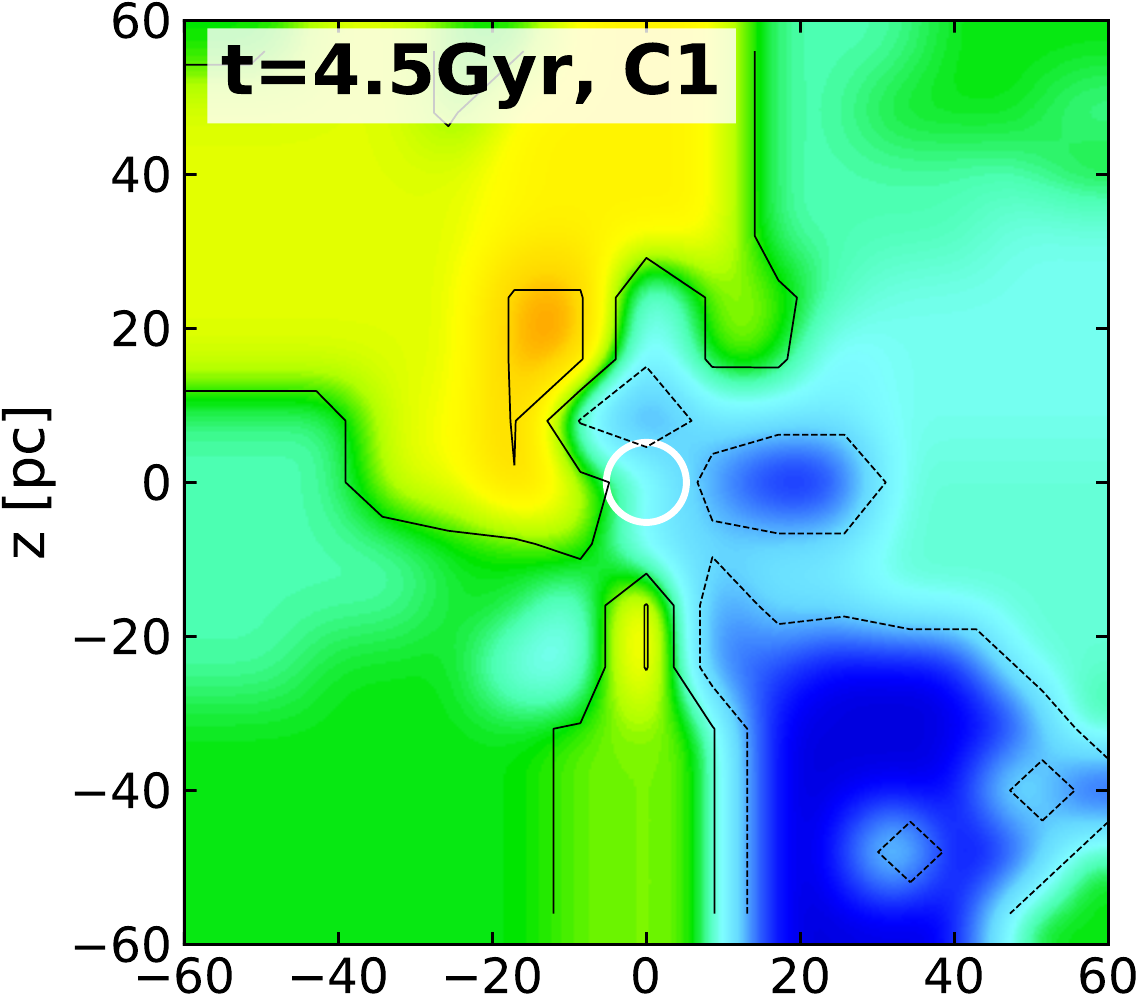}~\includegraphics[width=0.282\textwidth]{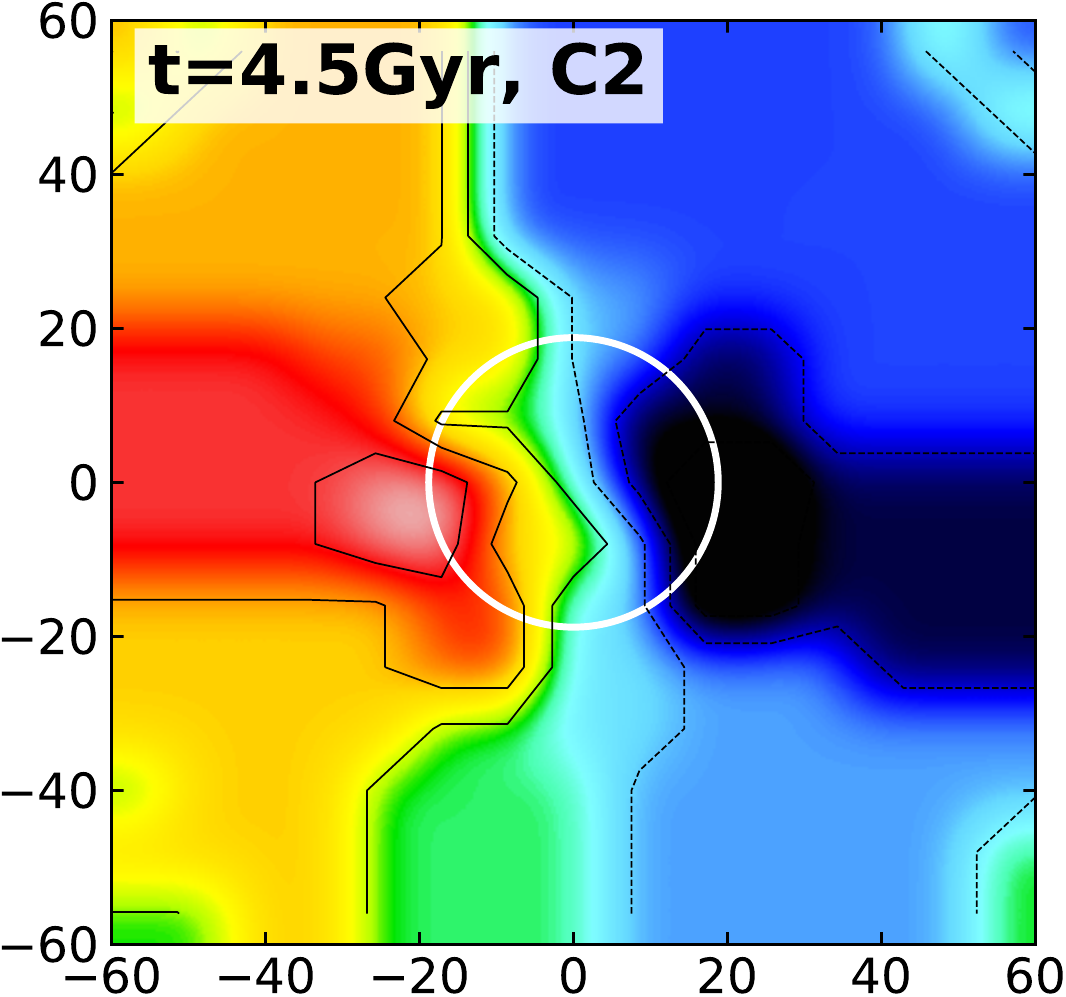}~\includegraphics[width=0.345\textwidth]{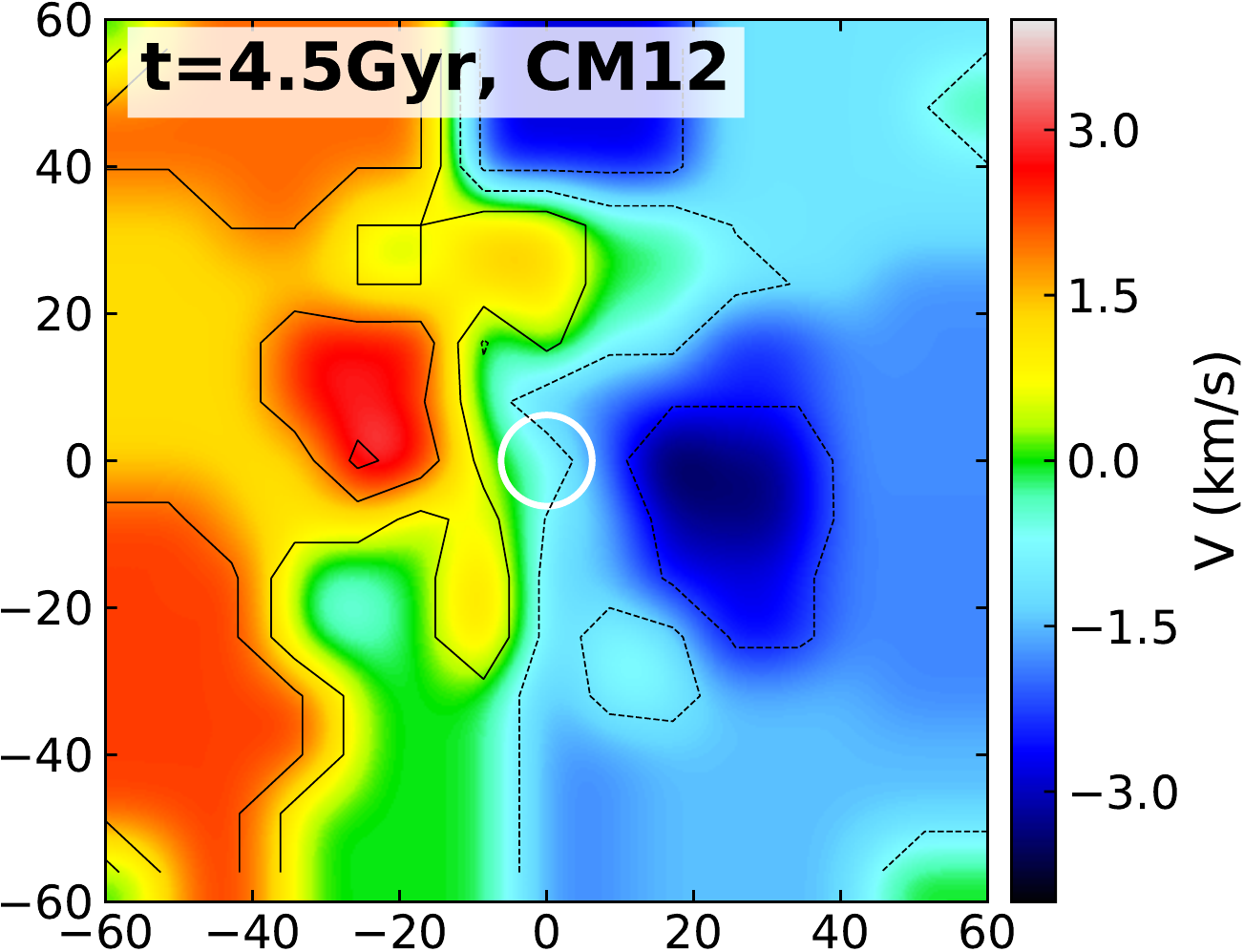}\\
\includegraphics[width=0.305\textwidth]{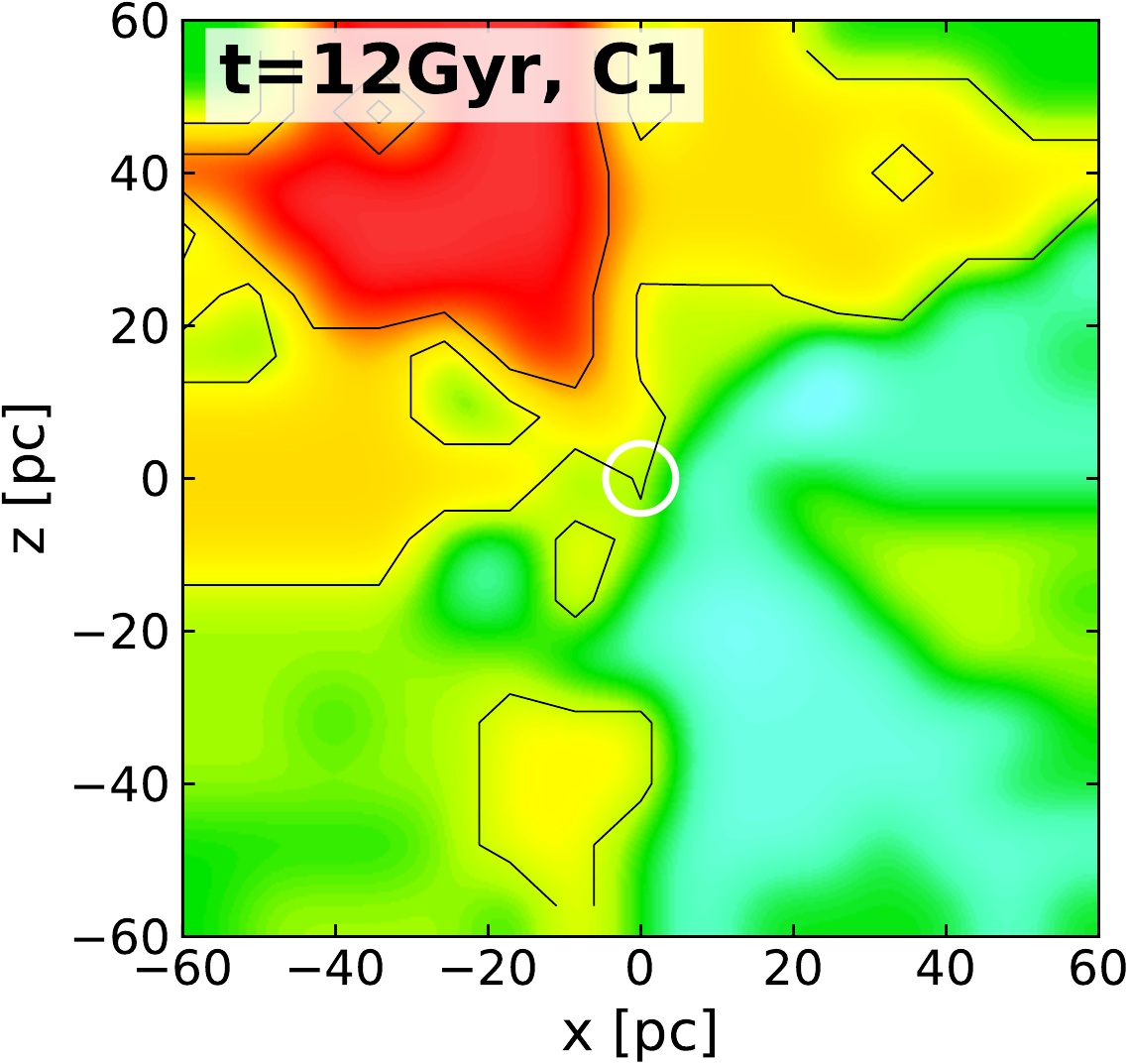}~\includegraphics[width=0.282\textwidth]{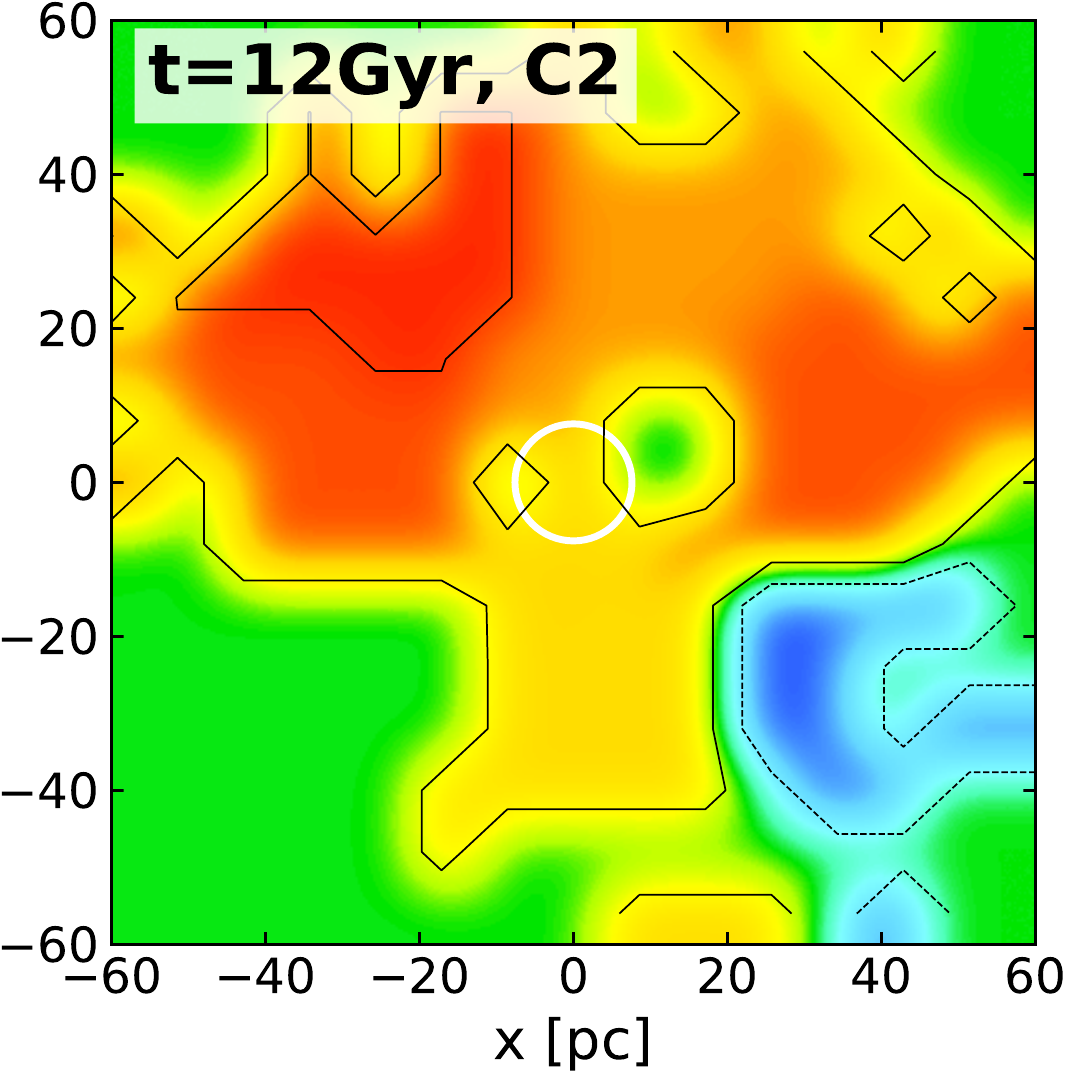}~\includegraphics[width=0.345\textwidth]{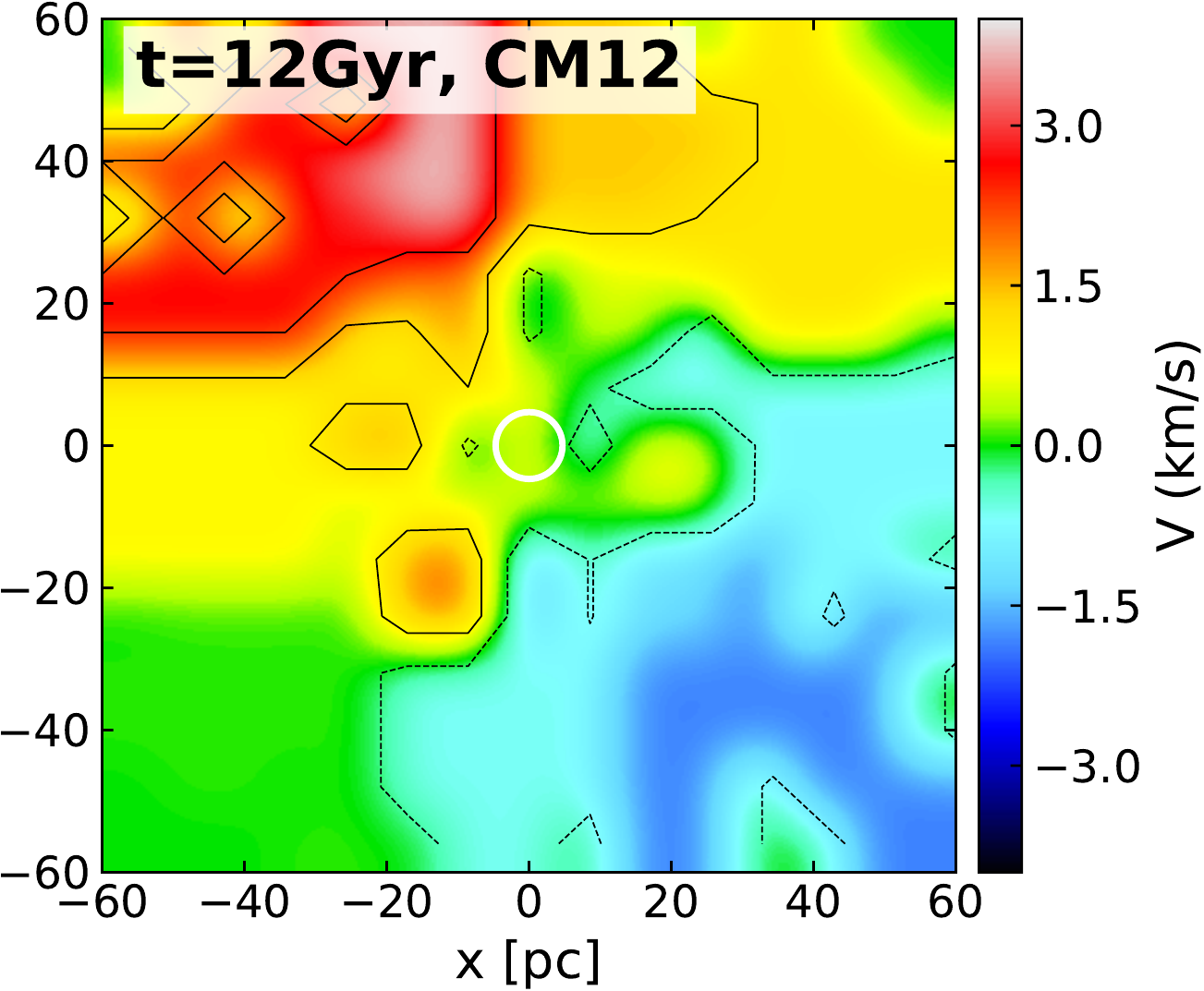}
\caption{Velocity maps of the two { populations} forming CM12 ({ i.e. the stars initially belonging to C1 and C2, identified within the merger product  CM12}), left and middle panels) after 4.5 and 12 Gyr (top and bottom panels). The velocity map of the whole CM12 is shown after 4.5 Gyr (top right panel) and after 12 Gyr (bottom right panel). The cluster and { each individual population coming from a different progenitor} are seen edge on, i.e. perpendicularly to the maximum angular momentum of the system. The white circle represents the half mass radius of each subsystem.}\label{fig:rotCM12}
\end{figure*}

\section{Results}\label{sec:res}
\subsection{Mergers between initially massive clusters}
{ 
In K18, we investigated the statistics of the encounters, finding that a population of 100 { massive} disc GCs undergoes 1.8 physical collisions per Gyr, { as well as several cases of mass exchange, during close passages}.
 Here we follow-up these results, focusing on single merger or mass exchange cases aiming at a detailed study of the long-lasting dynamical imprints of these interactions.  
Among the several simulated close encounters between different combinations of $N$-body models (C1, C2, CS1, CS2, CS3, see Table \ref{tab1}) moving on different orbits, }
three made of $10^7M_\odot$ clusters merged forming a new GC. One of these pairs was initially composed by two C1 clusters, one by two C2 clusters and one had a C1 and a C2 model (see Sect. \ref{sec:nbody}). 
%All these clusters are modelled using $51151$ $N$-body particles.
We will refer to the cluster resulting from the first merger as CM1, the one resulting from the second merger as CM2 and the one coming from the third (composite) merger as CM12. { The orbital initial conditions for the different simulated pairs and the IDs of the resulting clusters are listed in Table \ref{tab2}.}
The structural { and orbital} parameters for the final clusters are summarised in Table \ref{tab3}.
In all cases the merger takes place almost immediately, and in less than $100$Myr, after the beginning of the simulation { see Sect. \ref{sec:nbody}}. The relative distance between the centre of density of the two progenitors rapidly decreases from the initial value of $\sim50$pc to $<1$pc (see left panel of Figure \ref{fig:reldv}). 
At the same time, the relative velocity decreases from $\sim100$ km/s to $\approx 2$km/s (see right panel of Figure \ref{fig:reldv}).
%We then expect slight different centres for the two populations originating from the merger, accompanied by a negligible difference in their kinematics.
In the first phase of the process, the two clusters start interacting and developing a tidal bridge (see top panels of Figure \ref{fig:merger}).
After few close passages and within $1$\,Gyr from the first close encounter the two clusters have completely merged to form a { mixed} system, that then evolves like a new, 
independent, GC. 
While the central density of the final cluster increases with time, its outskirts become less bound
and outer stars are lost by the cluster leading to a smaller tidal radius.
This is quantitatively seen in Figure \ref{fig:merger}, where the evolution of the two C2 clusters that merge to form CM2 is shown from the moment of the closest approach 
up to 12 Gyr of evolution. 
Figure \ref{fig:gcdiscCM12} shows the final density contours plots for the whole system (left panel) and the stars initially belonging to each of the progenitors (middle and right panels).
While in the case of CM2 (top panels, CM1 is analog to this case) the two progenitors provide the final cluster with a similar amount of mass and they are almost completely spatially mixed after $12$ Gyr, in the case of CM12 (lower panels) the stars inherited from  C1  account for a significantly smaller mass and are less concentrated and extended than the ones coming from C2 (see Table \ref{tab3}). This final configuration reflects the initial profiles of the progenitors and lead to potentially observable spatial differences between the two populations.
{ The same  close encounters described above (see Table \ref{tab2} for the details on the orbital parameters) have been simulated replacing C1 with CS1 and C2 with CS2. However, these ``less massive'' encounters did not lead to any merger or mass exchange. Since the orbital conditions are the same as the ones adopted for the more massive clusters, we speculate that the main factor driving the mergers is the gravitational focusing exerted by the GCs, which has to be strong enough to overcome the action of the Galactic tidal field. Nevertheless, since the parameter space to explore is extremely large, more studies are necessary to find out if and in which conditions less massive clusters can interact in the Galactic disc.}

\subsection{Mass loss and density profile evolution over a Hubble time} 
We calculated the GC mass as the amount of bound (i.e. with total energy $E<0$) particles that are within 2 times the initial tidal radius of the cluster (see Table \ref{tab1}).
Figure \ref{fig:massloss1} shows the mass as a function of time (left panel) and the fraction of mass lost (right panel) by CM1, the first merger event we encountered. This merger that happens between two dense standard C1 clusters leads to a final cluster of mass
$9.4\times10^6M_\odot$. Each of the progenitors contributes for $50$\% of this mass. The cluster is relatively massive because of the high initial density of its progenitors. CM1 that moves on an orbit with apocenter $3.9$kpc and pericenter $0.68$kpc, which correspond to an eccentricity $e=0.7$. 
%CM1 moves on a new orbit with apocenter $3.9$kpc and pericenter $0.68$kpc and eccentricity $e=0.7$.
In the case of CM2, which results from the merger of two C2 less dense and more extended clusters with respect to the C1 model, the mass, immediately after the merger, is 12\% smaller than the total initial mass of the progenitors. The violent interaction causes an expansion of the interacting systems with a consequent
increase of the mass loss rate to a level not observed for an isolated cluster moving on the same orbit (green dashed line in Figure \ref{fig:massloss1}). 
After this violent initial phase, {  that lasts for about 100 Myr}, the cluster virialises and becomes stable, however the mass loss rate remains significant due to the low density of the cluster and on its orbital high  eccentricity ($e=0.9$) and low pericentre ($0.12$ kpc, the apocentre is $1.84$ kpc). In $\sim3$ Gyr the cluster loses half of their initial mass (see  Figure \ref{fig:massloss1}). { Afterwards} the mass loss rate slows down; in the following $9$\,Gyr of evolution the system loses more mass and reaches a final mass of $5.1\times10^6$\,M$_\odot$ ($\approx 2.5\times 10^6$\,M$_\odot$ in each population), comparable to the most massive clusters observed in the Milky Way ($\omega~Cen$, M54, Terzan 5).

A large fraction of the total mass of the cluster (more than 75\%, i.e. $1.5\times10^7M_\odot$) is lost in the disc, as shown in Figure \ref{fig:gcdisc}. However, since the
stars are redistributed on a large volume, their density is low compared to that of the merger remnant. 
We notice that if, as suggested in K18, every Gyr two $10^7M_\odot$ clusters had a merger, we predict that up to $1.5\times10^8M_\odot$ of the disc/bulge, which means over 1/100 of the total mass of the bulge could come from disrupted GCs.

The result of the collision and merger of a C1 and of a C2 cluster, CM12, loses approximately the same amount of mass lost by CM1 but at different rate. Within $\approx 2$ Gyr, CM12 loses $\approx 40\%$ of its initial mass. %At later times the mass loss rate decreases significantly because the cluster has stabilised after the merger.  
The final mass of this cluster is $8.9\times10^6$M$_\odot$ and it comes mostly ($8.0\times10^6$M$_\odot$) from the initially more concentrated { progenitor} (C1). The population inherited from C2 accounts only for 10\% of the total mass of CM12. This cluster moves on orbit with apocenter equal to $2.2$kpc and pericenter of $0.17$kpc, corresponding to $e=0.6$.

%In the following we focus on the results concerning CM2 (analogue to one of the mergers found by K18) and CM12.
Figure \ref{fig:den_t}  shows  the evolution of the  density of  CM2, the case of CM1 is analogue to this one. The central density slightly increases with time as a consequence of the cluster contraction.
The two populations composing the cluster are characterised by similar density profiles (see Figure \ref{fig:den_popsCM2}). The mixing is quick and already effective after $6$\,Gyr of evolution. 
%However, our cluster is composed by particles more massive than stars. This decreases the relaxation time of a factor ..., thus underestimating the mixing time.

%In the second case of merger, which involves two C1 clusters, the density evolution is similar to what found for CM2 and the two components become quickly mixed. 
In the case of the merger between two different clusters, a C1 and a C2, the result of the process is significantly different from what is illustrated above.  
As can be seen in Figure \ref{fig:denCM12}, the two populations forming CM12 have significantly different masses and spatial distributions. The stars coming from C1, which was initially the densest cluster, are more centrally concentrated than the ones coming from C2, that contributes only to 10\% of the mass of CM12.
{ As a comparison, we simulated a non-interacting C2, non rotating, model with $10^7$M$_\odot$, and another one with the same initial mass of the merger products ($2\times10^7$M$_\odot$). These clusters move on the same orbit as CM2 and their centres of mass are in (-1.40, -0.14, 0.23)kpc, with velocity (-130, 43.9, 1.60)km/s.
The non-interacting C2 models, after $12$ Gyr of evolution, have lost only
between $\sim10$\% and $\sim20$\% of their initial mass (dotted orange line in both panels of Figure \ref{fig:massloss1}). The violent interaction and the lower cluster density following a merger enhances by a factor $\sim3.5$ the mass loss rate on the same orbit.}

\subsection{Mass exchange and mutual contamination}
We  simulated a different case with two ``standard'' clusters with masses  of $3\times10^6M_\odot$ (CS3) and $10^7M_\odot$  (C1) respectively at their closest encounter. 
This case did not lead to a full merger, however, we observe a stripping operated by C1 at the expenses of CS3. The captured mass accounts for a fraction of percent of the total mass of C1 and it redistributes in the outskirts of the accreting cluster. This is in agreement with what found in K18. 
%The fraction of accreted mass depends both on the geometry of the encounter and on the structure of the clusters that interact. 
% The extra-tidal material of the final cluster is an equal-density mixture of material originating from both clusters. 
During the interaction, CS3 accretes a smaller fraction of stars that remain in the outskirt of the cluster. 

We notice that, although this is the only stripping case we simulated, many more are expected to happen between a full population of disc GCs. As shown by K18, mass exchanges are more frequent than full mergers and could be an important source of contaminations in Galactic GCs. 
%Because of computational limitations, we only simulated one encounter involving standard clusters, however we expect that in the case of two C2-like clusters more mass would be stripped from the less massive cluster which is also characterised by a more extended and less dense envelope. 

\subsection{Morphology, anisotropy and rotation}
In order to provide possible merger signatures we explored the morphology and kinematics of the clusters produced in our simulations.
The left panel of Figure \ref{fig:ev_ax_ratios} shows the minor axial ratio, $c/a$\footnote{$b/a$ is never significantly different from $1$.}, for CM2 calculated at different radii ($r_c$, $r_h$ and $2r_h$) { using the moment-of-inertia
tensor \citep[see][]{Ka91},} as a function of time. We are only showing this case because there are no relevant differences between the three simulated mergers. 
During the initial phases of the merger, the cluster is highly  
triaxial.  At later evolutionary stages the cluster becomes oblate (see axial ratios reported in Table \ref{tab3} and Figure \ref{fig:ev_ax_ratios}). 
At the core radius the cluster remains highly flattened through all its evolution. The flattening is $\approx0.9$ throughout all the evolution at the half mass radius, while at twice this radius the cluster is initially flattened and it becomes almost spherical in $2$-$4$ Gyr.
The stellar component belonging to the first progenitor is less centrally flattened than the one belonging to the second progenitor, and are in a more prolate configuration in the outskirt of the cluster (see right panel of Figure \ref{fig:ev_ax_ratios}). However these differences are negligible and would be observationally challenging to detect.
{ We note that the progenitors, evolved individually on the same orbit, without any encounter and not including any initial internal rotation, would have remained approximately spherical.} 
Therefore, the flattening observed for CM2 can be mostly attributed to the merger process. We, therefore, expect the clusters that went through mergers to show larger ellipticities than those that never experienced this kind of interactions. 

The clusters involved in the stripping remain almost perfectly spherical even after $12$ Gyr, and, as shown in Figure \ref{fig:ax_ratios_stripping}, the stars captured by C1 have a negligible effect on its morphology.

After $12$ Gyr of evolution there is no signature of significant rotation in any of the two populations forming CM1 and CM2. However, these systems are slightly tangentially anisotropic ($\beta\sim-0.1$\footnote{The anisotropy parameter is defined as  $\beta=1-\frac{\sigma_t^2}{2\sigma_r^2}$}). This anisotropy is observed in only one of the two populations, while the other one is almost isotropic and shows only a minor radial anisotropy ($\beta\sim0.01$). These features are also find in the case of CM12, which results from the merger of clusters with different initial densities.

Figure \ref{fig:rotCM12}  shows the velocity maps of CM12 and of its { two populations, inherited from the progenitors C1 and C2,} after $4.5$ Gyr (upper panels) and $12$ Gyr (lower panels) of evolution. Each of system is projected perpendicularly to its maximum angular momentum vector. The maps are obtained  by applying the Voronoi binning procedure described by \cite{Cap03} with a fixed S/N ratio of 15 in each bin\footnote{We assume poissonian noise; the signal to noise is $\sqrt{N}$, where $N$ is the number of particles in the bin.}. { Due to the merger, both progenitors, which were initially non-rotating, gain a different amounts of angular momentum.} The rotation { consequently acquired by CM12} is still clear after $4.5$\,Gyr and it decreases with time, from a peak velocity larger than $4.5$ km/s to a more irregular rotation pattern with a peak around $2$\,km/s (bottom right panel). { While at $4.5$ Gyr the  population initially belonging to C2 rotates faster than the one initially belonging to C1, after $12$ Gyr the situation is reversed. C2, indeed, efficiently loses its  angular momentum because of the strong mass loss affecting this stellar component. Additionally, part of the angular momentum gained during the merger by C2 is redistributed among C1 stars. The final rotation signal is observable only outside the half-mass radius, where the stars are less crowded, possibly allowing more detailed observations. This misalignment could be a distinctive signature of the merger, which, however, will need to be confirmed through the exploration of a larger parameter space for the mergers. CM12 velocity maps, produced projecting the cluster randomly respect to the LOS are presented in the Appendix \ref{AAp}.}
The system has no significant velocity anisotropy, however, due to the large amount of mass being lost from the cluster, C2 stars show a slightly larger radial anisotropy than the C1 component.

\section{Discussion} \label{sec:disc}
Globular clusters host several stellar populations, showing chemical anomalies in light elements
\citep[see][and references therein]{Gra12}. As a further complication to this picture, a small fraction
of Galactic GCs show metallicity spreads of the order of $0.1$ dex \citep{Mar15}. The origin of these clusters, which are among the most massive ones in the Galaxy, is still highly debated.
While $\omega$ Cen and M54 are strongly suspected to be the former nuclei of dwarf galaxies that contributed to the assembly of the Galaxy \citep[see e.g.][]{Be08, BF03}, mergers in dwarf galaxies
later accreted by the Milky Way have been proposed as a possible explanation for the presence of iron spreads in the remaining clusters \citep{Gav16}.

In K18 we found that close passages between two or more clusters belonging to a primordial population of Galactic disc GCs could lead to $1.8$ { physical collisions per Gyr. These collisions can lead to mergers, fly-bys and mass exchanges.}
This analytic result motivated { the} short-term $N$-body simulations run using 128 massive ($10^7$M$_\odot$) C2-like clusters, each modelled with $10^5$ particles {  presented in K18}. In { this simulation, in} $1.5$ Gyr the GC system experienced two major mergers and several interactions with consequent mass contaminations between two or more clusters. { K18 found} that only clusters on similar orbits can merge or have fly-bys.
Here we followed-up these results, focusing on cluster pairs that, while orbiting a realistic Galactic potential, experience close passages. We modelled each simulated cluster using different King models with a range of masses.
In { the explored configurations} only massive clusters ($10^7$M$\odot$) can merge. As { also found} in K18, { where we considered all clusters were modelled using the same density profiles and masses}, if the two clusters have the same properties, the primordial populations are quickly spatially mixed  and no long-lasting { differential} signature is left in the kinematics or morphology of the cluster. 
However, if the clusters have different initial densities{, as in the case producing our CM12}, the more compact { progenitor} retains most of its mass after the merger and produces a centrally concentrated population. The progenitor with a lower initial density contributes only for 10\% of the total mass of the final cluster and gives rise to the least dense population. 
%The less dense cluster is completely destroyed after the merger and it contributes for only 10\% the  mass of the final GC.
The two populations rotate differentially and the new cluster shows a net rotation whose amount depends on the cluster evolutionary stage and on the parameters of the impact. In all analysed cases the outskirts of the cluster are slightly flattened, with the axial ratio reaching a minimum of $c/a=0.9$, while the internal regions show larger ellipticities (see Figures \ref{fig:ev_ax_ratios} and Table \ref{tab3}).
As in K18, some of the encounters lead to small contaminations between different clusters. The material accreted during these mass exchanges only accounts for a fraction of percent of the total mass of each of the involved clusters. These kind of events could have happened frequently in the disc and could have involved clusters of any masses  (see K18).
We therefore predict a wide range of contaminations, accounting for less than one percent up to half of the total mass of the clusters. 
%If the chemical composition, and in particular the iron content, depends on the Galactocentric radius at the epoch of formation, we expect that the interacting clusters would have only a slight iron difference. . 
We notice that we only analysed few cases and further work is necessary to investigate in detail the conditions under which a merger can happen.
The phase-space to explore is wide, however, we proved that mergers in the primordial disc of the Galaxy can be frequent and might generate clusters similar to the 
massive ones observed in the Milky Way (e.g. Terzan 5). We also predict a range of contaminated mass fractions in several large and intermediate-large mass clusters. 
There is no reported correlation between GCs showing disc kinematics and the presence of iron spreads, however part of the initial disc GC population could have been tidally destroyed or might have migrated to the Galactic halo because of disc heating caused by satellite accretions \citep{K15}. Long-term signatures like flattening { and, if the merging clusters have different structural parameters, differential} rotation and spatial distributions for each population could be potentially observable and provide hints on the past violent life of interacting disc GCs. 

\section{Conclusions}\label{sec:concl}
In this paper we illustrated how massive primordial thick disc GCs can merge or have mass exchanges, a fact that could explain the metallicity spreads observed in a small, but growing \citep{Ma18}, fraction of the massive Galactic GCs. 
The main results achieved by studying of the long term-evolution of interacting clusters can be summarised as follows:

\begin{itemize}
\renewcommand\labelitemi{--}
\item When considering clusters with masses of $10^7M_\odot$, we found three couples, one consisting of two C1 clusters, one formed by two C2 clusters and one including a C1 and a C2 cluster, that had a full merger leading to  new clusters, namely CM1, CM2 and CM12. Another pair of clusters with masses $3\times10^6M_\odot$ (CS3) and  $10^7M_\odot$ (C1) did not merge but rather stripped some material from each other, leading to a mutual contamination. This is the only case of stripping we simulated but many more are expected to happen (see K18).
\item The clusters merge quickly, showing tidal bridges and tails in the initial phase of the process, and become more centrally concentrated with time. After $12$ Gyr of evolution they are similar to the massive Galactic GCs.
\item CM1 loses $\sim50$\% of its initial mass while CM2, which orbits closer to the Galactic center on a more eccentric orbit, loses $\sim75$\% of its initial mass, producing a final cluster of $\sim5\times10^6M_\odot$. CM12 inherits 90\% of its mass from C1 and the rest from C2. 
\item If the clusters are similar their populations mix very quickly. If one of the clusters is denser it will destroy the other dense one accreting part of its mass.
The two populations will have distinct density profiles that mirror the initial properties of the progenitors. 
\item The mass lost by the clusters is distributed along their orbits and spreads all over the thick disc. Even if only few clusters went through a merger, the mass lost by them should account for a non negligible fraction of the disc mass (few percents) and should be observable with future astrometric/spectroscopic surveys.
%\item The density profiles of the two populations become almost immediately indistinguishable. Their centres of mass and velocity are slightly displaced. However the displacement is within observational errors.
\item Clusters that merge are significantly flattened up to the core mass radius. The cluster becomes almost spherical at the tidal radius.
\item Different populations  show different degrees of velocity anisotropy. When merging two clusters with different initial density profiles the final cluster rotates, and the
two populations rotate differentially. { The direction perpendicular to the rotation plain observed outside the half-mass radius is misaligned with respect to the direction of the total angular momentum of the system.}
\item In the case of the stripping the more massive and denser cluster accretes a fraction of percent of its initial mass during the fly-by with a less massive and less dense cluster, that in turn accretes a smaller fraction of mass from its companion. 
%The amount of accreted mass would have been more significant for less dense and more extended clusters.
\item The accreted stars are distributed outside the core of the cluster. The accreting GC remains almost spherical throughout its lifetime.
\end{itemize}
We find that mergers between Galactic GCs are possible
in the thick disc. This kind of events could explain the metallicity spread observed in few Galactic GCs. Massive clusters can merge and consequently lose most of their mass, leading to a cluster which is similar to the massive GCs currently observed in the Milky Way. 
%Moreover, we expect correlations between the chemical and spatial/kinematical distributions of stars in Galactic GCs with metallicity spreads. 
Following the results described here and  K18 we can robustly conclude that mergers or mass exchanges could have happened in the past, possibly explaining in a  straightforward way the origin of metallicity spreads in  massive GCs.

%%%%%%%%%%%%%%%%%%%%%%%%%%%%%%%%%%%%%%%%%%%%%%%%%%
\begin{acknowledgements} 
The authors thank Nadine Neumayer and the anonymous referee for their useful comments.
AMB acknowledges support by Sonderforschungsbereich
(SFB) 881 ``The Milky Way System'' of the German Research Foundation
(DFG). This work has been supported by ANR (Agence Nationale de la Recherche) through the MOD4Gaia project (ANR-15-CE31-0007, P.I.: P. Di Matteo). 
This work was granted access to the HPC resources of CINES under the allocation 2017-040507 (PI : P. Di Matteo) made by GENCI.
%PDM thanks the Max Planck Institute f\"ur Astronomie for their hospitality
%during several visits while this Paper was in progress.
\end{acknowledgements}



\bibliographystyle{aa}
\bibliography{gcm} % if your bibtex file is called example.bib

\begin{thebibliography}{77}
\expandafter\ifx\csname natexlab\endcsname\relax\def\natexlab#1{#1}\fi

\bibitem[{{Allen} \& {Santillan}(1991)}]{AS91}
{Allen}, C. \& {Santillan}, A. 1991, RMxAA, 22, 255

\bibitem[{{Amaro-Seoane} {et~al.}(2013){Amaro-Seoane}, {Konstantinidis},
  {Brem}, \& {Catelan}}]{AS13}
{Amaro-Seoane}, P., {Konstantinidis}, S., {Brem}, P., \& {Catelan}, M. 2013,
  \mnras, 435, 809

\bibitem[{{Arnold} {et~al.}(2017){Arnold}, {Goodwin}, {Griffiths}, \&
  {Parker}}]{Ar17}
{Arnold}, B., {Goodwin}, S.~P., {Griffiths}, D.~W., \& {Parker}, R.~J. 2017,
  \mnras, 471, 2498

\bibitem[{{Bekki} \& {Freeman}(2003)}]{BF03}
{Bekki}, K. \& {Freeman}, K.~C. 2003, \mnras, 346, L11

\bibitem[{{Bekki} \& {Tsujimoto}(2016)}]{Bek16}
{Bekki}, K. \& {Tsujimoto}, T. 2016, ApJ, 831, 70

\bibitem[{{Bellazzini} {et~al.}(2008){Bellazzini}, {Ibata}, {Chapman},
  {Mackey}, {Monaco}, {Irwin}, {Martin}, {Lewis}, \& {Dalessandro}}]{Be08}
{Bellazzini}, M., {Ibata}, R.~A., {Chapman}, S.~C., {et~al.} 2008, \aj, 136,
  1147

\bibitem[{{B{\"o}ker}(2008)}]{Bo08}
{B{\"o}ker}, T. 2008, \apjl, 672, L111

\bibitem[{{Cappellari} \& {Copin}(2003)}]{Cap03}
{Cappellari}, M. \& {Copin}, Y. 2003, \mnras, 342, 345

\bibitem[{{Capuzzo-Dolcetta} {et~al.}(2011){Capuzzo-Dolcetta},
  {Mastrobuono-Battisti}, \& {Maschietti}}]{CMBM11}
{Capuzzo-Dolcetta}, R., {Mastrobuono-Battisti}, A., \& {Maschietti}, D. 2011,
  New Astron., 16, 284

\bibitem[{{Carretta}(2015)}]{Car15}
{Carretta}, E. 2015, ApJ, 810, 148

\bibitem[{{Carretta} {et~al.}(2009){Carretta}, {Bragaglia}, {Gratton}, \&
  {Lucatello}}]{Car09a}
{Carretta}, E., {Bragaglia}, A., {Gratton}, R., \& {Lucatello}, S. 2009, A\&A,
  505, 139

\bibitem[{{Carretta} {et~al.}(2010{\natexlab{a}}){Carretta}, {Bragaglia},
  {Gratton}, {Lucatello}, {Bellazzini}, {Catanzaro}, {Leone}, {Momany},
  {Piotto}, \& {D'Orazi}}]{Ca10}
{Carretta}, E., {Bragaglia}, A., {Gratton}, R.~G., {et~al.} 2010{\natexlab{a}},
  \aap, 520, A95

\bibitem[{{Carretta} {et~al.}(2007){Carretta}, {Bragaglia}, {Gratton},
  {Momany}, {Recio-Blanco}, {Cassisi}, {Fran{\c c}ois}, {James}, {Lucatello},
  \& {Moehler}}]{Car07}
{Carretta}, E., {Bragaglia}, A., {Gratton}, R.~G., {et~al.} 2007, A\&A, 464,
  967

\bibitem[{{Carretta} {et~al.}(2010{\natexlab{b}}){Carretta}, {Bragaglia},
  {Gratton}, {Recio-Blanco}, {Lucatello}, {D'Orazi}, \& {Cassisi}}]{Car10}
{Carretta}, E., {Bragaglia}, A., {Gratton}, R.~G., {et~al.} 2010{\natexlab{b}},
  \aap, 516, A55

\bibitem[{{Carretta} {et~al.}(2010{\natexlab{c}}){Carretta}, {Gratton},
  {Lucatello}, {Bragaglia}, {Catanzaro}, {Leone}, {Momany}, {D'Orazi},
  {Cassisi}, {D'Antona}, \& {Ortolani}}]{CGL10}
{Carretta}, E., {Gratton}, R.~G., {Lucatello}, S., {et~al.} 2010{\natexlab{c}},
  \apjl, 722, L1

\bibitem[{{Carretta} {et~al.}(2011){Carretta}, {Lucatello}, {Gratton},
  {Bragaglia}, \& {D'Orazi}}]{CGL11}
{Carretta}, E., {Lucatello}, S., {Gratton}, R.~G., {Bragaglia}, A., \&
  {D'Orazi}, V. 2011, \aap, 533, A69

\bibitem[{{Da Costa} {et~al.}(2014){Da Costa}, {Held}, \& {Saviane}}]{DC14}
{Da Costa}, G.~S., {Held}, E.~V., \& {Saviane}, I. 2014, \mnras, 438, 3507

\bibitem[{{D'Antona} \& {Caloi}(2004)}]{Dan04}
{D'Antona}, F. \& {Caloi}, V. 2004, \apj, 611, 871

\bibitem[{{de Mink} {et~al.}(2009){de Mink}, {Pols}, {Langer}, \&
  {Izzard}}]{DM09}
{de Mink}, S.~E., {Pols}, O.~R., {Langer}, N., \& {Izzard}, R.~G. 2009, \aap,
  507, L1

\bibitem[{{de Oliveira} {et~al.}(1998){de Oliveira}, {Dottori}, \&
  {Bica}}]{DO98}
{de Oliveira}, M.~R., {Dottori}, H., \& {Bica}, E. 1998, \mnras, 295, 921

\bibitem[{{Decressin} {et~al.}(2007){Decressin}, {Meynet}, {Charbonnel},
  {Prantzos}, \& {Ekstr{\"o}m}}]{De07}
{Decressin}, T., {Meynet}, G., {Charbonnel}, C., {Prantzos}, N., \&
  {Ekstr{\"o}m}, S. 2007, \aap, 464, 1029

\bibitem[{{Di Matteo}(2016)}]{Dim16}
{Di Matteo}, P. 2016, PASA, 33, e027

\bibitem[{{Dinescu} {et~al.}(1999){Dinescu}, {Girard}, \& {van Altena}}]{Di99}
{Dinescu}, D.~I., {Girard}, T.~M., \& {van Altena}, W.~F. 1999, \aj, 117, 1792

\bibitem[{{Ferraro} {et~al.}(2009){Ferraro}, {Dalessandro}, {Mucciarelli},
  {Beccari}, {Rich}, {Origlia}, {Lanzoni}, {Rood}, {Valenti}, {Bellazzini},
  {Ransom}, \& {Cocozza}}]{Fe09}
{Ferraro}, F.~R., {Dalessandro}, E., {Mucciarelli}, A., {et~al.} 2009, \nat,
  462, 483

\bibitem[{{Ferraro} {et~al.}(2016){Ferraro}, {Massari}, {Dalessandro},
  {Lanzoni}, {Origlia}, {Rich}, \& {Mucciarelli}}]{Fer16}
{Ferraro}, F.~R., {Massari}, D., {Dalessandro}, E., {et~al.} 2016, ApJ, 828, 75

\bibitem[{{Freeman}(1993)}]{Fr93}
{Freeman}, K.~C. 1993, in Astronomical Society of the Pacific Conference
  Series, Vol.~48, The Globular Cluster-Galaxy Connection, ed. G.~H. {Smith} \&
  J.~P. {Brodie}, 608

\bibitem[{{Gavagnin} {et~al.}(2016){Gavagnin}, {Mapelli}, \& {Lake}}]{Gav16}
{Gavagnin}, E., {Mapelli}, M., \& {Lake}, G. 2016, MNRAS, 461, 1276

\bibitem[{{Gratton} {et~al.}(2004){Gratton}, {Sneden}, \& {Carretta}}]{Gra04}
{Gratton}, R., {Sneden}, C., \& {Carretta}, E. 2004, A\&ARr, 42, 385

\bibitem[{{Gratton} {et~al.}(2012){Gratton}, {Carretta}, \&
  {Bragaglia}}]{Gra12}
{Gratton}, R.~G., {Carretta}, E., \& {Bragaglia}, A. 2012, A\&ARr, 20, 50

\bibitem[{{Harris}(1996)}]{H96}
{Harris}, W.~E. 1996, \aj, 112, 1487

\bibitem[{{Hesser} {et~al.}(1977){Hesser}, {Hartwick}, \& {McClure}}]{HHM77}
{Hesser}, J.~E., {Hartwick}, F.~D.~A., \& {McClure}, R.~D. 1977, \apjs, 33, 471

\bibitem[{{Hong} {et~al.}(2017){Hong}, {de Grijs}, {Askar}, {Berczik}, {Li},
  {Wang}, {Deng}, {Kouwenhoven}, {Giersz}, \& {Spurzem}}]{HDG17}
{Hong}, J., {de Grijs}, R., {Askar}, A., {et~al.} 2017, \mnras, 472, 67

\bibitem[{{Hughes} \& {Wallerstein}(2000)}]{Hu00}
{Hughes}, J. \& {Wallerstein}, G. 2000, \aj, 119, 1225

\bibitem[{{Johnson} {et~al.}(2015){Johnson}, {Rich}, {Pilachowski}, {Caldwell},
  {Mateo}, {Bailey}, \& {Crane}}]{Jo15}
{Johnson}, C.~I., {Rich}, R.~M., {Pilachowski}, C.~A., {et~al.} 2015, \aj, 150,
  63

\bibitem[{{Katz}(1991)}]{Ka91}
{Katz}, N. 1991, \apj, 368, 325

\bibitem[{{Kayser} {et~al.}(2008){Kayser}, {Hilker}, {Grebel}, \&
  {Willemsen}}]{Kay08}
{Kayser}, A., {Hilker}, M., {Grebel}, E.~K., \& {Willemsen}, P.~G. 2008, A\&A,
  486, 437

\bibitem[{{Khoperskov} {et~al.}(2018){Khoperskov}, {Mastrobuono-Battisti}, {Di
  Matteo}, \& {Haywood}}]{K18}
{Khoperskov}, S., {Mastrobuono-Battisti}, A., {Di Matteo}, P., \& {Haywood}, M.
  2018, ArXiv e-prints [\eprint[arXiv]{1809.04350}]

\bibitem[{{Khoperskov} {et~al.}(2014){Khoperskov}, {Vasiliev}, \&
  {Lubimov}}]{Kho14}
{Khoperskov}, S.~A., {Vasiliev}, E.~O.and~{Khoperskov}, A.~V., \& {Lubimov},
  V.~N. 2014, Journal of Physics: Conference Series, 510, conference 1

\bibitem[{{King}(1966)}]{King66}
{King}, I.~R. 1966, AJ, 71, 64

\bibitem[{{Kruijssen}(2015)}]{K15}
{Kruijssen}, J.~M.~D. 2015, \mnras, 454, 1658

\bibitem[{{Lardo} {et~al.}(2013){Lardo}, {Pancino}, {Mucciarelli},
  {Bellazzini}, {Rejkuba}, {Marinoni}, {Cocozza}, {Altavilla}, \&
  {Ragaini}}]{La13}
{Lardo}, C., {Pancino}, E., {Mucciarelli}, A., {et~al.} 2013, \mnras, 433, 1941

\bibitem[{{Lee}(2015)}]{Lee15}
{Lee}, J.-W. 2015, \apjs, 219, 7

\bibitem[{{Lee}(2016)}]{Lee16}
{Lee}, J.-W. 2016, \apjs, 226, 16

\bibitem[{{Leigh} {et~al.}(2014){Leigh}, {Mastrobuono-Battisti}, {Perets}, \&
  {B{\"o}ker}}]{Le14}
{Leigh}, N.~W.~C., {Mastrobuono-Battisti}, A., {Perets}, H.~B., \& {B{\"o}ker},
  T. 2014, \mnras, 441, 919

\bibitem[{{Makino} {et~al.}(1991){Makino}, {Akiyama}, \& {Sugimoto}}]{Mak91}
{Makino}, J., {Akiyama}, K., \& {Sugimoto}, D. 1991, \apss, 185, 63

\bibitem[{{Marino} {et~al.}(2015){Marino}, {Milone}, {Karakas}, {Casagrande},
  {Yong}, {Shingles}, {Da Costa}, {Norris}, {Stetson}, {Lind}, {Asplund},
  {Collet}, {Jerjen}, {Sbordone}, {Aparicio}, \& {Cassisi}}]{Mar15}
{Marino}, A.~F., {Milone}, A.~P., {Karakas}, A.~I., {et~al.} 2015, MNRAS, 450,
  815

\bibitem[{{Marino} {et~al.}(2013){Marino}, {Milone}, \& {Lind}}]{Ma13}
{Marino}, A.~F., {Milone}, A.~P., \& {Lind}, K. 2013, \apj, 768, 27

\bibitem[{{Marino} {et~al.}(2009){Marino}, {Milone}, {Piotto}, {Villanova},
  {Bedin}, {Bellini}, \& {Renzini}}]{Ma09}
{Marino}, A.~F., {Milone}, A.~P., {Piotto}, G., {et~al.} 2009, \aap, 505, 1099

\bibitem[{{Marino} {et~al.}(2012){Marino}, {Milone}, {Sneden}, {Bergemann},
  {Kraft}, {Wallerstein}, {Cassisi}, {Aparicio}, {Asplund}, {Bedin}, {Hilker},
  {Lind}, {Momany}, {Piotto}, {Roederer}, {Stetson}, \& {Zoccali}}]{Ma12}
{Marino}, A.~F., {Milone}, A.~P., {Sneden}, C., {et~al.} 2012, \aap, 541, A15

\bibitem[{{Marino} {et~al.}(2011){Marino}, {Sneden}, {Kraft}, {Wallerstein},
  {Norris}, {da Costa}, {Milone}, {Ivans}, {Gonzalez}, {Fulbright}, {Hilker},
  {Piotto}, {Zoccali}, \& {Stetson}}]{Ma11}
{Marino}, A.~F., {Sneden}, C., {Kraft}, R.~P., {et~al.} 2011, \aap, 532, A8

\bibitem[{{Marino} {et~al.}(2018){Marino}, {Yong}, {Milone}, {Piotto},
  {Lundquist}, {Bedin}, {Chen{\'e}}, {Da Costa}, {Asplund}, \& {Jerjen}}]{Ma18}
{Marino}, A.~F., {Yong}, D., {Milone}, A.~P., {et~al.} 2018, \apj, 859, 81

\bibitem[{{Massari} {et~al.}(2014){Massari}, {Mucciarelli}, {Ferraro},
  {Origlia}, {Rich}, {Lanzoni}, {Dalessandro}, {Valenti}, {Ibata}, {Lovisi},
  {Bellazzini}, \& {Reitzel}}]{MMF14}
{Massari}, D., {Mucciarelli}, A., {Ferraro}, F.~R., {et~al.} 2014, \apj, 795,
  22

\bibitem[{{Mastrobuono-Battisti} {et~al.}(2012){Mastrobuono-Battisti}, {Di
  Matteo}, {Montuori}, \& {Haywood}}]{Mas12}
{Mastrobuono-Battisti}, A., {Di Matteo}, P., {Montuori}, M., \& {Haywood}, M.
  2012, \aap, 546, L7

\bibitem[{{Meylan}(1987)}]{Me87}
{Meylan}, G. 1987, \aap, 184, 144

\bibitem[{{Milone} {et~al.}(2013){Milone}, {Marino}, {Piotto}, {Bedin},
  {Anderson}, {Aparicio}, {Bellini}, {Cassisi}, {D'Antona}, {Grundahl},
  {Monelli}, \& {Yong}}]{Mil13}
{Milone}, A.~P., {Marino}, A.~F., {Piotto}, G., {et~al.} 2013, \apj, 767, 120

\bibitem[{{Milone} {et~al.}(2012){Milone}, {Marino}, {Piotto}, {Bedin},
  {Anderson}, {Aparicio}, {Cassisi}, \& {Rich}}]{Mil12}
{Milone}, A.~P., {Marino}, A.~F., {Piotto}, G., {et~al.} 2012, \apj, 745, 27

\bibitem[{{Milone} {et~al.}(2015){Milone}, {Marino}, {Piotto}, {Bedin},
  {Anderson}, {Renzini}, {King}, {Bellini}, {Brown}, {Cassisi}, {D'Antona},
  {Jerjen}, {Nardiello}, {Salaris}, {Marel}, {Vesperini}, {Yong}, {Aparicio},
  {Sarajedini}, \& {Zoccali}}]{Mil15}
{Milone}, A.~P., {Marino}, A.~F., {Piotto}, G., {et~al.} 2015, \mnras, 447, 927

\bibitem[{{Milone} {et~al.}(2010){Milone}, {Piotto}, {King}, {Bedin},
  {Anderson}, {Marino}, {Momany}, {Malavolta}, \& {Villanova}}]{Mil10}
{Milone}, A.~P., {Piotto}, G., {King}, I.~R., {et~al.} 2010, \apj, 709, 1183

\bibitem[{{Milone} {et~al.}(2009){Milone}, {Stetson}, {Piotto}, {Bedin},
  {Anderson}, {Cassisi}, \& {Salaris}}]{Mi09}
{Milone}, A.~P., {Stetson}, P.~B., {Piotto}, G., {et~al.} 2009, \aap, 503, 755

\bibitem[{{Mucciarelli} {et~al.}(2015){Mucciarelli}, {Lapenna}, {Massari},
  {Pancino}, {Stetson}, {Ferraro}, {Lanzoni}, \& {Lardo}}]{Muc15}
{Mucciarelli}, A., {Lapenna}, E., {Massari}, D., {et~al.} 2015, ApJ, 809, 128

\bibitem[{{Nataf} {et~al.}(2013){Nataf}, {Gould}, {Pinsonneault}, \&
  {Udalski}}]{Na13}
{Nataf}, D.~M., {Gould}, A.~P., {Pinsonneault}, M.~H., \& {Udalski}, A. 2013,
  \apj, 766, 77

\bibitem[{{Norris} \& {Da Costa}(1995)}]{Nor95}
{Norris}, J.~E. \& {Da Costa}, G.~S. 1995, ApJ, 447, 680

\bibitem[{{Pancino} {et~al.}(2010){Pancino}, {Rejkuba}, {Zoccali}, \&
  {Carrera}}]{Pan10}
{Pancino}, E., {Rejkuba}, M., {Zoccali}, M., \& {Carrera}, R. 2010, \aap, 524,
  A44

\bibitem[{{Perets} {et~al.}(2018){Perets}, {Mastrobuono-Battisti}, {Meiron}, \&
  {Gualandris}}]{Pe18}
{Perets}, H.~B., {Mastrobuono-Battisti}, A., {Meiron}, Y., \& {Gualandris}, A.
  2018, ArXiv e-prints [\eprint[arXiv]{1802.00012}]

\bibitem[{{Piotto} {et~al.}(2012){Piotto}, {Milone}, {Anderson}, {Bedin},
  {Bellini}, {Cassisi}, {Marino}, {Aparicio}, \& {Nascimbeni}}]{Pi12}
{Piotto}, G., {Milone}, A.~P., {Anderson}, J., {et~al.} 2012, \apj, 760, 39

\bibitem[{{Plummer}(1911)}]{P11}
{Plummer}, H.~C. 1911, MNRAS, 71, 460

\bibitem[{{Portegies Zwart} \& {Rusli}(2007)}]{Po07}
{Portegies Zwart}, S.~F. \& {Rusli}, S.~P. 2007, \mnras, 374, 931

\bibitem[{{Pouliasis} {et~al.}(2017){Pouliasis}, {Di Matteo}, \&
  {Haywood}}]{PDMH17}
{Pouliasis}, E., {Di Matteo}, P., \& {Haywood}, M. 2017, \aap, 598, A66

\bibitem[{{Priyatikanto} {et~al.}(2016){Priyatikanto}, {Kouwenhoven},
  {Arifyanto}, {Wulandari}, \& {Siregar}}]{Pr16}
{Priyatikanto}, R., {Kouwenhoven}, M.~B.~N., {Arifyanto}, M.~I., {Wulandari},
  H.~R.~T., \& {Siregar}, S. 2016, \mnras, 457, 1339

\bibitem[{{Renaud} {et~al.}(2017){Renaud}, {Agertz}, \& {Gieles}}]{Re17}
{Renaud}, F., {Agertz}, O., \& {Gieles}, M. 2017, \mnras, 465, 3622

\bibitem[{{Rodionov} {et~al.}(2009){Rodionov}, {Athanassoula}, \&
  {Sotnikova}}]{Ro09}
{Rodionov}, S.~A., {Athanassoula}, E., \& {Sotnikova}, N.~Y. 2009, \mnras, 392,
  904

\bibitem[{{Sarajedini} \& {Layden}(1995)}]{Sa95}
{Sarajedini}, A. \& {Layden}, A.~C. 1995, \aj, 109, 1086

\bibitem[{{Saviane} {et~al.}(2012){Saviane}, {da Costa}, {Held}, {Sommariva},
  {Gullieuszik}, {Barbuy}, \& {Ortolani}}]{Sa12}
{Saviane}, I., {da Costa}, G.~S., {Held}, E.~V., {et~al.} 2012, \aap, 540, A27

\bibitem[{{Sollima} {et~al.}(2012){Sollima}, {Nipoti}, {Mastrobuono Battisti},
  {Montuori}, \& {Capuzzo-Dolcetta}}]{So12}
{Sollima}, A., {Nipoti}, C., {Mastrobuono Battisti}, A., {Montuori}, M., \&
  {Capuzzo-Dolcetta}, R. 2012, \apj, 744, 196

\bibitem[{{van den Bergh}(1996)}]{VdB96}
{van den Bergh}, S. 1996, \apjl, 471, L31

\bibitem[{{Ventura} {et~al.}(2001){Ventura}, {D'Antona}, {Mazzitelli}, \&
  {Gratton}}]{Ve01}
{Ventura}, P., {D'Antona}, F., {Mazzitelli}, I., \& {Gratton}, R. 2001, \apj,
  550, L65

\bibitem[{{Yong} \& {Grundahl}(2008)}]{Yong08}
{Yong}, D. \& {Grundahl}, F. 2008, \apjl, 672, L29

\end{thebibliography}

\begin{appendix}
\section{Velocity maps}\label{AAp}
\begin{figure*}
\centering
\includegraphics[width=0.3\textwidth]{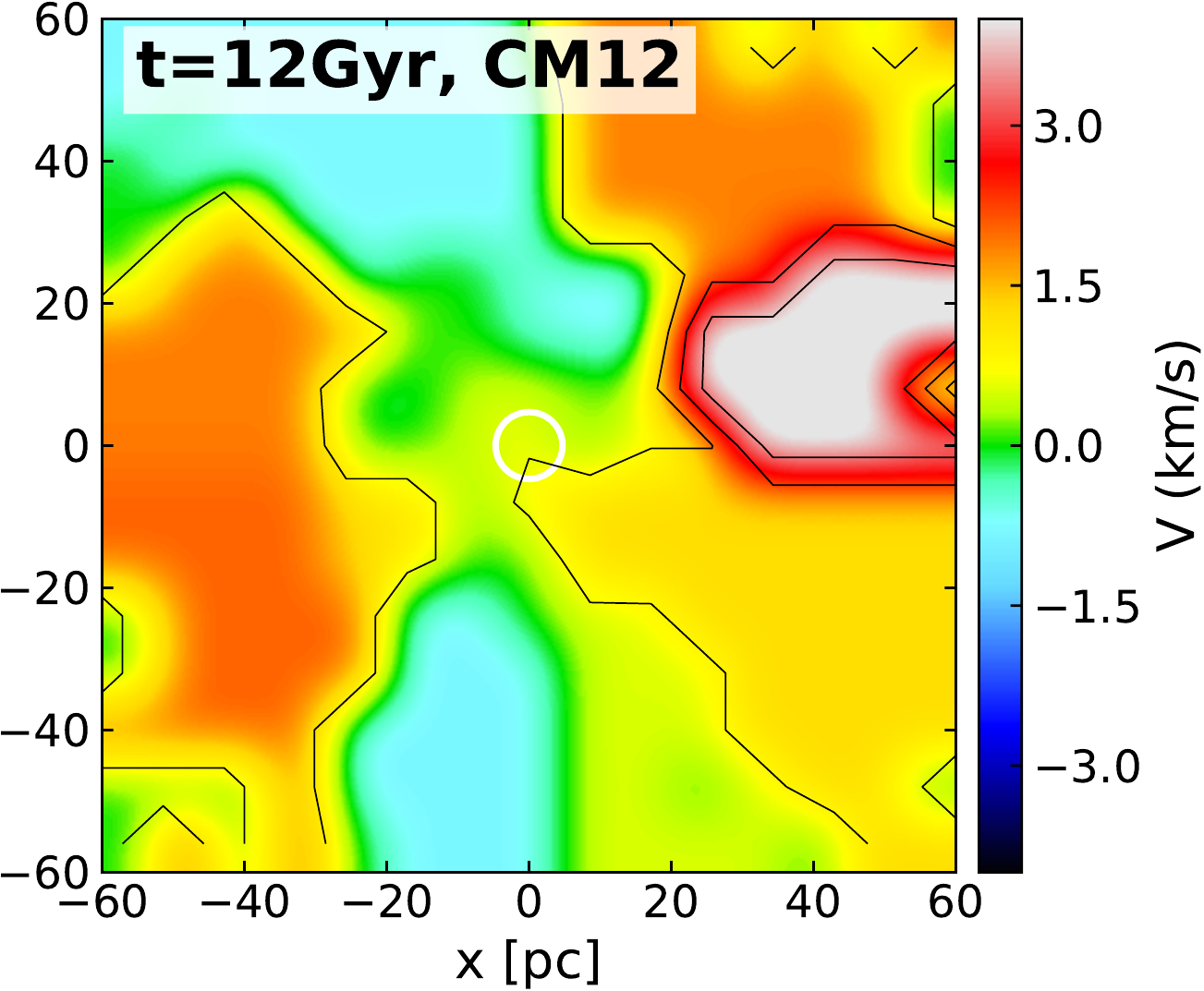}
\includegraphics[width=0.3\textwidth]{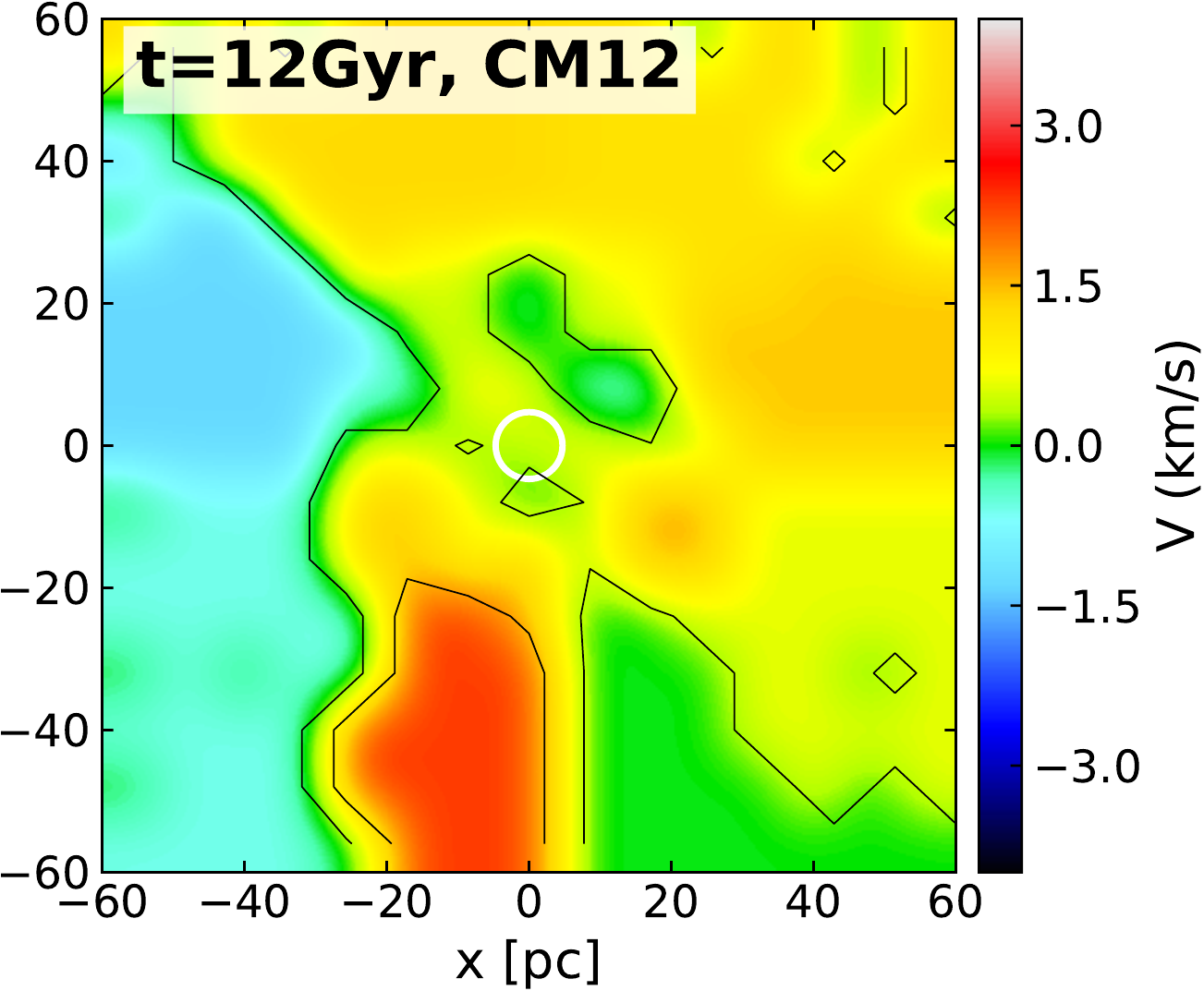}
\includegraphics[width=0.3\textwidth]{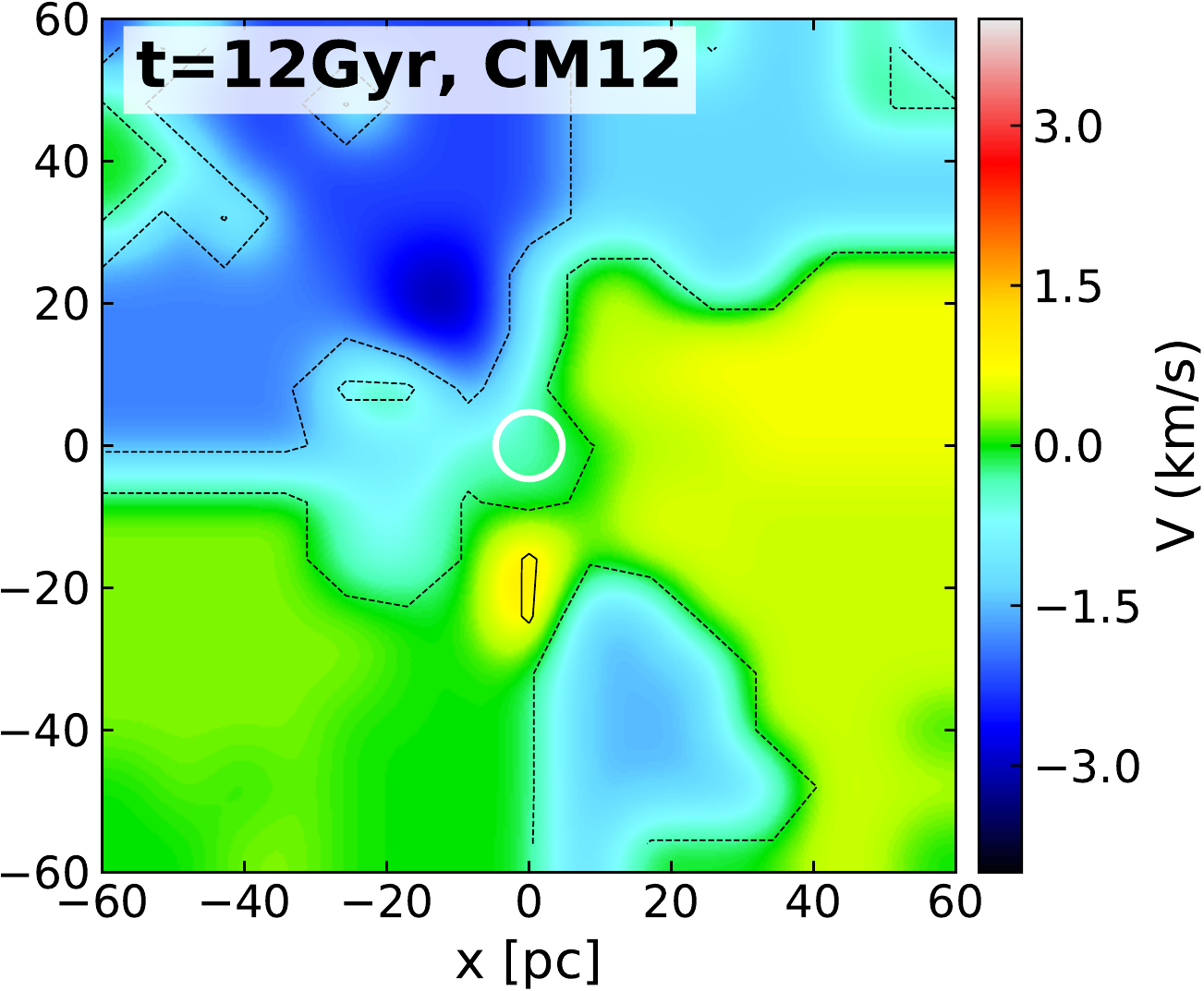}
\caption{Velocity maps of CM12, randomly oriented with respect to the LOS (see text for details). The observed rotation signal depends on the observation angle.}\label{fig:appa}
\end{figure*}

We present here the velocity maps of CM12 as seen from random LOS directions.
In the three panels of Figure \ref{fig:appa}the observation direction has been chosen randomly inclining the total angular moment of the cluster .
We present three cases with projection angles ($\phi$, $\theta$) respectively equal to (101, 28) deg, (360, 101) deg and (12.2, 5.9) deg.
The observed signal depends on the LOS direction, however, if observed, the rotation is always misaligned with respect to the total angular momentum.

\end{appendix}

% Don't change these lines
%\bsp	% typesetting comment
%\label{lastpage}
\end{document}